\documentclass[numberedappendix]{emulateapj}

\usepackage{amsmath}
\usepackage{amssymb}
\usepackage{txfonts}
\usepackage{natbib}
\usepackage{graphicx}

\usepackage{color} 

\shorttitle{SASI activity as key to successful neutrino-driven SN explosions?}

\shortauthors{Hanke, Marek, M\"uller, \& Janka}

\begin{document}

\title{Is strong SASI activity the key to successful neutrino-driven supernova
explosions?}

\author{Florian Hanke,
        Andreas Marek,
        Bernhard M\"uller, and
        Hans-Thomas Janka
       }
\affil{Max-Planck-Institut f\"ur Astrophysik,
       Karl-Schwarzschild-Str. 1, D-85748 Garching, Germany;
       fhanke@mpa-garching.mpg.de, thj@mpa-garching.mpg.de} 




\begin{abstract}
Following a simulation approach of recent publications
we explore the viability of the neutrino-heating explosion
mechanism in dependence on the spatial dimension.
Our results disagree with
previous findings. While we also observe that two-dimensional
(2D) models explode for lower driving neutrino luminosity than
spherically symmetric (1D) models, we do not find that explosions
in 3D occur easier and earlier than in 2D. Moreover,
we find that the average entropy of matter in the gain layer
hardly depends on the dimension and thus is no good diagnostic
quantity for the readiness to explode. Instead, mass, integrated
entropy, total neutrino-heating rate, and nonradial kinetic energy
in the gain layer are higher when models are closer
to explosion. Coherent, large-scale mass motions as
typically associated with the standing accretion-shock instability
(SASI) are observed to be supportive for explosions because they
drive strong shock expansion and thus enlarge the gain layer.
While 2D models with better angular resolution explode clearly
more easily, the opposite trend is seen in 3D. We interpret this
as a consequence of the turbulent energy cascade, which transports
energy from small to large spatial scales in 2D, thus fostering
SASI activity. In contrast, the energy flow in 3D is in the opposite
direction, feeding fragmentation and vortex motions on smaller
scales and thus making the 3D evolution with finer grid resolution
more similar to 1D.
More favorable conditions for explosions in 3D may therefore be
tightly linked to efficient growth of low-order SASI modes
including nonaxisymmetric ones.
\end{abstract}


\section{Introduction}
Recent simulations in two dimensions (2D) with sophisticated neutrino transport
have demonstrated that the neutrino-driven mechanism, supported by hydrodynamic
instabilities in the postshock layer, can yield supernova explosions at least
for some progenitor stars (11.2 and 15\,$M_\odot$ ones in 
\citealp{Marek2009}; \citealp{Mueller2012}). The
explosions occur relatively late after bounce and tend to have fairly
low energy, being ``marginal'' or only slightly above the ``critical 
threshold'' in this sense. 
\cite{Suwa2010} obtained a similar explosion for a 13\,$M_\odot$
progenitor in their axisymmetric simulations. However, the Oak Ridge group
has found stronger and earlier explosions for a wider range of
progenitors \citep{Bruenn2009}, while in purely Newtonian simulations
with multi-dimensional neutrino diffusion (including energy dependence
but without energy-bin coupling) \cite{Burrows2006,Burrows2007}
could not see any success of the delayed neutrino-driven mechanism.

While the reason for the discrepant results of these simulations cannot be 
satisfactorily understood on the basis of published results, the marginality
of the 2D explosions of the Garching group and the lack of neutrino-driven
explosions in the simulations by \cite{Burrows2006,Burrows2007}
raises the important question about the influence of the third spatial
dimension on the post-bounce evolution of collapsing stellar cores.
Three-dimensional (3D) fluid dynamics with their inverse turbulent energy
cascade compared to the 2D case are likely to change the flow pattern on
large scales as well as small scales. They could have an influence on the
existence and the growth rate of nonradial hydrodynamic instabilities 
in the supernova core even in the absence of stellar rotation (see, e.g., 
\citealp{Iwakami2008}) but in particular with a moderate amount of 
angular momentum in the progenitor star (e.g.,
\citealp{Blondin2006a,Iwakami2009,Fernandez2010}), 
and thus could lead to differences in the
hydrodynamic and thermodynamic conditions for the operation of the
neutrino-heating mechanism. In particular, 3D flows might cause important 
changes of the dwell time of postshock matter in the layer where
neutrinos deposit energy, which is a crucial aspect deciding about the 
viability and efficiency of the neutrino-driven supernova mechanism
(some aspects of this were discussed by \citealp{Murphy2008} and 
\citealp{Marek2009}).

Indeed, employing a simplified treatment of neutrino effects by
including local neutrino-cooling and heating terms for a chosen
value of the neutrino luminosity and spectral temperature instead
of solving the computationally intense neutrino transport,
\cite{Nordhaus2010} found considerably easier and earlier explosions
in 3D than in 2D. In the context of the concept of a critical 
value of the neutrino luminosity that (for a given mass accretion
rate onto the stalled supernova shock) has to be exceeded to obtain 
neutrino-driven explosions
(\citealp{Burrows1993,Janka1996,Janka2001,Yamasaki2005,Murphy2008,Pejcha2011,Fernandez2012}),
they quantified the improvement of 3D relative to 2D by 
a 15--25\% reduction of the critical luminosity value.
In particular, they observed that 3D postshock
convection leads to higher average entropies in the 
neutrino-heating layer, thus improving the conditions for shock revival
due to a significant stretching of the residence time of matter in the
layer where it gains energy from neutrinos. Very recently, 
\cite{Takiwaki2011} reported enhanced maximum dwell times of a small
fraction of the material in the gain region in a 3D simulation 
compared to the 2D case of an 11.2\,$M_\odot$
star, but could not unambiguously link this 
effect to an easier explosion of the 3D model. In particular, their
3D simulation showed a shock expansion that was more delayed than 
in the 2D run, and the 3D conditions did not appear more favorable 
for an explosion with respect to
a variety of quantities like the net heating rate, the heating 
timescale or the profiles of maximum and minimum entropies.

In this paper we present a comparative investigation for 11.2\,$M_\odot$ and
15\,$M_\odot$ progenitors in one, two, and three dimensions along the lines of
the study by \cite{Nordhaus2010}, varying the driving neutrino luminosities
used in time-dependent collapse simulations of the two stars. While our
results for spherically symmetric (1D) and 2D models basically confirm 
the dimension-dependent differences
found by \cite{Murphy2008} and \cite{Nordhaus2010}, our calculations
do neither exhibit a strict 1D-2D-3D hierarchy of the average entropy 
in the gain
layer, nor do they show any clear signs that 3D effects facilitate the
development of the explosion better than nonradial motions in 2D.
Attempting to understand the reason for this puzzling finding we vary the
resolution of the spherical coordinate grid used for our 2D and 3D simulations.
The outcome of these studies reflects the action of the energy flow within the
turbulent energy cascade. The latter transports the driving energy provided by
neutrino heating and gravitational energy release in the accretion flow
from small to large scales in 2D and opposite in 3D. Models in 2D show
growing large-scale asymmetry and quasi-periodic time variability and explode
clearly easier with higher resolution, whereas in 3D better resolved models are
observed to become more similar to the 1D case and thus to be farther away
from an explosion. This suggests that the success of the neutrino-driven
mechanism could be tightly linked to the initiation of strong non-radial 
mass motions in the neutrino-heated postshock layer on the largest possible 
scales, implying that the easier explosions of our 2D models with higher
resolution are a consequence of more violent activity due to the standing
accretion-shock instability (SASI; \citealp{Blondin2003}), whereas the 
better resolved 3D models 
for the employed artificial setup of supernova-core conditions
tend to reveal considerably reduced amplitudes of low-order
spherical-harmonics modes of nonradial deformation 
and thus behave more similar to the 1D case.

The paper is structured as follows. In Sect.~\ref{sec:num} we briefly
describe our numerics and implementation of neutrino source terms. In
Sect.~\ref{sec:models} we give an overview of the simulations presented in this
paper. Our investigations of the dependence of the critical luminosity on the
spatial dimension will be presented in Sect.~\ref{sec:crit_lum} and results
of resolution studies in Sect.~\ref{sec:high_res}. An interpretation of our
findings will follow in Sect.~\ref{sec:cond}. Section~\ref{sec:con} contains
the summary and conclusions. In App.~\ref{sec:appendix} we
present 1D simulations that document our efforts to reproduce
the results of previous, similar studies in the literature by 
straightforwardly applying the neutrino treatment described in 
these works.

\section{Numerical setup} 
\label{sec:num}

We solve the equations of hydrodynamics reflecting the conservation of
mass, momentum, and energy,
\begin{equation}
\label{eq:mas_cons}
\frac{\partial \rho}{\partial t}+\mathbf{\nabla}\cdot (\rho\mathbf{v})=0\, ,
\end{equation}
\begin{equation}
\label{eq:imp_cons}
\frac{\partial \rho \mathbf{v}}{\partial t}+
 \mathbf{\nabla} \cdot (\rho \mathbf{v} \otimes \mathbf{v}) +
 \mathbf{\nabla} P = -\rho \mathbf{\nabla} \Phi \, ,
\end{equation}
\begin{equation}
\label{eq:ene_cons}
\frac{\partial e}{\partial t}+
  \mathbf{\nabla} \cdot \left[\left(e + P\right) \mathbf{v} \right] =
  -\rho \mathbf{v} \cdot \mathbf{\nabla} \Phi + 
  \rho \left(Q_\nu^+ - Q_\nu ^-\right) \, ,
\end{equation}
where $\rho$ is the mass density, $\mathbf{v}$ the fluid velocity,
$\Phi$ the gravitational potential, $P$ the pressure, and $e$ the
total (internal+kinetic) fluid energy density.
These equations are integrated in a conservative form (for which reason
the energy equation is solved for the total energy density)
using the explicit, finite-volume,
higher-order Eulerian, multi-fluid hydrodynamics code {\sc Prometheus}
\citep{Fryxell1991,Mueller1991a,Mueller1991b}. It is a direct implementation of
the Piecewise Parabolic Method (PPM) of \cite{Collela1984} using the Riemann
solver for real gases developed by \cite{Collela1985} and the directional
splitting approach of \cite{Strang1968} to treat the multi-dimensional problem.
In order to prevent odd-even coupling \citep{Quirk1994} we switch from the
original PPM method to the HLLE solver of \cite{Einfeldt1988} in the vicinity
of strong shocks. The advection of nuclear species is treated by the Consistent
Multi-fluid Advection (CMA) scheme of \cite{Plewa1999}.

To facilitate comparison with \cite{Nordhaus2010} we also employ the
high-density equation of state (EoS) of \cite{Shen1998} and do not include
general relativistic corrections. We use the monopole approximation of the
Poisson equation to treat Newtonian self-gravity.

To make our extensive parameter study possible, 
we use the local source terms applied
by \cite{Murphy2008} and \cite{Nordhaus2010} instead of detailed neutrino
transport (see \citealp{Janka2001} for a derivation of these source terms). In
this approach the neutrino heating and cooling rates $Q^{+}_{\nu}$ and
$Q^{-}_{\nu}$ are given by
\begin{eqnarray}
\label{eq:heat}
Q^{+}_{\nu} & = & 
1.544\cdot10^{20} \left(\frac{L_{\nu_{e}}}{10^{52} \,\text{erg s}^{-1}}\right)
\left(\frac{T_{\nu_{e}}}{4\,\text{MeV}}\right)^{2} \\
& & \left(\frac{100\,\text{km}}{r}\right)^2 
\left( Y_{n}+Y_{p}\right) \, e^{-\tau_{\textrm{eff}}}
\left[\frac{\text{erg}}{\text{g s}}\right] \, , \notag
\end{eqnarray}
\begin{equation}
\label{eq:cool}
Q^{-}_{\nu} = 1.399\cdot10^{20} \left(\frac{T}{2\,\text{MeV}}\right)^{6} 
\left( Y_{n}+Y_{p}\right) \, e^{-\tau_{\textrm{eff}}}
\left[\frac{\text{erg}}{\text{g s}}\right] \, .
\end{equation}
These approximations depend on local quantities, namely the density $\rho$,
the temperature $T$, the distance from the center of the star $r$, and the
neutron and proton number fractions $Y_{n}$ and $Y_{p}$,
respectively.
In Eq.~(\ref{eq:heat}) the electron-neutrino luminosity $L_{\nu_{e}}$ is a
parameter and is assumed to be equal to the electron antineutrino
luminosity $L_{\bar{\nu}_{e}} = L_{\nu_{e}}$. The
neutrino temperature $T_{\nu_{e}}$ is set to 4$\,$MeV.

The employed source terms, Eqs.~(\ref{eq:heat}) and (\ref{eq:cool}), without
the factors $e^{-\tau_{\textrm{eff}}}$ are valid for optically thin regions
only. In order to model the transition to neutrino trapping at high optical
depths, we follow \cite{Nordhaus2010} and multiply the heating and cooling terms
by $e^{-\tau_{\textrm{eff}}}$ to suppress them in the inner opaque core of the
proto-neutron star. Here, the optical depth for electron neutrinos and
antineutrinos is defined as
\begin{equation}
\label{eq:depth}
\tau_{\textrm{eff}}(r) = \int_{r}^{\infty} 
\kappa_{\textrm{eff}}(r') \, \mathrm{d}r' \, .
\end{equation}
The effective opacity $\kappa_{\textrm{eff}}$ was derived by
\cite{Janka2001} and given in \cite{Murphy2009}:
\begin{equation}
\label{eq:opa}
\kappa_{\textrm{eff}} \approx 1.5\cdot10^{-7}\cdot X_{n,p}
\left(\frac{\rho}{10^{10}\,\text{g cm}^{-3}}\right)\left(\frac{T_{\nu_{e}}}{4
\,\text{MeV}}\right)^{2} \text{cm}^{-1},
\end{equation}
where $X_{n,p}=\frac{1}{2}\left(Y_n + Y_p\right)$ accounts for composition
averaging. In multi-dimensional simulations we evaluate the radial 
integration for the optical depth $\tau_{\textrm{eff}}$ independently
for each latitude $\theta$ in 2D and each pair of latitudinal and azimuthal
angles ($\theta$,$\phi$) in 3D.
Note that in \cite{Murphy2008} the exponential suppression factor is
absent in the heating and cooling terms (or was not mentioned), 
which otherwise agree with
ours, while no definition of the effective optical depth $\tau_{\textrm{eff}}$
is given in \cite{Nordhaus2010}. 
The factor $\left( Y_{n}+Y_{p}\right)$ in Eqs.~(\ref{eq:heat}) and
(\ref{eq:cool}) is included in \cite{Nordhaus2010}, but not
in \cite{Murphy2008} and \cite{Murphy2009}.

The time period from the onset of the collapse until 15\,ms after bounce was
tracked with the {\sc Prometheus-Vertex} code \citep{Rampp2002} in 1D
including its detailed, multi-group neutrino transport, all relevant
neutrino reactions and a ``flashing treatment'' for an approximative 
description of nuclear burning during infall. This means that
until 15\,ms after bounce we describe neutrino effects
including the evolution of the electron fraction $Y_e$ with high
sophistication. At 15\,ms after core
bounce we switch to the simple neutrino heating and cooling terms and
upon mapping from 1D to 2D also impose (on the whole computational grid)
random zone-to-zone seed perturbations with an amplitude of 1\% of the
density to break spherical symmetry.

Although during the subsequent evolution we apply the heating and cooling 
expressions
of Eqs.~(\ref{eq:heat}) and (\ref{eq:cool}) following \cite{Nordhaus2010} and
\cite{Murphy2008}, we refrain from adopting their treatment of changes of
the electron fraction $Y_e$. Following a suggestion by \cite{Liebendorfer2005},
they prescribed $Y_e$ simply as a function of density, $Y_e(\rho)$, instead of
solving a rate equation with source terms for electron neutrino and antineutrino
production and destruction consistently with the expressions employed for
neutrino heating and cooling. \cite{Liebendorfer2005} found that such a
parametrization, supplemented by a corresponding entropy source term (and a
neutrino pressure term in the equation of motion), yields results in good
agreement with 1D simulations including neutrino transport during the collapse
phase until the moment of bounce-shock formation. A universal
$Y_e$-$\rho$-relation, which serves as the basis of this approximation and
can be inferred for the infalling matter during the homologous collapse phase,
however, applies neither for the evolution of the shocked accretion flow in the
post-bounce phase nor for expanding neutrino-heated gas (see, e.g., Fig.~4.9
in \citealp{Thielemann2011}). For example, a comparison with supernova models 
with detailed neutrino transport shows that the $Y_e$-$\rho$-relation fitted to
the homologous phase overestimates the deleptonization of the gas in most of
the gain layer but underestimates the lepton loss of matter in the cooling
layer and neutrinospheric region. Moreover, the question arises 
how the lepton evolution shall be treated in matter that reexpands and thus
moves from high to low density?
Even more, the entropy source term introduced in Eq.~(5) of
\cite{Liebendorfer2005} is designed to specifically account for gas-entropy
changes due to neutrino production by electron captures and subsequent energy
transfers in (neutrino-electron) scatterings. The corresponding energy loss or
gain rate of the medium through the escape or capture of electron neutrinos with
mean energy $E_{\nu_e}$, which is given by $\delta Q_{\nu_e}/\delta t =
E_{\nu_e}\,\delta Y_e/\delta t$, is not included in the heating and cooling
terms $Q_\nu^+$ and $Q_\nu^-$ of Eqs.~(\ref{eq:heat},\ref{eq:cool}) of the 
present work. Adding it as an
extra term would imply partial double-counting of the energy exchange via
electron neutrinos, and omitting it means to overestimate the entropy increase
in infalling, deleptonizing matter and to underestimate entropy gains of 
decompressed gas with growing $Y_e$.
Because of the long list of such inconsistencies, whose implications are
hard to judge or control, we decided to ignore $Y_e$ changes of the stellar
medium in our post-bounce simulations 
(except for the models discussed in App.~\ref{sec:appendix}).

This choice can be justified, but it is certainly not a perfect approach
because it may also exclude effects of importance in real supernova cores, whose
physical processes require the inclusion of neutrino transport for an accurate
description of the energy and lepton-number evolution. 
One of the undesired consequences of keeping $Y_e$ fixed in the
accretion flow after core bounce is an overestimation of the electron pressure
in the gas settling onto the forming neutron star. In order to enforce more
compression and thus to ensure close similarity of our results to the 1D models
studied by \cite{Nordhaus2010} and \cite{Murphy2008}, we have to enhance the net
cooling of the accreted matter by reducing the effective opacity
$\kappa_{\mathrm{eff}}$ by a factor of 2.7 compared to the value given in
\cite{Murphy2009} and in Eq.~(7). This reduction factor is chosen such that our
simulations reproduce the minimum values of the critical luminosity 
found to be necessary for triggering 
explosions of the 15\,$M_\odot$ progenitor ($2.6\cdot
10^{52}$\,erg\,s$^{-1}$) and for the 11.2\,$M_\odot$ star ($1.3\cdot
10^{52}$\,erg\,s$^{-1}$) in the 1D simulations of \cite{Murphy2008}. Without
the reduction factor of $\kappa_{\mathrm{eff}}$, our models turn out to explode
too easily because of weak cooling (see the results in 
App.~\ref{sec:appendix}). We stress that any exponential suppression
factor of the heating and cooling rates in Eqs.~(\ref{eq:heat},\ref{eq:cool})
is a pragmatic and ad hoc procedure to bridge the transition from the optically
thin to the optically thick regime, where neutrino transport is most 
complicated. From transport theory neither the exponential factor nor the
exact definition of the optical depth of the exponent can be rigorously
derived.

In our reference set of standard simulations, we employ 400 non-equidistant 
radial zones, which are distributed from the center to an outer boundary
at 9000\,km. The latter is sufficiently far out to ensure that the gas there
remains at rest during the simulated evolution periods. 
The radial zones are chosen such that the
resolution $\Delta r/r$ is typically better than 2\%.
In the multi-dimensional models with standard resolution, we employ a
polar grid with an angular bin size of 3$^{\circ}$ (60 $\theta$- and
120 $\phi$-zones). For the high-resolution 1D and multi-dimensional
models of Sects.~\ref{sec:crit_lum} and \ref{sec:high_res} we compute with
up to 800 radial zones and in 2D with an angular resolution down to 
0.5$^{\circ}$, in 3D down to 1.5$^{\circ}$.
The additional radial grid zones are distributed such 
that the region between 20 and 400\,km, i.e., not only the cooling layer
around the neutrinosphere but also the gain layer between 
gain radius and shock, are resolved significantly better. 
While full convergence of the 1D results requires 600 radial
zones or more, most of the plots comparing models for different dimensions
show simulations with our standard resolution of only 400 radial cells
(unless stated otherwise), because this is the limit for which we could 
perform a larger set of 3D runs in acceptable time. To avoid an extremely
restrictive Courant-Friedrich-Lewy (CFL) timestep in our 3D
calculations, we simulate the inner core above a density of
$\rho=10^{12}\,\textrm{g/cm}^3$ in spherical symmetry.

\begin{figure}
\plotone{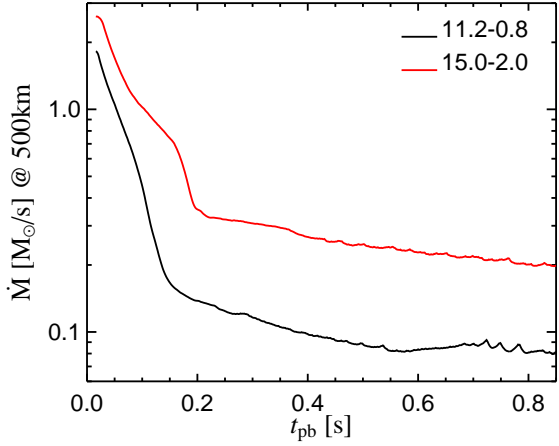}
\caption{Time evolution of the mass accretion rate, $\dot{M}(r) = 4\pi r^2
\rho(r)\left|v(r)\right|$, evaluated at 500\,km for the 11.2\,$M_{\odot}$ 
and the 15\,$M_{\odot}$ progenitors in nonexploding models.}
\label{fig:mdot_time}
\end{figure}

\begin{deluxetable}{ccccccc}
\tablecolumns{7}
\tabletypesize{\scriptsize}
\tablecaption{
11.2\,$M_\odot$ and 15\,$M_\odot$ results with standard grid of 400 radial
zones and 3 degrees angular resolution.
\label{tbl:models_std}
}
\tablewidth{0pt}
\tablehead{
\colhead{} &
\multicolumn{2}{c}{1D} &
\multicolumn{2}{c}{2D} &
\multicolumn{2}{c}{3D} \\
\colhead{$L_{\nu_e}$\tablenotemark{a}} &
\colhead{$t_{\textrm{exp}}$\tablenotemark{b}} &
\colhead{$\dot{M}_{\textrm{exp}}$\tablenotemark{c}} &
\colhead{$t_{\textrm{exp}}$} &
\colhead{$\dot{M}_{\textrm{exp}}$} &
\colhead{$t_{\textrm{exp}}$} &
\colhead{$\dot{M}_{\textrm{exp}}$} \\
\colhead{($10^{52}$ erg/s)} &
\colhead{(ms)} &
\colhead{($M_{\odot}$/s)} &
\colhead{(ms)} &
\colhead{($M_{\odot}$/s)} &
\colhead{(ms)} &
\colhead{($M_{\odot}$/s)} }
\startdata
\cutinhead{11.2\,$M_{\odot}$}
0.8 &  $-$       & $-$     &  $-$       &  $-$    &  $-$       &  $-$   \\
0.9 &  $-$       & $-$     &  $-$       &  $-$    &  731       &  0.085 \\
1.0 &  $-$       & $-$     &  563       &  0.082  &  537       &  0.086 \\
1.1 &  $-$       & $-$     &  461       &  0.091  &            &        \\
1.2 &  $-$       & $-$     &  357       &  0.104  &  319       &  0.112 \\
1.3 &  819       & 0.082   &  307       &  0.114  &            &        \\
1.4 &  499       & 0.088   &  241       &  0.126  &  232       &  0.130 \\
1.5 &  380       & 0.100   &  232       &  0.130  &            &        \\
1.6 &  345       & 0.106   &  203       &  0.137  &            &        \\
\cutinhead{15\,$M_{\odot}$}
2.0 &  $-$       &  $-$    &  $-$       &  $-$    &  $-$       &  $-$   \\
2.1 &  $-$       &  $-$    &  $-$       &  $-$    &  $-$       &  $-$   \\
2.2 &  $-$       &  $-$    &  876       &  0.197  &  612       &  0.226 \\
2.3 &  $-$       &  $-$    &  428       &  0.261  &  426       &  0.261 \\
2.4 &  $-$       &  $-$    &  442       &  0.255  &            &        \\
2.5 &  $-$       &  $-$    &  283       &  0.313  &  281       &  0.314 \\
2.6 &  710       &  0.215  &  285       &  0.312  &            &        \\
2.7 &  489       &  0.247  &  262       &  0.316  &            &        \\
2.8 &  390       &  0.271  &  242       &  0.322  &  236       &  0.324 \\
2.9 &  281       &  0.314  &  235       &  0.325  &            &        \\
3.0 &  258       &  0.318  &  236       &  0.324  &            &        \\
3.1 &  248       &  0.320  &  220       &  0.327  &            &        
\enddata
\tablenotetext{a}{Electron-neutrino luminosity.}
\tablenotetext{b}{Time after bounce of onset of explosion. 
A ``$-$'' symbol indicates 
that the model does not explode during the simulated period of evolution.}
\tablenotetext{c}{Mass accretion rate at onset of explosion.}
\end{deluxetable}

\section{Investigated progenitors and models} 
\label{sec:models}

Our models are based on the 15\,$M_\odot$ progenitor star s15s7b2 of
\cite{Woosley1995} and an 11.2\,$M_\odot$ progenitor of
\cite{Woosley2002}. The calculations for these
progenitors with our standard angular resolution of 3$^{\circ}$ are summarized
in Table~\ref{tbl:models_std}. This table is arranged such that
horizontal rows have the same driving luminosities for simulations performed in
different dimensions. Varying the prescribed driving luminosity $L_{\nu_e}$
from run to run we present for each of the 11.2\,$M_\odot$ and 15\,$M_\odot$
progenitors several 1D, 2D, and 3D simulations. All of the 1D and 2D simulations
cover at least 1s after bounce. The nonexploding 3D simulations with standard
angular resolution of 3 degrees were not stopped until at least 600\,ms
after bounce. For models with higher resolution the simulation times
are given in Table \ref{table:models_res}.

In Table~\ref{tbl:models_std} the time of the onset of an explosion,
$t_\mathrm{exp}$, and the mass accretion rate at that time,
$\dot M_\mathrm{exp}$, are listed as characteristic quantities of
the models. The beginning of the explosion is defined as the moment of
time $t_\mathrm{exp}$ when the shock reaches an average radius of 400\,km 
(and does not return lateron), while nonexploding models are
denoted by a ``$-$'' symbol. In multidimensional simulations the
corresponding shock position is defined as the surface average over all
angular directions, $\langle R_\mathrm{S} \rangle \equiv  \frac{1}{4\pi}
\oint\mathrm{d}\Omega\,R_\mathrm{S}(\vec{\Omega})$. The lowest driving
luminosity yielding an explosion for a given value of the mass accretion
rate is termed the critical luminosity \citep{Burrows1993,Murphy2008}.
We determine the mass accretion rate $\dot{M}(r) = 4\pi
r^2\rho(r) \left|v(r)\right|$ at the time of the onset of the explosion
just exterior to the shock, i.e.\ at a radius of 500\,km, where the infalling
envelope is spherical (except for the small seed perturbations imposed
on the multi-dimensional models). In Fig.~\ref{fig:mdot_time} the mass 
accretion rates of the 11.2\,$M_{\odot}$ and 15\,$M_{\odot}$ progenitors 
are depicted for the nonexploding runs with the lowest driving neutrino
luminosities. Because the shock can be largely deformed in multi-dimensional 
simulations and its outermost parts can extend beyond 500\,km (and thus
impede a clean determination of the mass accretion rate) when its average
radius just begins to exceed 400\,km, we refer to the functions plotted
in Fig.~\ref{fig:mdot_time} for defining the mass accretion rate at the
time when the explosion sets in.

\begin{figure*}
\plottwo{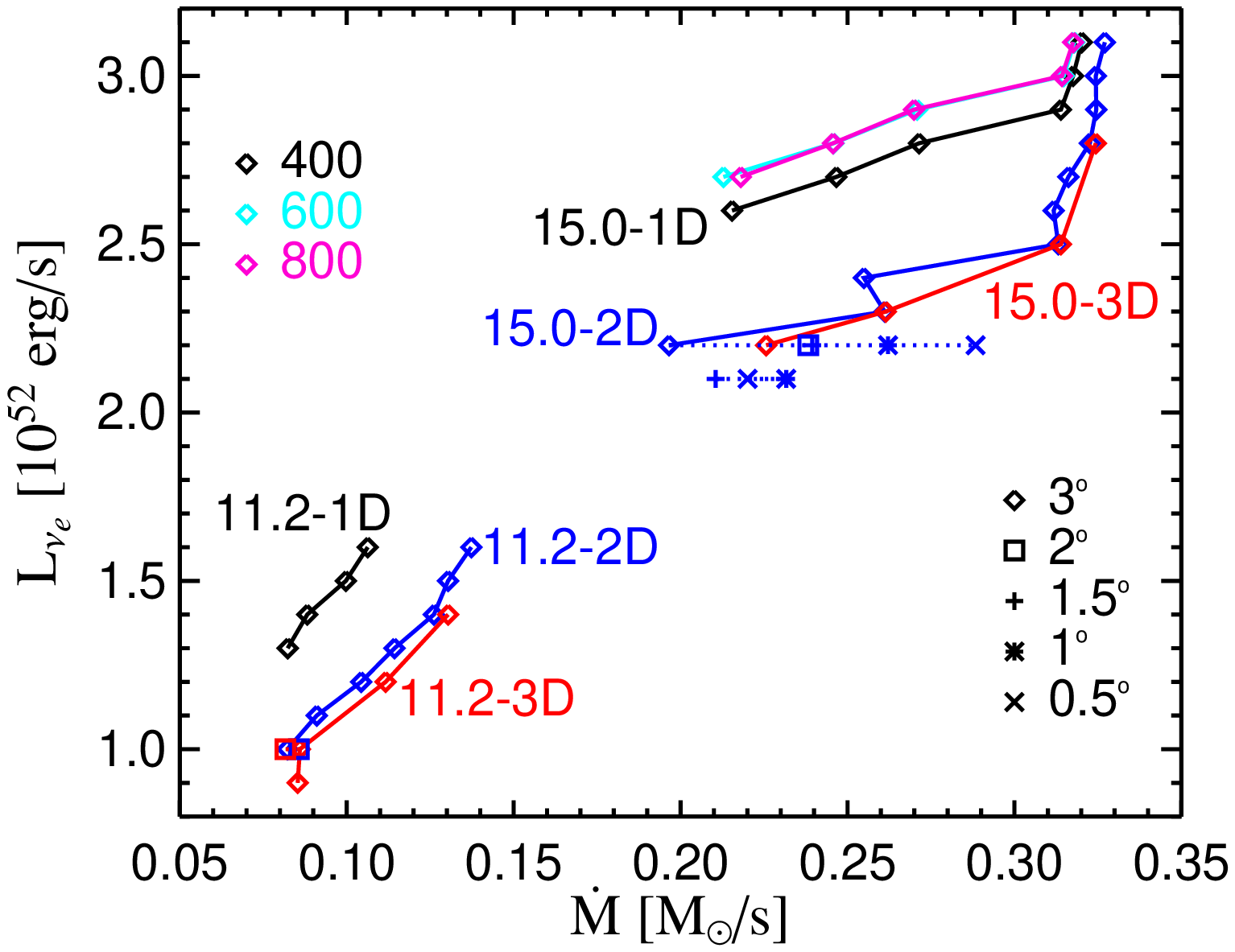}{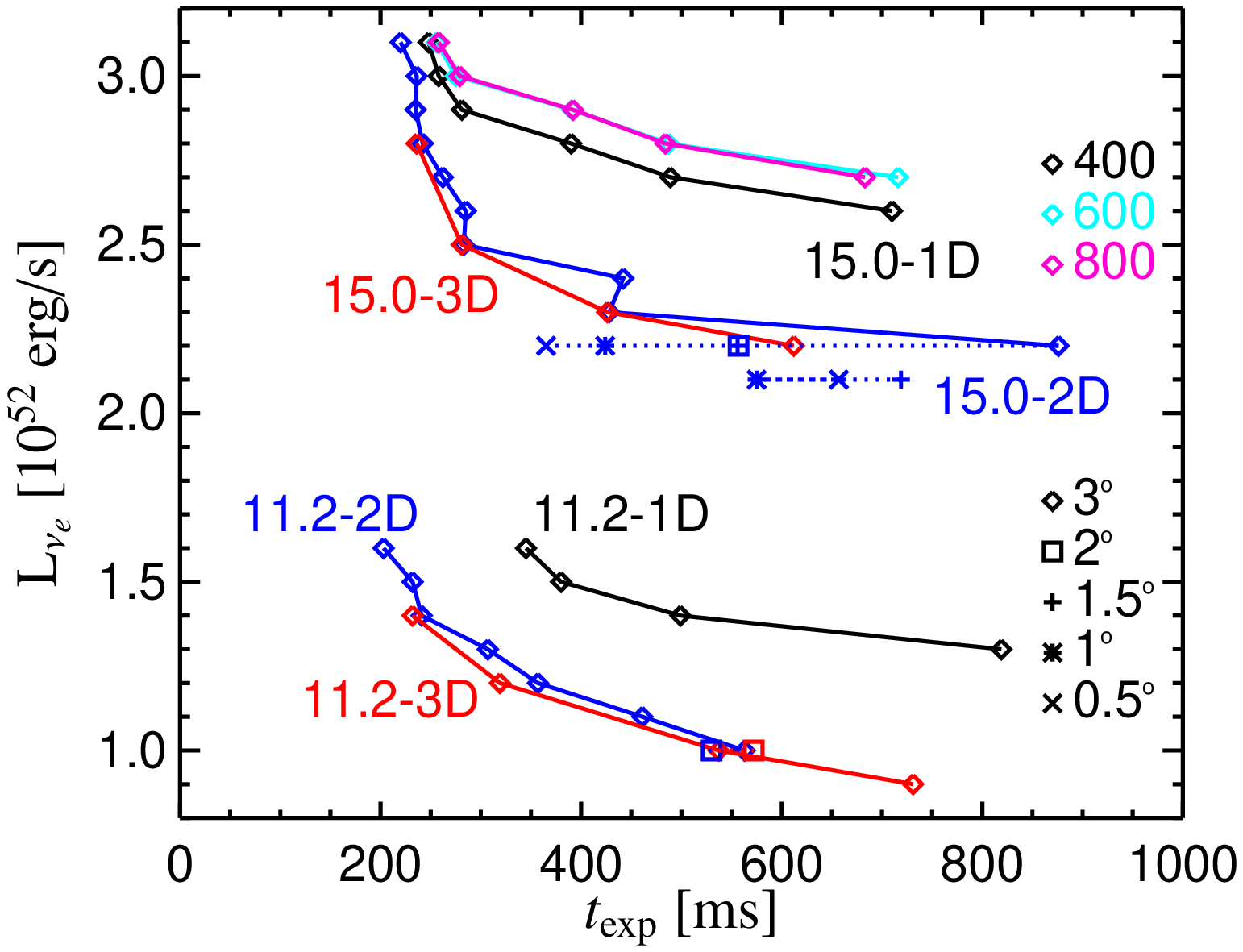}
\caption{Critical curves for the electron-neutrino luminosity ($L_{\nu_e}$)
versus mass accretion rate ($\dot{M}$) (left plot) and versus
explosion time $t_{\textrm{exp}}$ (right plot) for simulations in 1D
(black), 2D (blue), and 3D (red) with standard resolution. The accretion rate is
measured just outside of the shock at the time $t_{\textrm{exp}}$ when the
explosion sets in. For the 15\,$M_{\odot}$ progenitor 1D 
results are displayed for 400, 600, and 800 radial zones.
Higher radial resolution compared to the standard 1D runs
makes explosions slightly more difficult; convergence is achieved for
$\ge$600 radial bins. Multi-dimensional results are shown for different angular
resolutions, where available, but always computed with 400 radial zones.
Note that 2D simulations with improved angular zoning explode more easily,
whereas in 3D only one case was computed (the 11.2\,$M_\odot$ simulation for
$L_{\nu_e} = 1.0\cdot 10^{52}\,$erg\,s$^{-1}$) that developed an explosion 
also with better angular resolution.}
\label{fig:lum_mdot}
\end{figure*}

\begin{figure}
\plotone{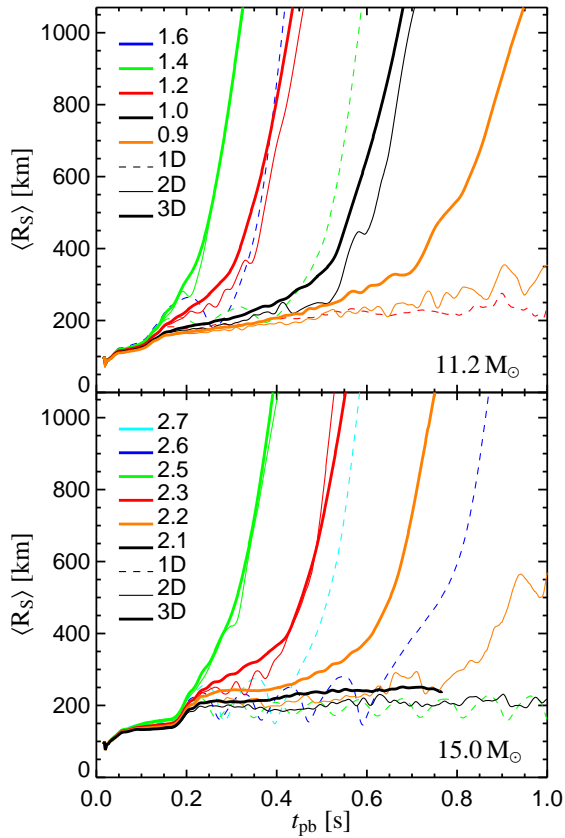}
\caption{Time evolution of the average shock radius as function of the
post-bounce time $t_\mathrm{pb}$ for simulations in one (thin dashed lines), two
(thin solid lines), and three dimensions (thick lines). The shock position is
defined as the surface average over all angular directions. The top panel
shows results for the 11.2\,$M_{\odot}$ progenitor and the bottom
panel for the 15\,$M_{\odot}$ progenitor, all obtained with our standard
resolution. Different electron-neutrino luminosities (labelled in the plots
in units of $10^{52}$ erg s$^{-1}$) are displayed by different colors.}
\label{fig:spos}
\end{figure}

\section{Critical luminosity as function of dimension}
\label{sec:crit_lum}

Based on steady-state solutions of neutrino-heated and \hbox{-cooled}
accretion flows between the stalled shock
and the proto-neutron star surface, \cite{Burrows1993}
identified a critical condition that can be considered to 
separate exploding from nonexploding
models. They found that for a given value of the mass infall rate $\dot M$ 
onto the accretion shock steady-state solutions cannot exist
when the neutrino-heating rate
in the gain layer is sufficiently large, i.e., for neutrino luminosities above
a threshold value $L_\nu$. This result can be coined in terms of
a critical condition $L_\nu(\dot M)$ expressing the fact that either
the driving luminosity has to be sufficiently high or the damping mass
accretion rate enough low for an explosion to become possible.  
The critical $\dot M$-$L_\nu$-curve was interpreted by \cite{Burrows1993}
as a separating line between the region above,
where due to the non-existence of steady-state accretion solutions
explosions are expected to take place, from the region below, where neutrino 
energy input behind the shock is insufficient to accelerate the stalled
shock outwards and thus to trigger an explosion.

This interpretation of the steady-state results 
was consistent with hydrodynamical simulations of
collapsing and exploding stars in 1D and 2D by \cite{Janka1996}. 
Performing a more extended parameter study than the latter work,
\cite{Murphy2008} explored the concept of a critical 
condition systematically with time-dependent hydrodynamical
models. They showed that a critical luminosity indeed
separates explosion from accretion and confirmed that this value 
is lowered by $\sim$30\%
when going from spherical symmetry to two dimensions, at least
when a fixed driving luminosity is adopted and feedback effects of
the hydrodynamics on the neutrino emission are ignored.
Some 2D effects including possible consequences of rotation were 
discussed before on grounds of steady-state models by 
\cite{Yamasaki2005,Yamasaki2006}, while \cite{Janka2001} tried to
include time-dependent aspects of the shock-revival problem and
took into account an accretion component of the neutrino luminosity
in addition to the fixed core component. The influence of such
an accretion contribution was more recently also estimated by 
\cite{Pejcha2011}, who solved the one-dimensional steady-state 
accretion problem between the neutron star and the accretion shock
along the lines of \cite{Burrows1993}, but attempted to obtain a deeper
understanding of the critical condition by comprehensively analysing
the structure of the
accretion layer and of the limiting steady-state solution in dependence
of the stellar conditions. They found that the critical value for
the neutrino luminosity is linked to an ``antesonic condition''
in which the ratio of the adiabatic sound speed to the local escape 
velocity in the postshock layer reaches a critical value above which 
steady-state solutions of neutrino-heated accretion flows cannot be
obtained. By performing high-resolution
hydrodynamic simulations \cite{Fernandez2012} found 
that radial instability is a sufficient 
condition for runaway expansion of an initially stalled core-collapse 
supernova shock if the neutrinospheric parameters 
do not vary with time and if heating by the accretion luminosity is 
neglected. However, the threshold neutrino luminosities for the
transition to runaway instability are in general different from the 
limiting values for steady-state solutions of the kind discussed by 
\cite{Burrows1993} and \cite{Pejcha2011}. \cite{Nordhaus2010} 
generalized the hydrodynamic investigations of \cite{Murphy2008}
to include 3D models and found another reduction of the threshold 
luminosity for explosion by 15--25\% compared to the 2D case. 

Despite the basic agreement of the outcome of these investigations
it should be kept in mind that it is not ultimately clear whether
the simple concept of a critical threshold condition separating
explosions from failures (and the dependences of this threshold 
on dimension and rotation for example) holds beyond the highly idealized 
setups considered in the mentioned works.
None of the mentioned systematic studies by steady-state or 
hydrodynamic models was able to include adequately the complexity
of the feedback between hydrodynamics and neutrino transport physics. 
In particular, none of these studies could yield the proof that 
the non-existence of a steady-state accretion solution for a given
combination of mass accretion rate and neutrino luminosity is 
equivalent to the onset of an explosion. The latter requires the 
persistence of sufficiently strong energy input by neutrino heating
for a suffiently long period of time. This is especially important
because \cite{Pejcha2011} showed that the total energy in the gain
layer is still negative even in the case of the limiting accretion 
solution that corresponds to the critical 
luminosity\footnote{Note that \cite{Fernandez2012}
demonstrated that transition to runaway occurs when the fluid in 
the gain region reaches positive specific energy.}. Within the 
framework of simplified modeling setups, however, the question
cannot be answered whether such a persistent energy input can be
maintained in the environment of the supernova core.

Following the previous investigations by \cite{Murphy2008} and 
\cite{Nordhaus2010} we performed hydrodynamical simulations that 
track the post-bounce evolution of collapsing stars for different,
fixed values of the driving neutrino luminosity.
Since the mass accretion rate decreases with time
according to the density profile that is characteristic of the 
initial structure of the progenitor core (see Fig.~\ref{fig:mdot_time}
for the 11.2 and 15\,$M_\odot$ stars considered in this work),
each model run probes the critical value of $\dot M_{\mathrm{exp}}$
at which the explosion becomes possible for the chosen value of
$L_\nu = L_{\nu_e} = L_{\bar \nu_e}$. The collection of 
value pairs ($\dot{M}_{\mathrm{exp}}$,$L_{\nu_e}$) defines a
critical curve $L_{\nu}(\dot M)$. These are shown for our 1D, 2D, 
and 3D studies with standard resolution for both progenitor
stars in the left panel of Fig.~\ref{fig:lum_mdot} and in the case
of the 15\,$M_\odot$ star can be directly compared with
Fig.~1 of \cite{Nordhaus2010}. Table~\ref{tbl:models_std}
lists, as a function of the chosen $L_{\nu_e}$,
the corresponding times $t_{\mathrm{exp}}$ when the onset of the
explosion takes place and the mass accretion rate has the value of 
$\dot M_{\mathrm{exp}}$. The post-bounce evolution of a collapsing
star proceeds from high to low mass accretion rate
(Fig.~\ref{fig:mdot_time}), i.e., from right to left on the
horizontal axis of the left panel of Fig.~\ref{fig:lum_mdot}. 
When $\dot M$ reaches
the critical value for the given $L_{\nu_e}$, the model develops
an explosion. The right panel of 
Fig.~\ref{fig:lum_mdot} visualizes the functional
relations between the neutrino luminosities $L_{\nu_e}$ and the
explosion times $t_{\mathrm{exp}}$ for both progenitors and for
the simulations with different dimensions.

At first glance Fig.~\ref{fig:lum_mdot} reproduces basic trends 
that are visible in Figs.~17 of \cite{Murphy2008} and in
Fig.~1 of \cite{Nordhaus2010}. For example, the critical luminosity 
increases for higher mass accretion rate and the values for 
spherically symmetric models are clearly higher than those for
2D simulations. However, a closer inspection reveals interesting 
differences compared to the previous works. 
\begin{itemize}
\item
In general the slopes of our critical $L_{\nu_e}(\dot M)$-relations
appear to be considerably steeper and in the case of the
15\,$M_\odot$ star they exhibit a very steep
rise above $\dot M\approx 0.31\,M_\odot$\,s$^{-1}$. This means that 
explosions for higher mass accretion rates are much harder to obtain
and therefore the tested driving luminosities in our simulations
do not lead to explosions earlier than about 200\,ms after bounce, 
independent of whether the modeling is performed in 1D, 2D or 3D. 
In contrast, the critical curves given by \cite{Nordhaus2010}
show a moderately steep increase over the whole range of plotted
mass accretion rates between about 0.1\,$M_\odot$\,s$^{-1}$ and more 
than 0.5\,$M_\odot$\,s$^{-1}$.
\item
Nonradial flows in the 2D case, by which the residence time of 
accreted matter in the gain layer could be extended or more matter
could be kept in the gain region, reduce the critical luminosities
by at most $\sim$15\% of the 1D-values for the 15\,$M_\odot$ star
and $\lesssim$25\% for the 11.2\,$M_\odot$ model, which is a somewhat 
smaller difference than found by \cite{Murphy2008} and
\cite{Nordhaus2010}. 
\item
Most important, however, is the fact that we cannot confirm the
observation by \cite{Nordhaus2010} that 3D provides considerably
more favorable conditions for an explosion than 2D. Our critical
curves for the 2D and 3D cases nearly lie on top of each other. 
There are minor improvements of the explosion conditions
in 3D visible in both panels of Fig.~\ref{fig:lum_mdot} and the 
numbers of Table~\ref{tbl:models_std}, e.g., a 10\% reduction
of the smallest value of $L_{\nu}$ for which an explosion can be 
obtained for the 11.2\,$M_\odot$ star, a $\sim$260\,ms earlier 
explosion for the lowest luminosity driving the 15\,$M_\odot$ 
explosion ($2.2\cdot 10^{52}$\,erg\,s$^{-1}$), 
and a tendency of slightly faster 3D explosions for
all tested luminosities (see Fig.~\ref{fig:spos} and 
Table~\ref{tbl:models_std}).
All of these more optimistic 3D features, however, will disappear 
for simulations with higher resolution as we will see in
Sect.~\ref{sec:high_res}.
\end{itemize}
%

\begin{figure}
 \plotone{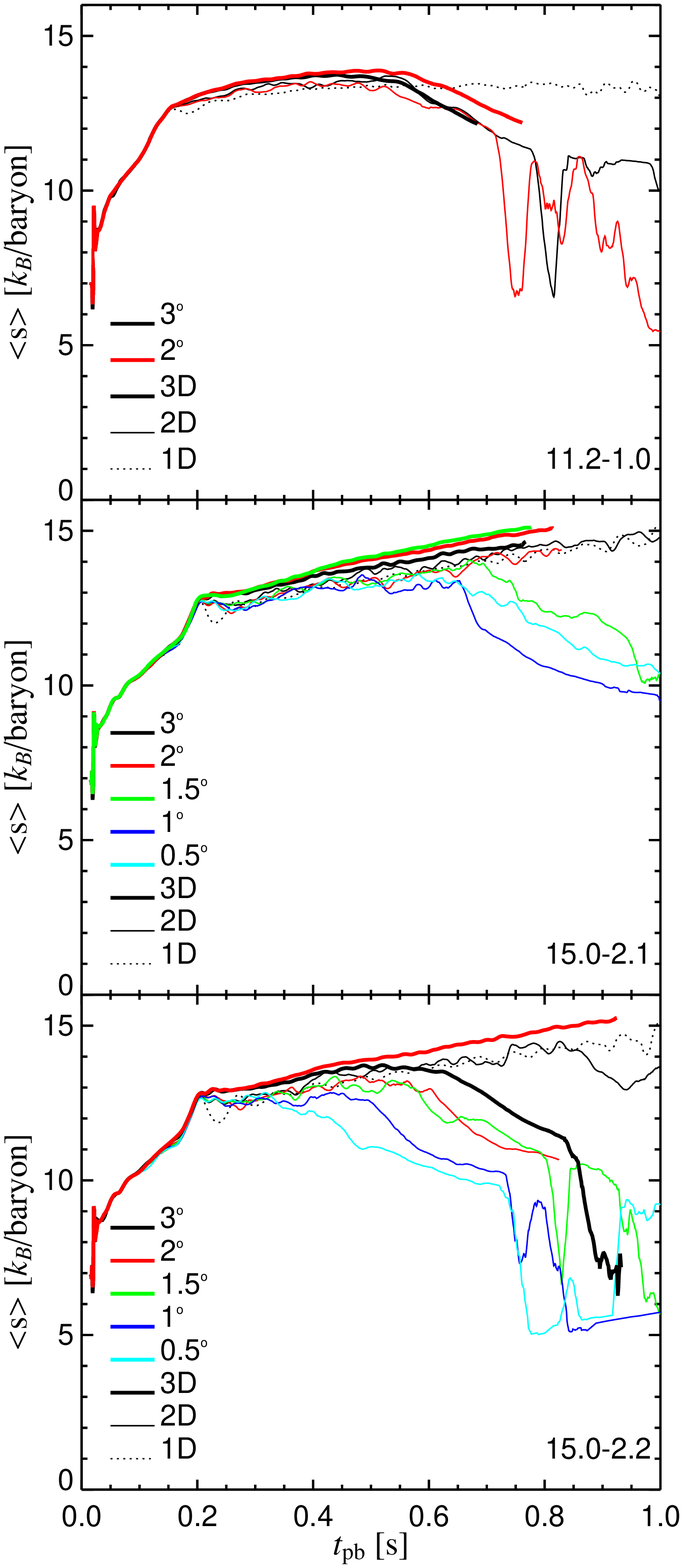}
 \caption{Time evolution of the mass-weighted average entropy in the gain
region for one-dimensional (thin dotted lines), two-dimensional
(thin solid lines), and three-dimensional (thick lines) simulations
with different angular resolutions (corresponding to different colors). 
The top panel
displays the 11.2\,$M_{\odot}$ results for an electron-neutrino luminosity
of $L_{\nu_e} = 1.0\cdot10^{52}$\,erg\,s$^{-1}$, the middle panel shows the
15\,$M_{\odot}$ runs for an electron-neutrino luminosity of $L_{\nu_e} =
2.1\cdot10^{52}$\,erg\,s$^{-1}$, and the bottom panel the 15\,$M_{\odot}$ 
models for $L_{\nu_e} = 2.2\cdot10^{52}$\,erg\,s$^{-1}$. 
The strong decrease of the average entropy 
that terminates a phase of continuous, slow increase signals the onset of the
explosion when a growing mass of cooler (low-entropy) gas is added into the
gain layer after being swept up by the expanding and accelerating shock wave.} 
\label{fig:ave_sto_res}
\end{figure}

\begin{figure*}
 \plottwo{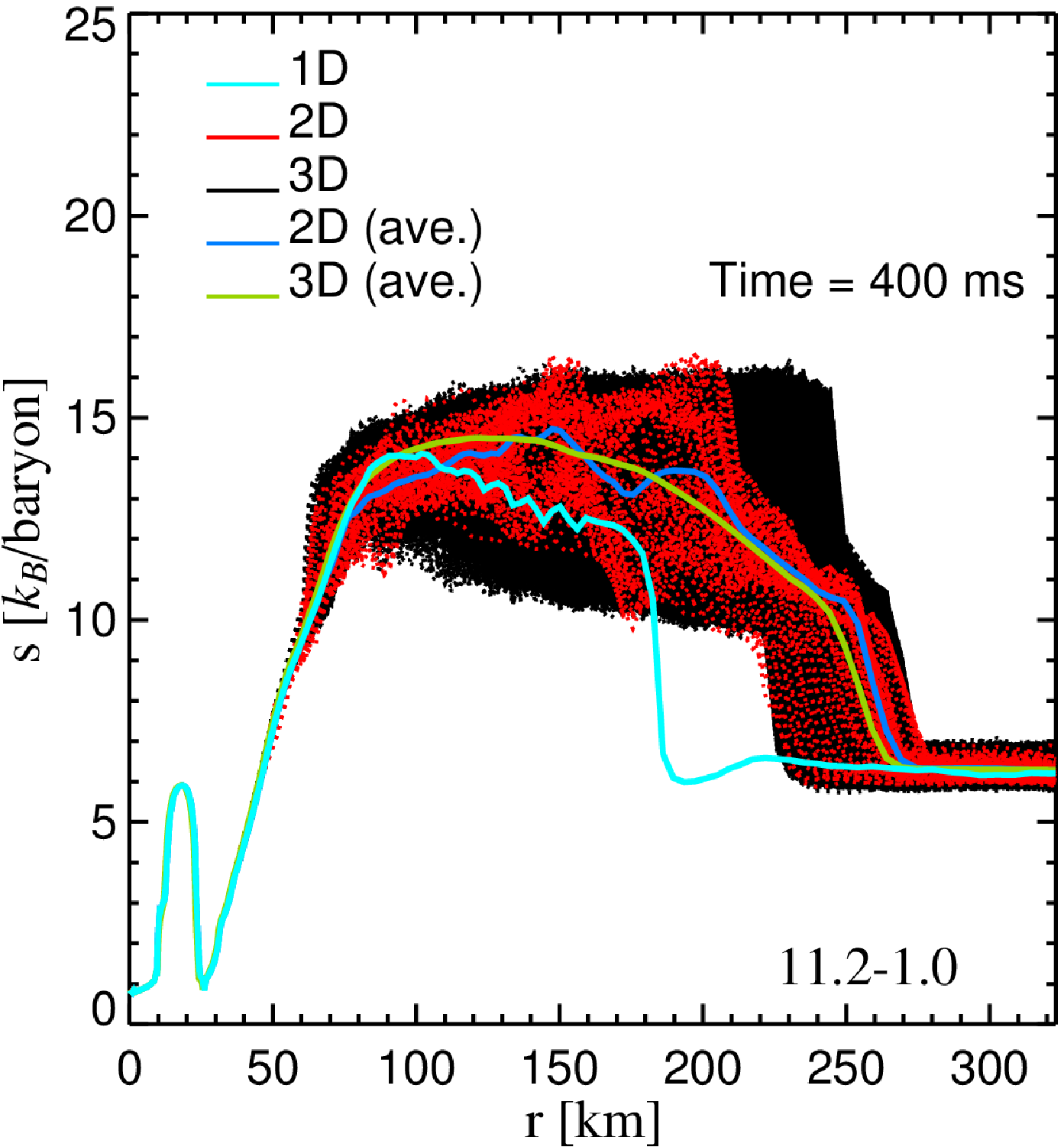}{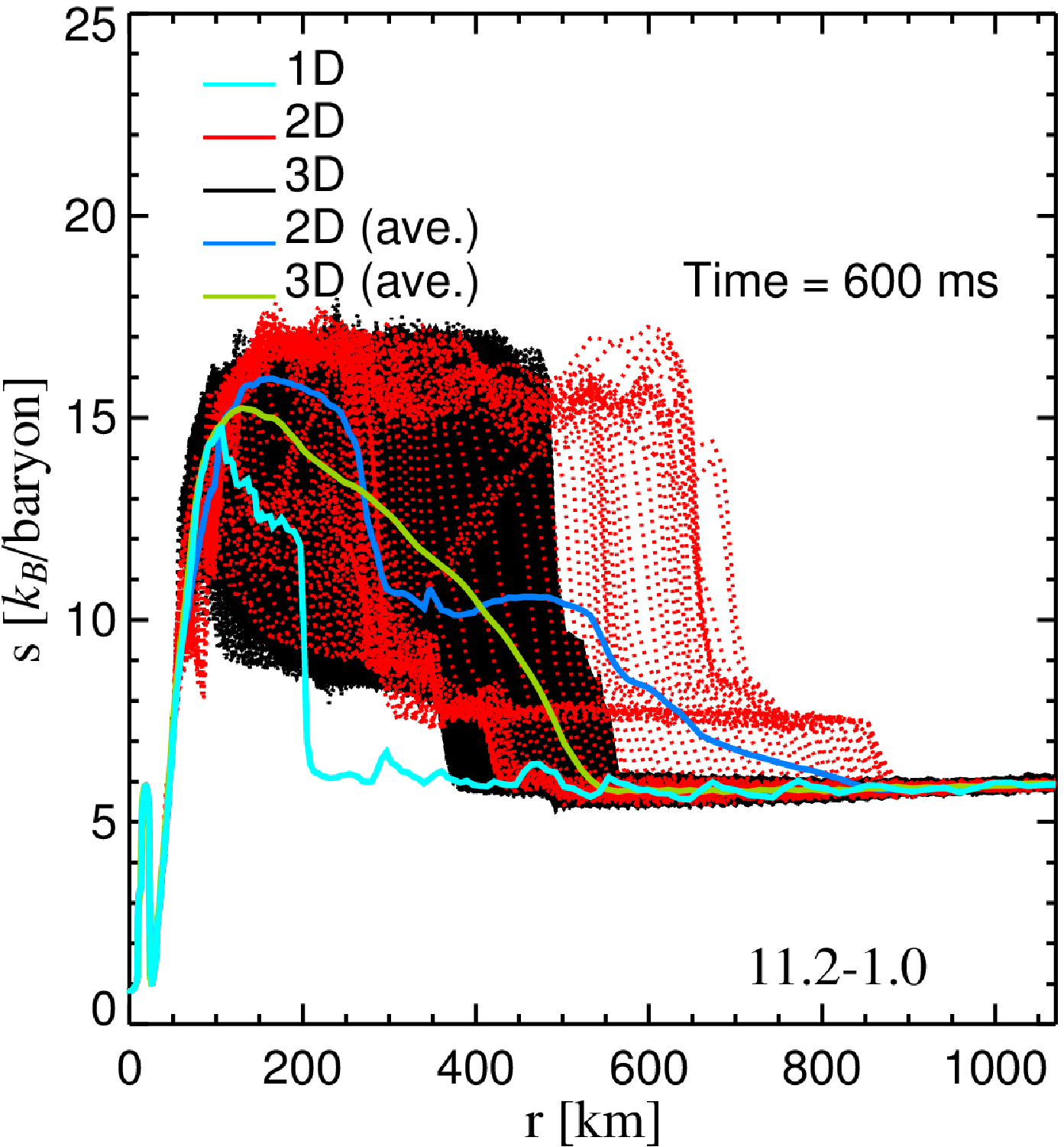}
 \caption{Scatter-plots of the entropy structure as function of radius for 
simulations of the 11.2\,$M_{\odot}$ progenitor with an electron-neutrino
luminosity of $L_{\nu_e} = 1.0\cdot10^{52}$\,erg\,s$^{-1}$ at 400\,ms (left) and
600\,ms (right) after core bounce. The red dots correspond to the 2D results,
black ones to 3D, the light blue line is the entropy profile of the 1D simulation,
and the dark-blue and green curves are mass-weighted angular averages of the 
2D and 3D models, respectively. Both multi-dimensional simulations were
performed with an angular resolution of two degrees and both yield explosions
(at $\sim$530\,ms p.b.\ in 2D and $\sim$570\,ms p.b.\ in 3D; 
see Table~\ref{table:models_res}). Note that different from 
Fig.~\ref{fig:ave_sto_res}, unshocked material at a given radius is included
when computing angular averages. The dispersion of entropy values in the 
unshocked flow of 2D and 3D simulations is a consequence of the imposed 
density-seed perturbations (cf.\ Sect.~\ref{sec:num}), which grow in the
supersonical infall regime (see \citealp{Buras2006b}).}
\label{fig:sc_s11_hr}
\end{figure*}

\begin{figure*}
 \plottwo{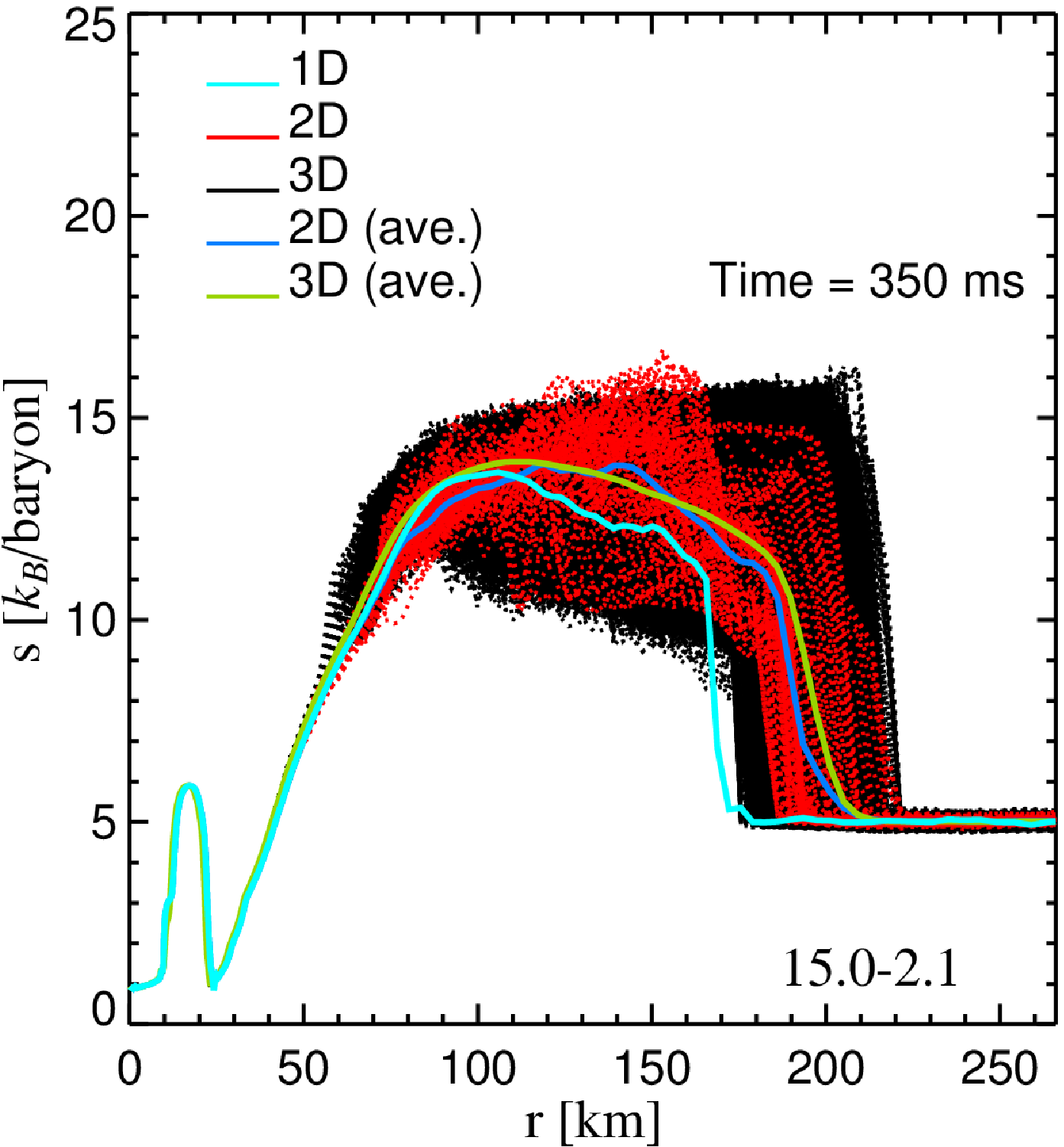}{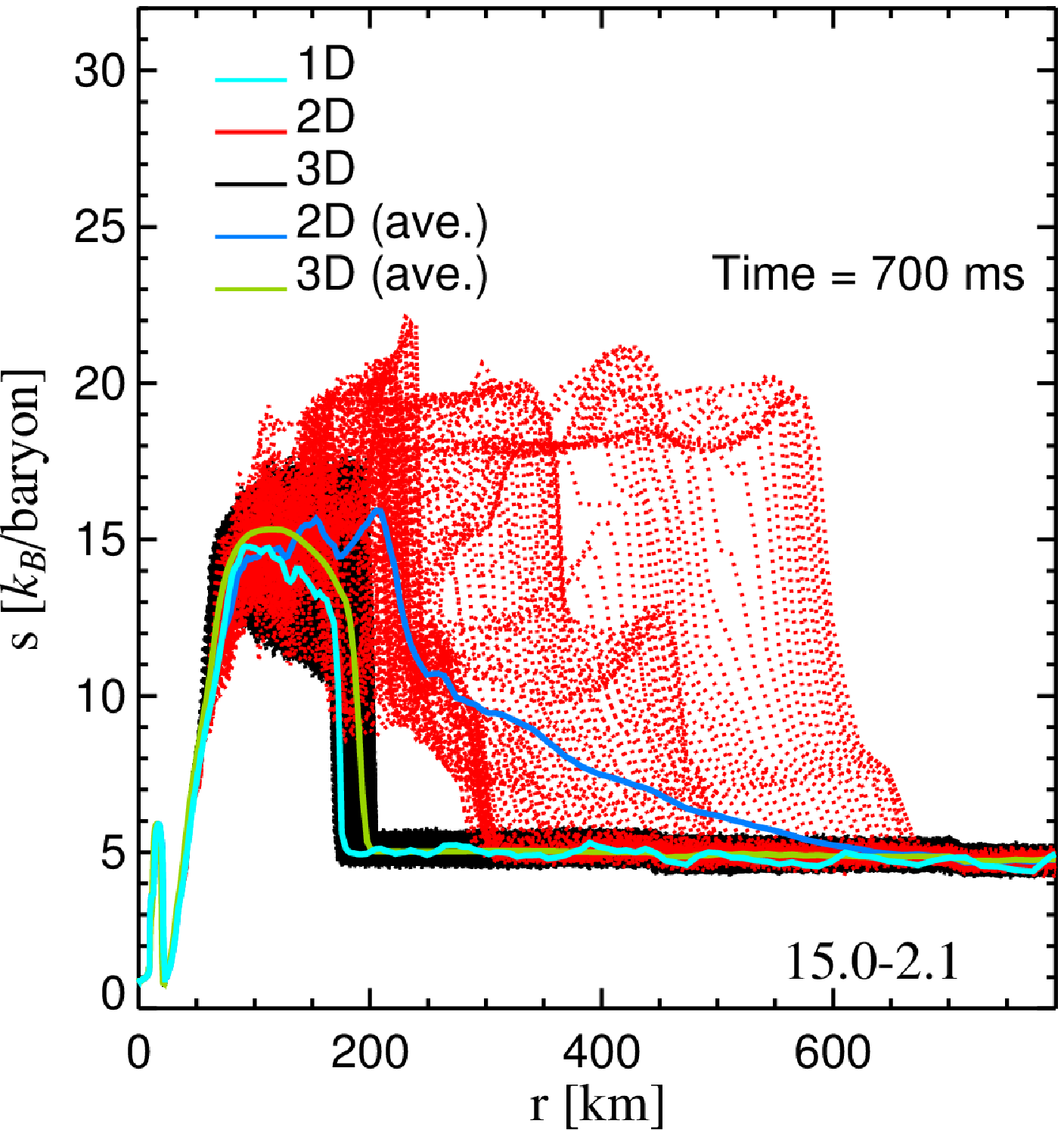}
 \caption{Scatter-plots of the entropy structure as function of radius for 
simulations of the 15\,$M_{\odot}$ progenitor with an electron-neutrino
luminosity of $L_{\nu_e} = 2.1\cdot10^{52}$\,erg\,s$^{-1}$ at 350\,ms (left) and
700\,ms (right) after core bounce. The red dots correspond to the 2D results,
black ones to 3D, the light blue line is the entropy profile of the 1D simulation,
and the dark-blue and green curves are mass-weighted angular averages of the
2D and 3D models, respectively. Both multi-dimensional simulations were
performed with an angular resolution of 1.5 degrees. While the 2D model
develops an explosion setting in $\sim$720\,ms after bounce, the 3D model 
does not produce an explosion (Table~\ref{table:models_res}).
Note that different from
Fig.~\ref{fig:ave_sto_res}, unshocked material at a given radius is included
when computing angular averages. The dispersion of entropy values in the
unshocked flow of 2D and 3D simulations is a consequence of the imposed
density-seed perturbations (cf.\ Sect.~\ref{sec:num}), which grow in the
supersonical infall regime (see \citealp{Buras2006b}).}
\label{fig:sc_s15_hr}
\end{figure*}

Before we discuss the origin of the differences between our results
and those of \cite{Murphy2008} and \cite{Nordhaus2010} we would like
to remark that the kinks and even nonmonotonic parts of the
curves shown in Fig.~\ref{fig:lum_mdot} in particular for the 
multi-dimensional cases are connected to our definition of the explosion
time as being the moment when the mean shock radius exceeds 400\,km.
Especially in cases where the shock deformation is large (which
is an issue mainly in some of the 2D simulations) this definition
is associated with significant uncertainty in the determination of
the exact explosion time $t_{\mathrm{exp}}$ and therefore also of the
corresponding mass accretion rate $\dot{M}_{\mathrm{exp}}$.

In addition to the results for our standard resolution,
Fig.~\ref{fig:lum_mdot} presents 1D models with higher radial resolution 
(600 and 800 radial zones). The critical curves with better zoning exhibit
the same overall slopes as those of the standard runs, but there is a
slight shift towards higher values of the critical luminosity (or,
equivalently, a small shift to lower $\dot{M}_{\mathrm{exp}}$ and later 
$t_{\mathrm{exp}}$). This is
caused by somewhat larger neutrino energy losses from the cooling layer
with better radial resolution, an effect which makes explosions more
difficult. We stress that this resolution-dependent cooling effect 
is a consequence of the simplified neutrino-loss treatment. The employed
cooling rate is not able to reproduce real transport behavior and
leads to the development of a pronounced gaussian-like
density peak (with density excess of a few compared to its surroundings)
in the neutrino-cooling layer. This local density maximum
has a strong influence on the integrated energy loss by neutrino
emission. With better zoning the peak becomes better resolved and even
grows in size. 
Convergence of 1D results seems to be achieved for $\ge$600 radial zones,
but in 2D and 3D the artificial density peak prevents numerical 
convergence for all employed radial
zonings (cf.\ Sect.~\ref{sec:high_res}). 
Figure~\ref{fig:lum_mdot} also displays some data points for 
multi-dimensional models with better angular resolution 
(all of them, however, computed with 400 radial 
mesh points). These will be discussed in Sect.~\ref{sec:high_res}.

A more detailed analysis, which we will report on below and
in App.~\ref{sec:appendix}, reveals
that the exact values of the critical luminosities as well as the 
detailed slope of the critical curves seem to depend strongly on
the employed description of neutrino effects, whose implementation
is subject to a significant degree of arbitrariness if detailed 
neutrino transport is not included in the modeling (cf.\ the discussion
in Sect.~\ref{sec:num}). The fact that \cite{Murphy2008} found
fairly good overall agreement between their 
critical relations $L_\nu(\dot M)$ and those obtained
by \cite{Burrows1993} is likely to be linked to a basically similar
treatment of the neutrino effects.

Before $\sim$0.2\,s after bounce the mass accretion
rate in the case of the 15\,$M_\odot$ progenitor changes much more 
rapidly than during the subsequent evolution (Fig.~\ref{fig:mdot_time}).
For the corresponding $\dot M$ values in excess of
$\sim$0.31\,$M_\odot$\,s$^{-1}$, the accretion shock is therefore not
as close to steady-state conditions as later on.
We see a steep rise of our critical curves at $\dot M \gtrsim
0.31\,M_\odot$\,s$^{-1}$ ($t_\mathrm{exp}\lesssim 0.25$\,s), 
which is a very prominent difference 
compared to the results of \cite{Burrows1993}, who assumed steady-state
accretion, but in particular also compared to the hydrodynamic results
of \cite{Murphy2008}, and \cite{Nordhaus2010} even in the 1D case.
In order to explore possible reasons for this difference, we performed 
1D simulations in which the neutrino treatment is
copied from \cite{Nordhaus2010} as closely as possible (i.e.,
the reduction factor of 2.7 in the exponent of $e^{-\tau_{\mathrm{eff}}}$
is not applied and deleptonization is taken into account by using
a $Y_e(\rho)$ relation, but not the corresponding entropy changes
proposed by \citealp{Liebendorfer2005}; for more details 
on these results, see App.~\ref{sec:appendix}). These runs reveal that
the steep rise of our $L_\nu(\dot M)$-curves is caused by a
strong increase of the neutrino-cooling rate with higher values
of $\dot M$, in particular when we apply our
neutrino treatment. The corresponding energy losses inhibit 
explosions for low values of the driving luminosity. 
The stronger cooling is linked mainly to our reduction
of the effective optical depth $\tau_{\mathrm{eff}}$, which we had
to apply in order to reconcile the mass accretion rates and 
explosion times with the lowest driving luminosities for which 
\cite{Murphy2008} had obtained explosions for the 11.2 
and 15\,$M_\odot$ stars (cf.\ Sect.~\ref{sec:num}).
For example, in the case of the 15\,$M_\odot$ progenitor a driving 
luminosity of $L_{\nu_e} = 3.1\cdot 10^{52}$erg\,s$^{-1}$ triggers
an explosion at $t_{\mathrm{exp}}\approx 250$\,ms p.b.\ and
$\dot M_{\mathrm{exp}}\approx 0.32\,M_\odot$\,s$^{-1}$ 
(Table~\ref{tbl:models_std} and Fig.~\ref{fig:lum_mdot}), whereas
with an implementation of neutrino effects closer to that of 
\cite{Nordhaus2010} the explosion sets 
in at $t_{\mathrm{exp}}\approx 120$\,ms p.b.\ and 
$\dot M_{\mathrm{exp}}\approx 0.8\,M_\odot$\,s$^{-1}$
(see Fig.~\ref{fig:lum_mdot_cmp} in 
App.~\ref{sec:appendix}). Shortly before this moment (at 75\,ms 
after bounce) the total energy loss by neutrino cooling
is about 10 times lower with the scheme of \cite{Nordhaus2010}
than with our neutrino implementation. The latter yields an integrated
energy-loss rate of $\sim$9$\cdot 10^{52}$\,erg$\,$s$^{-1}$ and
significant cooling even at densities between $10^{12}$ and
$10^{13}$\,g$\,$cm$^{-3}$, where the \cite{Nordhaus2010} treatment
shows essentially no cooling. 
Neither the magnitude of the total neutrino-energy
loss rate nor the region of energy extraction with our modeling 
approach are implausible and in disagreement with detailed transport
simulations during a stage when the mass accretion rate still exceeds
1\,$M_\odot$\,s$^{-1}$ (cf., e.g., Fig.~20 in
\citealp{Buras2006a}). In contrast, the \cite{Nordhaus2010}
treatment appears to massively underestimate the neutrino energy
extraction from the accretion flow during this time.

These findings demonstrate that the results of the critical
$L_{\nu_e}(\dot M)$-relation in 1D can be quantitatively as well as 
qualitatively different with different approximations of neutrino 
heating and in particular of neutrino cooling. Moreover,
this gives reason for concern that the differences of the
critical explosion conditions for 2D and 3D simulations seen
by \cite{Nordhaus2010} might have been connected to their 
treatment of the neutrino physics, in particular also because the 
decrease of the critical luminosity from 2D to 3D they found was
only 15--25\%, which is a relatively modest change (much smaller
than the 1D-2D difference) and thus could easily be overruled by
other effects. Our results for 2D and 3D simulations with a 
different implementation of neutrino sources confirm this concern.

In the \cite{Nordhaus2010} paper the average entropy of the matter
in the gain region, $\langle s(t)\rangle$,
was considered to be a suitable diagnostic quantity
that reflects the crucial differenes of 1D, 2D, and 3D simulations
concerning their relevance for the supernova dynamics.
While in the spherically symmetric case accreted matter moves 
through the gain region on the shortest, radial paths,
nonradial motions can increase the time that shock-accreted
plasma can stay in the gain layer and absorb energy from neutrinos.
This can raise the mean entropy, internal energy, and pressure
in the postshock region and thus support the revival of the stalled
supernova shock. This seems to happen in the simulations by 
\cite{Nordhaus2010}, who found that turbulent mass motions in 3D
can even improve the conditions for the neutrino-heating mechanism
compared to the 2D case. A crude interpretation of this difference
can be given by means of random-walk arguments,
considering the mass motions in convective and turbulent structures
as diffusive process in the gain layer \citep{Murphy2008}. 
The question, however, is whether this effect is a robust 2D-3D
difference and whether it is the crucial key to successful 
explosions by the neutrino-heating mechanism.

Our results at least raise doubts. Figure~\ref{fig:ave_sto_res}
displays the time evolution of the mean entropy in the gain layer
for 11.2 and a 15\,$M_\odot$ models computed with driving luminosities
near the minimum value for which we obtained explosions. While the 
11.2\,$M_\odot$ model explodes with a luminosity of 
$1.0\cdot 10^{52}\,$erg$\,$s$^{-1}$ for all tested 
resolutions in more than one dimension despite minimal differences
between the values of $\langle s(t)\rangle$ compared to the 1D 
counterpart, the 15\,$M_\odot$ progenitor develops an explosion
for the chosen luminosity of $2.1\cdot 10^{52}\,$erg$\,$s$^{-1}$ 
only in the case of higher-resolution 2D runs 
(this will be further discussed in Sect.~\ref{sec:high_res}). 
These successful cases, however, do not stick out by especially
high values of $\langle s(t)\rangle$. On the contrary, they even
have lower mean entropies than the unsuccessful 3D models! 
It is obvious that Fig.~\ref{fig:ave_sto_res} does not
exhibit the clear 1D-2D-3D hierarchy visible in Fig.~5 of the 
\cite{Nordhaus2010} paper, which was found there to closely 
correlate with the explosion behavior of their models. 
Instead, the differences between simulations in the different
dimensions are fairly small, and even 
two-dimensional flows, which unquestionably allow for explosions 
also when none happen in 1D, do not appear more promising than the
1D case in terms of the average entropy of the matter in the gain
layer\footnote{We stress that our basic findings are independent of
the exact way how the gain radius of the multi-dimensional models
is determined, i.e., whether the evaluation is performed with an
angularly averaged gain radius or a direction-dependent gain radius.
The outer boundary of the integration volume is defined by the 
shock position, which usually forms a non-spherical surface in 
the multi-dimensional case.}. Similarly, 3D models possess slightly
(insignificantly?) higher values of $\langle s(t)\rangle$ but do
not show a clear tendency of easier explosions, in particular not
the better resolved simulations (see Sect.~\ref{sec:high_res}).

The radial entropy structures of 1D, 2D, and 3D runs for both 
progenitors, shown in Figs.~\ref{fig:sc_s11_hr} 
and \ref{fig:sc_s15_hr} once before an 
explosion begins and another time around the onset of an explosion
in at least one of the runs, demonstrate that low-entropy accretion
downdrafts and high-entropy rising plumes of neutrino-heated plasma 
lead to large local variations of the entropy per baryon of the matter 
in the gain layer (scatter regions in the plots). However, the 
mass-weighted angular averages of the entropies reveal much smaller
differences between the 1D and 2D cases than visible in Fig.~4 of
the \cite{Nordhaus2010} paper and in Fig.~13 of \cite{Murphy2008}, 
and exhibit no obvious signs of more advantageous
explosion conditions in the 3D cases compared to 2D. The noticeable
differences in the radial profiles seem to be insufficient to 
cause major differences in the mean entropies computed by additional
radial averaging (see Fig.~\ref{fig:ave_sto_res}).

How can this discrepancy compared to \cite{Nordhaus2010} and 
\cite{Murphy2008} be explained, and how can one understand the 
fact that 2D effects play a supportive role for neutrino-driven
explosions? We will return to these questions in Sect.~\ref{sec:cond},
but before that we shall present our results of multi-dimensional 
simulations with varied resolution in the following section.

\begin{deluxetable}{cccccccc}
\tablecolumns{8}
\tabletypesize{\scriptsize}
\tablecaption{
Multidimensional models with different resolution.
\label{table:models_res}
}
\tablewidth{0pt}
\tablehead{
  \colhead{Mass\tablenotemark{a}}                                        &
  \colhead{Dim\tablenotemark{b}}                                         &
  \colhead{$L_{\nu_e}$\tablenotemark{c}}                                 &
  \colhead{Ang.\tablenotemark{d}}                                        &
  \colhead{$N_{r}$\tablenotemark{e}}                                     &
  \colhead{$t_{\textrm{exp}}$\tablenotemark{f}}                          &
  \colhead{$\dot{M}_{\textrm{exp}}$\tablenotemark{g}}                    &
  \colhead{$t_{\textrm{sim}}$\tablenotemark{h}}                          \\
  \colhead{($M_{\sun}$)}                                                 &
  \colhead{}                                                             &
  \colhead{($10^{52}$}                                                   &
  \colhead{Res.}                                                         &
  \colhead{}                                                             &
  \colhead{(ms)}                                                         &
  \colhead{}                                                             &
  \colhead{(ms)}                                                         \\
  \colhead{}                                                             &
  \colhead{}                                                             &
  \colhead{erg/s)}                                                       &
  \colhead{}                                                             &
  \colhead{}                                                             &
  \colhead{}                                                             &
  \colhead{}                                                             &
  \colhead{}                                                             }
\startdata
11.2  & 2D   & 0.8         & 3$^{\circ}$    & 400         & $-$ & $-$ & 1017 \\
11.2  & 2D   & 0.8         & 2$^{\circ}$    & 400         & $-$ & $-$ & 979  \\
\textbf{11.2}  & \textbf{3D}  & \textbf{0.8} & \textbf{3$^{\circ}$} &
\textbf{400}   & \textbf{$-$} & \textbf{$-$} & \textbf{915} \\
\textbf{11.2}  & \textbf{3D}  & \textbf{0.8} & \textbf{2$^{\circ}$} &
\textbf{400}   & \textbf{$-$} & \textbf{$-$} & \textbf{758} \\
      &      &             &                &            &      &     &      \\
11.2  & 2D   & 0.9         & 3$^{\circ}$    & 400        & $-$  & $-$ & 1006 \\
11.2  & 2D   & 0.9         & 2$^{\circ}$    & 400        & $-$  & $-$ & 985 \\
\textbf{11.2}  & \textbf{3D}  & \textbf{0.9} & \textbf{3$^{\circ}$} &
\textbf{400}   & \textbf{731} & \textbf{0.085} & \textbf{954} \\
\textbf{11.2}  & \textbf{3D}  & \textbf{0.9} & \textbf{2$^{\circ}$} &
\textbf{400}   & \textbf{$-$} & \textbf{$-$} & \textbf{819} \\
      &      &             &                &            &     &     &     \\
11.2  & 2D   & 1.0         & 3$^{\circ}$    & 400        & 563 & 0.082 & 1053\\
11.2  & 2D   & 1.0         & 2$^{\circ}$    & 400        & 527 & 0.086 & 1053\\
\textbf{11.2}  & \textbf{3D}  & \textbf{1.0} & \textbf{3$^{\circ}$} &
\textbf{400}   & \textbf{537} & \textbf{0.086} & \textbf{684} \\
\textbf{11.2}  & \textbf{3D}  & \textbf{1.0} & \textbf{2$^{\circ}$} &
\textbf{400}   & \textbf{572} & \textbf{0.082} & \textbf{761} \\
      &      &             &                &            &      &     &      \\
15    & 2D   & 2.0         & 3$^{\circ}$    & 400        & $-$  & $-$ & 1016 \\
15    & 2D   & 2.0         & 2$^{\circ}$    & 400        & $-$  & $-$ & 829  \\
15    & 2D   & 2.0         & 1.5$^{\circ}$  & 400        & $-$  & $-$ & 1016 \\
15    & 2D   & 2.0         & 1$^{\circ}$    & 400        & $-$  & $-$ & 1016 \\
\textbf{15}    & \textbf{3D}  & \textbf{2.0} & \textbf{3$^{\circ}$}  &
\textbf{400}   & \textbf{$-$} & \textbf{$-$} & \textbf{723} \\
\textbf{15}    & \textbf{3D}  & \textbf{2.0} & \textbf{2$^{\circ}$} &
\textbf{400}   & \textbf{$-$} & \textbf{$-$} & \textbf{524} \\
      &      &             &                &            &      &     &     \\
15    & 2D   & 2.1         & 3$^{\circ}$    & 400        & $-$  & $-$ & 1016 \\
15    & 2D   & 2.1         & 2$^{\circ}$    & 400        & $-$  & $-$ & 829 \\
15    & 2D   & 2.1         & 1.5$^{\circ}$  & 400        & 719  & 0.210 & 1016\\
15    & 2D   & 2.1         & 1$^{\circ}$    & 400        & 575  & 0.232 & 1016\\
15    & 2D   & 2.1         & 0.5$^{\circ}$  & 400        & 657  & 0.220 & 1016\\
\textbf{15}    & \textbf{3D}  & \textbf{2.1} & \textbf{3$^{\circ}$} &
\textbf{400}   & \textbf{$-$} & \textbf{$-$} & \textbf{767} \\
\textbf{15}    & \textbf{3D}  & \textbf{2.1} & \textbf{2$^{\circ}$} &
\textbf{400}   & \textbf{$-$} & \textbf{$-$} & \textbf{815} \\
\textbf{15}    & \textbf{3D}  & \textbf{2.1} & \textbf{1.5$^{\circ}$} &
\textbf{400}   & \textbf{$-$} & \textbf{$-$} & \textbf{777} \\
      &      &             &                &             &     &     &      \\
15    & 2D   & 2.2         & 3$^{\circ}$    & 400         & 876 & 0.197 & 1016\\
15    & 2D   & 2.2         & 2$^{\circ}$    & 400         & 557 & 0.238 & 825 \\
15    & 2D   & 2.2         & 1.5$^{\circ}$  & 400         & 556 & 0.239 & 1016\\
15    & 2D   & 2.2         & 1$^{\circ}$    & 400         & 424 & 0.262 & 1016\\
15    & 2D   & 2.2         & 0.5$^{\circ}$  & 400         & 365 & 0.288 & 1016\\
\textbf{15}    & \textbf{3D}  & \textbf{2.2} & \textbf{3$^{\circ}$} &
\textbf{400}   & \textbf{612} & \textbf{0.226} & \textbf{932} \\
\textbf{15}    & \textbf{3D}  & \textbf{2.2} & \textbf{2$^{\circ}$} &
\textbf{400}   & \textbf{$-$} & \textbf{$-$} & \textbf{925} \\
      &      &             &                &             &      &     &      \\
15    & 2D   & 2.1         & 3$^{\circ}$    & 600         & $-$  & $-$ & 1016 \\
15    & 2D   & 2.1         & 3$^{\circ}$    & 800         & $-$  & $-$ & 1016 \\
15    & 2D   & 2.1         & 2$^{\circ}$    & 600         & $-$  & $-$ & 1016 \\
15    & 2D   & 2.1         & 2$^{\circ}$    & 800         & $-$  & $-$ & 1016 \\
15    & 2D   & 2.1         & 1.5$^{\circ}$  & 600         & $-$  & $-$ & 1016 \\
15    & 2D   & 2.1         & 1.5$^{\circ}$  & 800         & $-$  & $-$ & 1016 \\
15    & 2D   & 2.1         & 1$^{\circ}$    & 600         & 780  & 0.203 & 1016 \\
15    & 2D   & 2.1         & 1$^{\circ}$    & 800         & $-$  & $-$ & 1016 \\
15    & 2D   & 2.1         & 0.5$^{\circ}$  & 600         & 749  & 0.211 & 1016 \\
15    & 2D   & 2.1         & 0.5$^{\circ}$  & 800         & 886  & 0.198 & 1016 \\
\textbf{15}    & \textbf{3D}  & \textbf{2.1} & \textbf{3$^{\circ}$} &
\textbf{600}   & \textbf{$-$} & \textbf{$-$} & \textbf{946} \\
\textbf{15}    & \textbf{3D}  & \textbf{2.1} & \textbf{3$^{\circ}$} &
\textbf{800}   & \textbf{$-$} & \textbf{$-$} & \textbf{958} \\
      &      &             &                &             &     &     &     \\
15    & 2D   & 2.2         & 3$^{\circ}$    & 600         & $-$ & $-$ & 1016 \\
15    & 2D   & 2.2         & 3$^{\circ}$    & 800         & $-$ & $-$ & 1016 \\
15    & 2D   & 2.2         & 2$^{\circ}$    & 600         & 683 & 0.218 & 1016\\
15    & 2D   & 2.2         & 2$^{\circ}$    & 800         & $-$ & $-$ & 1016 \\
15    & 2D   & 2.2         & 1.5$^{\circ}$  & 600         & 713 & 0.215 & 1016 \\
15    & 2D   & 2.2         & 1.5$^{\circ}$  & 800         & 863 & 0.196 & 1016\\
15    & 2D   & 2.2         & 1$^{\circ}$    & 600         & 761 & 0.214 & 1016 \\
15    & 2D   & 2.2         & 1$^{\circ}$    & 800         & 961 & 0.194 & 1016\\
15    & 2D   & 2.2         & 0.5$^{\circ}$  & 600         & 588 & 0.232 & 1016 \\
15    & 2D   & 2.2         & 0.5$^{\circ}$  & 800         & 643 & 0.224 & 1016\\
\textbf{15}    & \textbf{3D}  & \textbf{2.2} & \textbf{3$^{\circ}$} &
\textbf{600}   & \textbf{803} & \textbf{0.202} & \textbf{975} \\
\textbf{15}    & \textbf{3D}  & \textbf{2.2} & \textbf{3$^{\circ}$} &
\textbf{800}   & \textbf{857} & \textbf{0.195} & \textbf{977} \\
\textbf{15}    & \textbf{3D}  & \textbf{2.2} & \textbf{2$^{\circ}$} &
\textbf{600}   & \textbf{$-$} & \textbf{$-$} & \textbf{988} \\
\textbf{15}    & \textbf{3D}  & \textbf{2.2} & \textbf{1.5$^{\circ}$} &
\textbf{800}   & \textbf{$-$} & \textbf{$-$} & \textbf{898} \\
\enddata
\tablenotetext{a}{Progenitor model.}
\tablenotetext{b}{Dimensionality.}
\tablenotetext{c}{Electron-neutrino luminosity.}
\tablenotetext{d}{Angular Resolution.}
\tablenotetext{e}{Number of radial zones.}
\tablenotetext{f}{Time of onset of explosion.}
\tablenotetext{g}{Mass accretion rate at onset of explosion.}
\tablenotetext{h}{Simulation time.}
\end{deluxetable}

\begin{figure}
\plotone{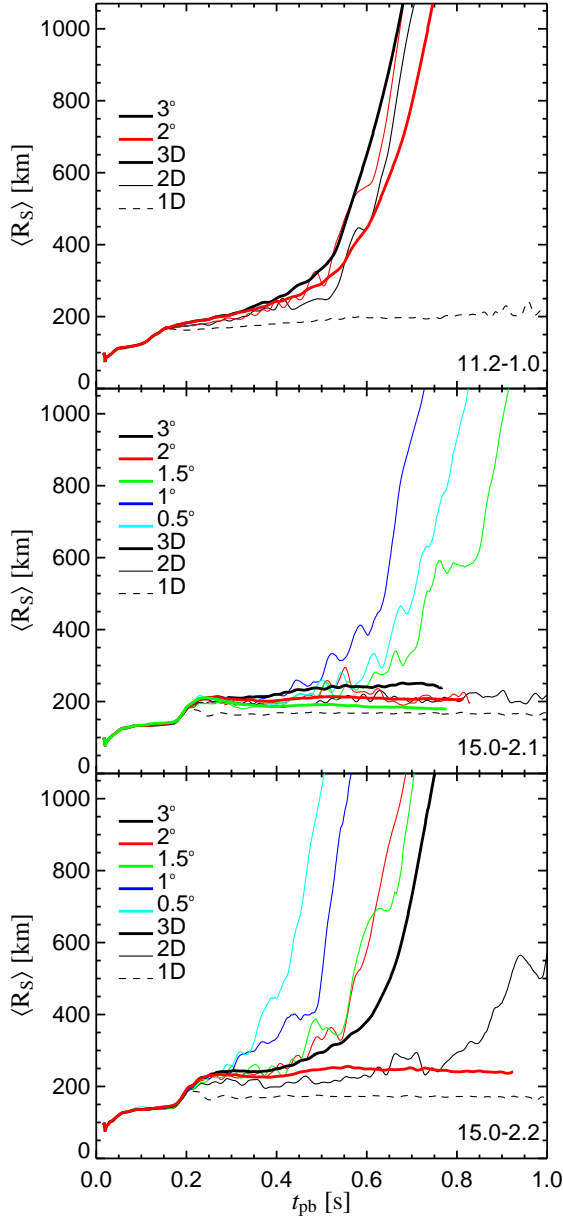}
\caption{Evolution of the average shock radius as a function of time (in
seconds after bounce) for one-dimensional (dashed lines), two-dimensional
(thin solid lines), and three-dimensional (thick solid lines) simulations
employing different angular resolutions (color coding). The top panel
displays the 11.2\,$M_{\odot}$ model for an electron-neutrino luminosity
of $L_{\nu_e} = 1.0\cdot10^{52}$\,erg\,s$^{-1}$, the middle panel shows
the 15\,$M_{\odot}$ star for an electron-neutrino luminosity of
$L_{\nu_e} = 2.1\cdot10^{52}$\,erg\,s$^{-1}$, and the bottom panel the
15\,$M_{\odot}$ results for $L_{\nu_e} = 2.2\cdot10^{52}$\,erg\,s$^{-1}$.}
\label{fig:spos_res}
\end{figure}

\begin{figure}
\plotone{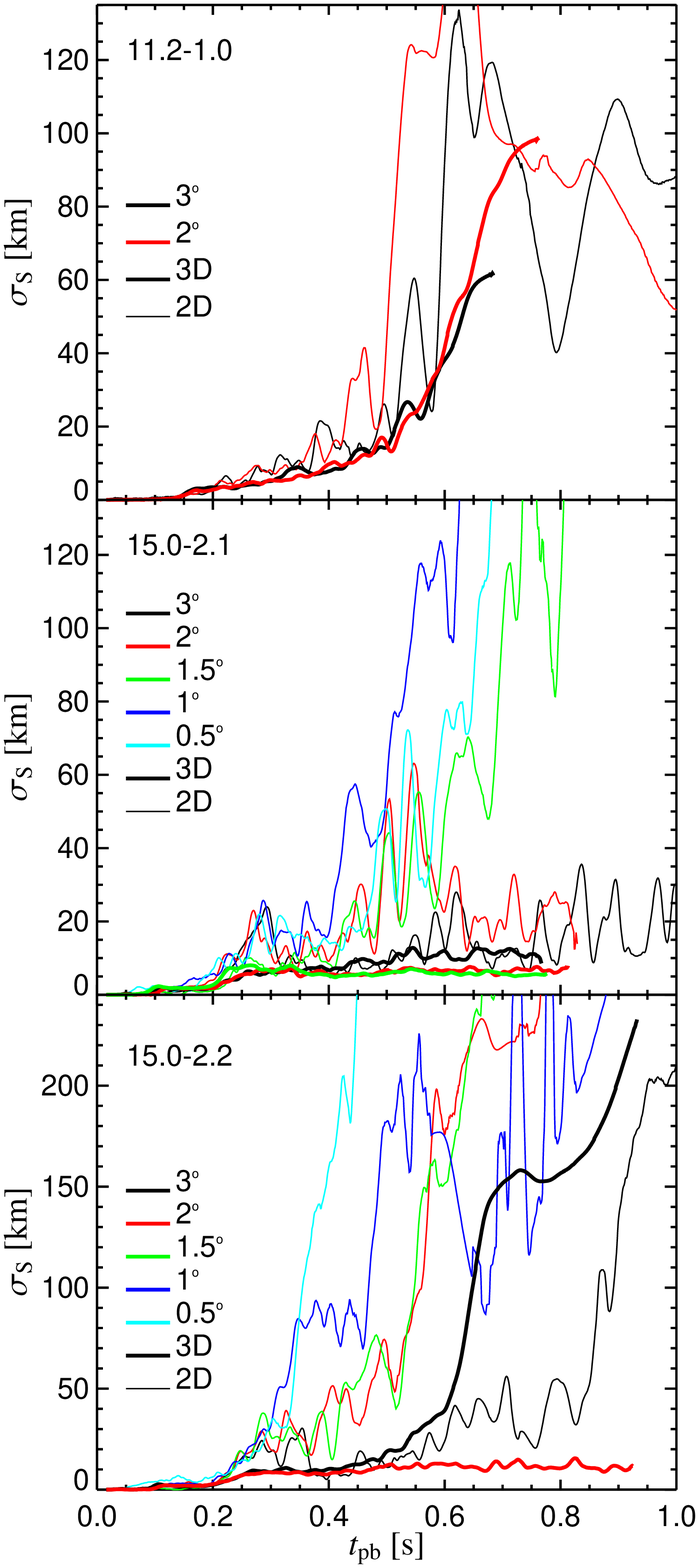}
\caption{Evolution of the standard deviation for the shock asphericity
as a function of post-bounce time for two-dimensional 
(thin solid lines) and three-dimensional (thick solid lines) simulations
employing different angular resolutions (color coding). As in 
Fig.~\ref{fig:spos_res} the top panel
displays the exploding 11.2\,$M_{\odot}$ models for an electron-neutrino 
luminosity of $L_{\nu_e} = 1.0\cdot10^{52}$\,erg\,s$^{-1}$. The
middle panel contains the results for the 15\,$M_{\odot}$ star with an
electron-neutrino luminosity of 
$L_{\nu_e} = 2.1\cdot10^{52}$\,erg\,s$^{-1}$, where 2D runs with higher
resolution lead to explosions while 3D runs do not. The bottom
panel shows the 15\,$M_{\odot}$ case for 
$L_{\nu_e} = 2.2\cdot10^{52}$\,erg\,s$^{-1}$. It is remarkable that
the 3D run with 2$^\circ$ angular resolution does not explode whereas
the one with angle bins of 3$^\circ$ explodes earlier
than its 2D counterpart and develops a very large shock deformation at
the time the explosion sets in.}
\label{fig:sigma_res}
\end{figure}

\begin{figure*}
 \plottwo{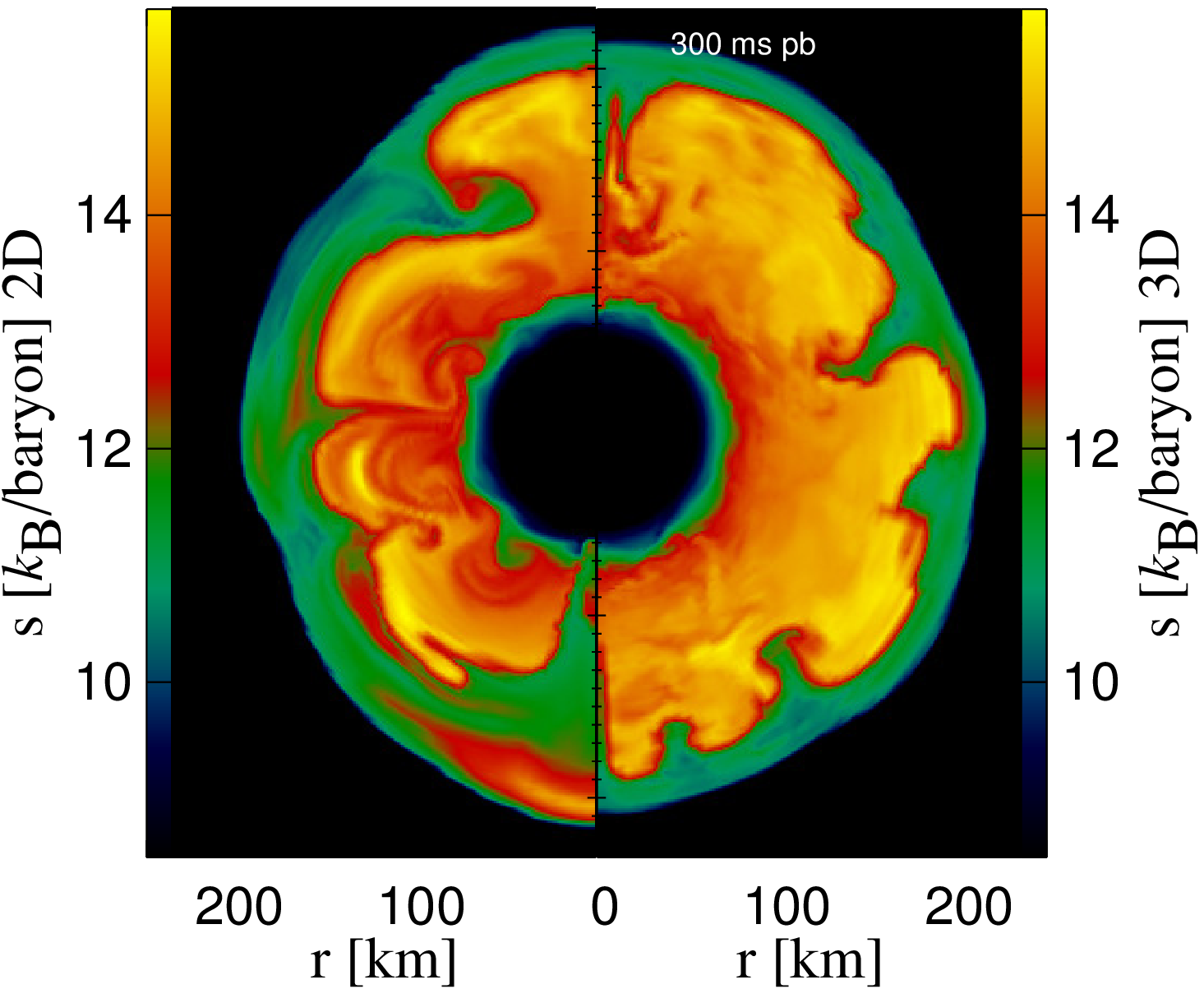}{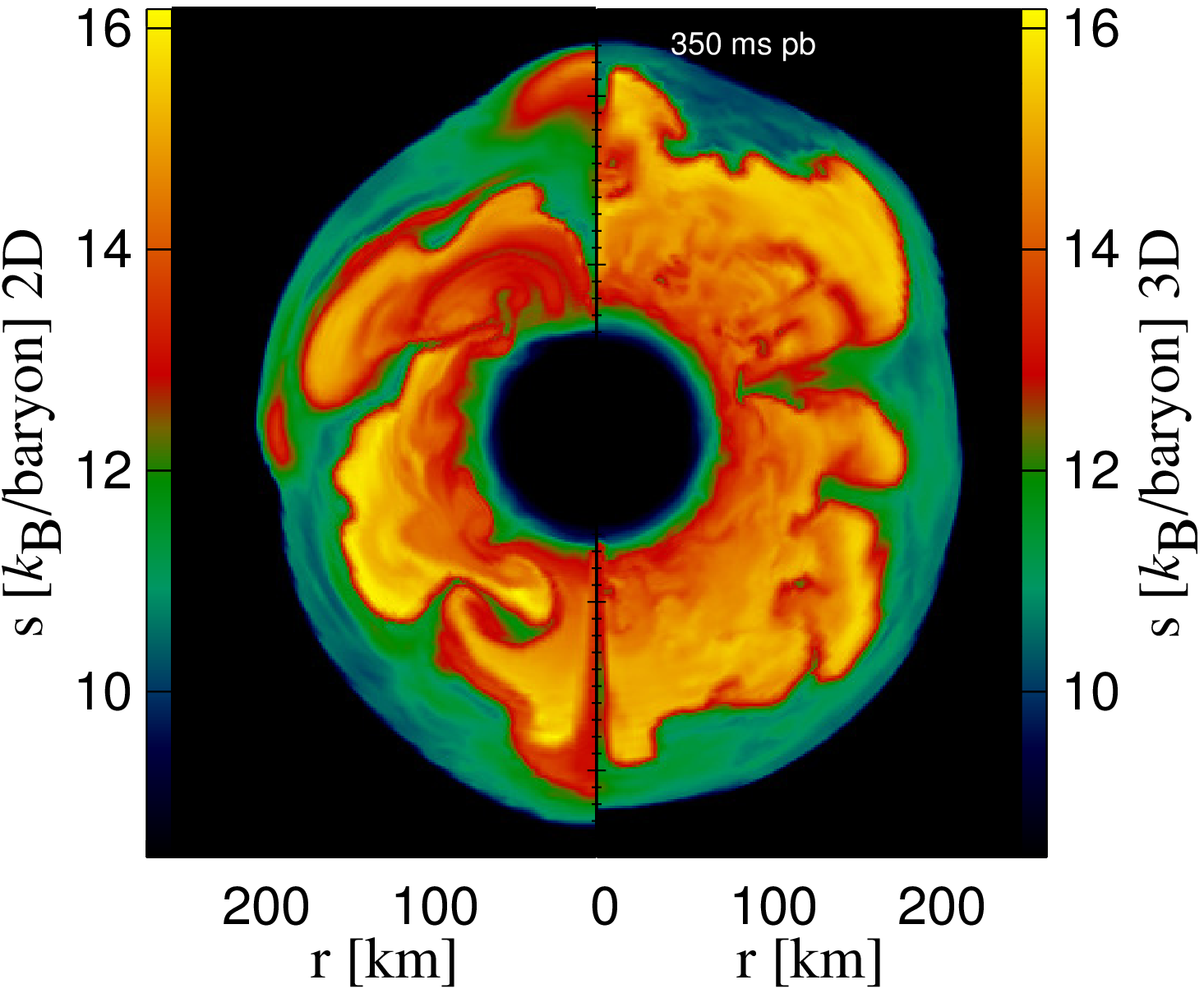}\\
 \plottwo{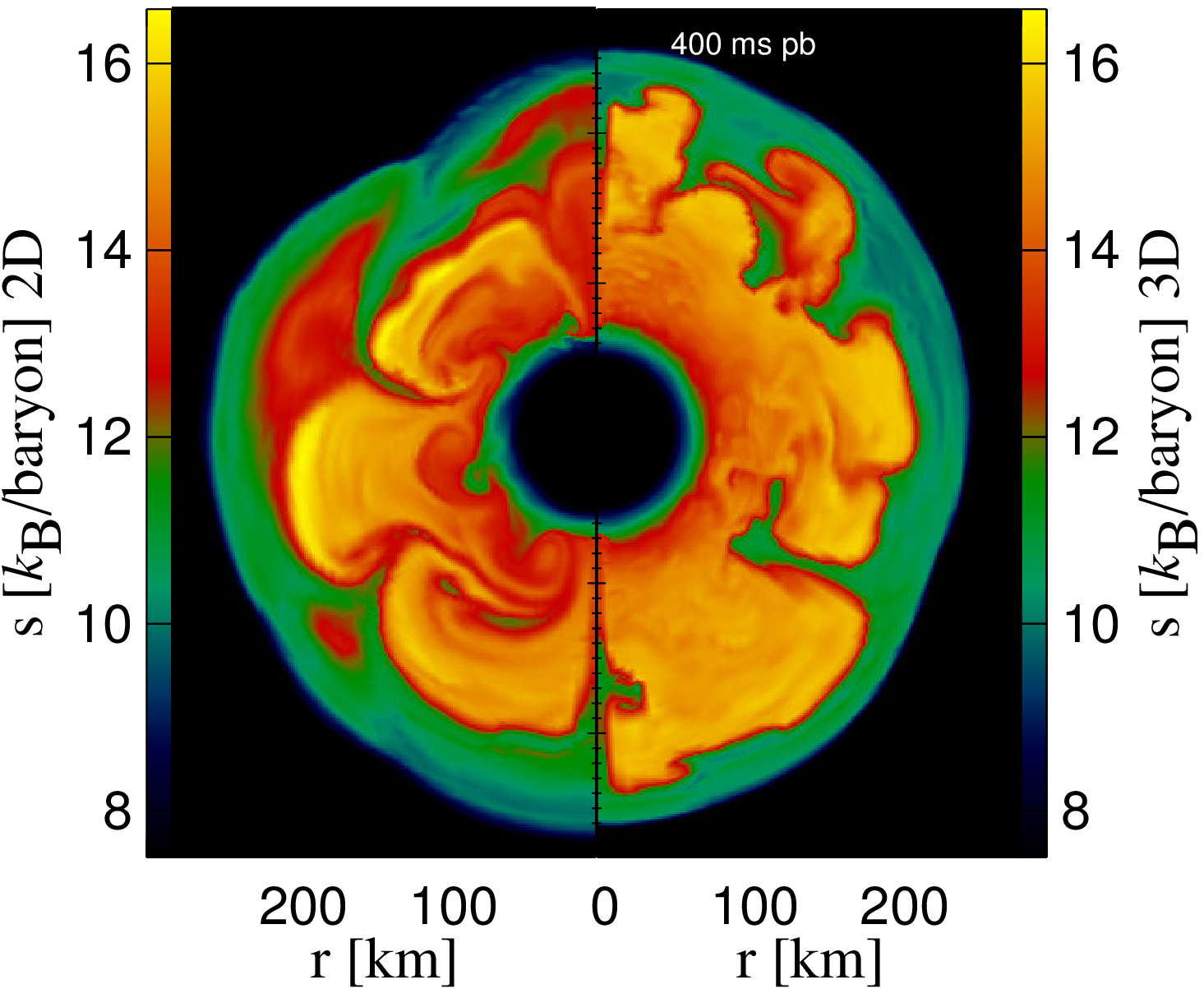}{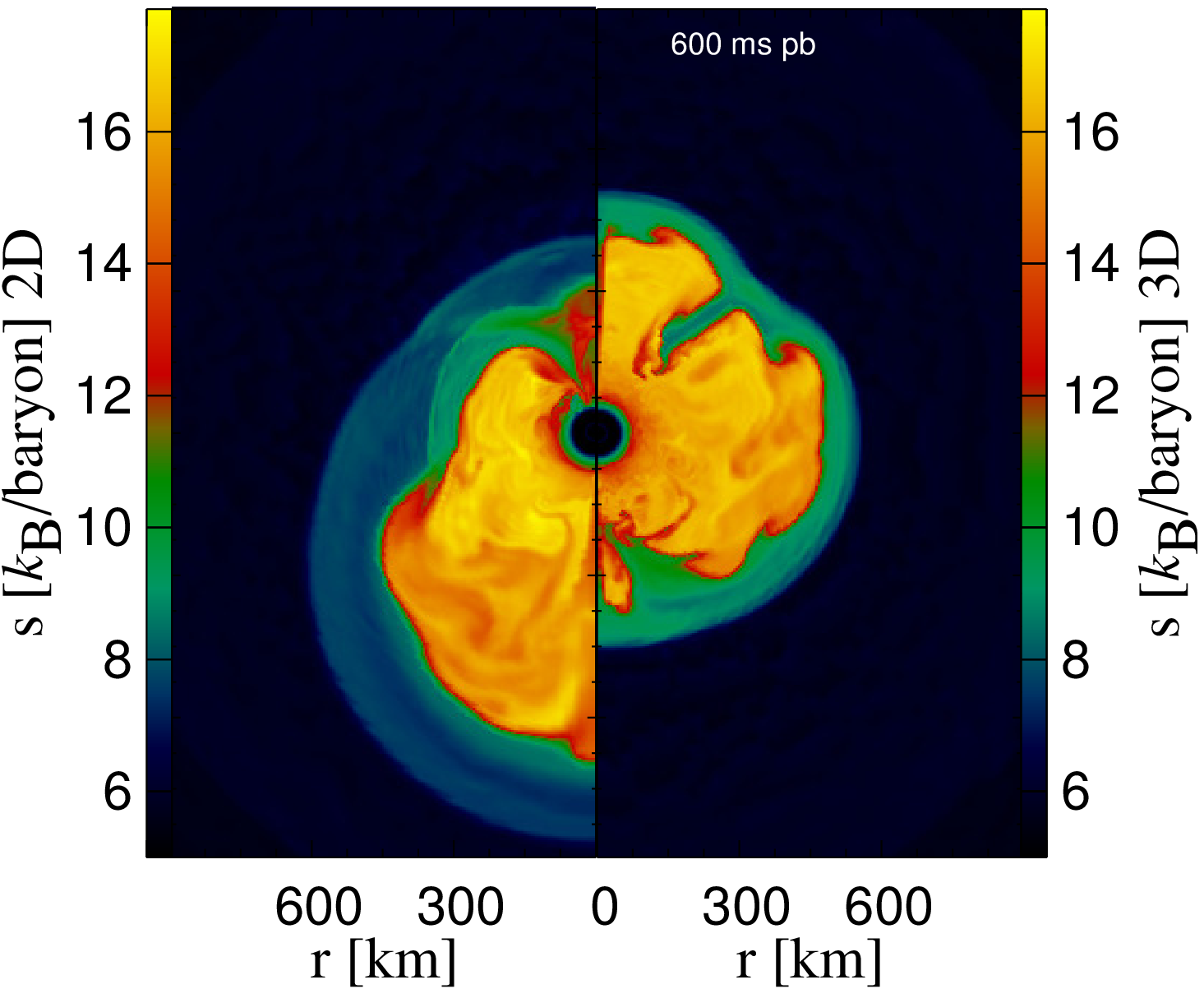}
 \caption{Snapshots of the evolution of the 11.2\,$M_{\odot}$
model with an electron-neutrino luminosity of 
$L_{\nu_e} = 1.0\cdot10^{52}$\,erg\,s$^{-1}$ and 2$^\circ$ angular resolution
at post-bounce times of $t_\mathrm{pb} =$\,300, 350, 400, and 600\,ms (from
top left to bottom right).
The color coding represents the entropy per nucleon of the stellar plasma.
The left half of each panel displays the entropy distribution for a 2D
(axisymmetric) simulation, the right half shows the structure in a
meridional cut of the corresponding 3D simulation. Both models explode
after roughly 550\,ms after bounce (see Fig.~\ref{fig:spos_res}, top panel, and 
Table~\ref{table:models_res}). Note that the structures of low-entropy downdrafts
and high-entropy plumes in the neutrino-heating region are rather similar for both
cases, despite considerably larger SASI sloshing motions of the shock and
postshock layer in 2D. When the explosion has set in, the 2D model exhibits
an apparent prolate deformation whose development is supported by the symmetry axis
defining a preferred direction of the 2D system. While the 3D explosion does 
not appear to be as strongly distorted (in particular the shock surface looks
more spherical), the postshock flow in this case also develops a pronounced
hemispheric (dipolar) asymmetry, which can be more clearly seen in the upper
and lower right panels of Fig.~\ref{fig:viz}.}
\label{fig:sn_s11_hr}
\end{figure*}

\begin{figure*}
 \plottwo{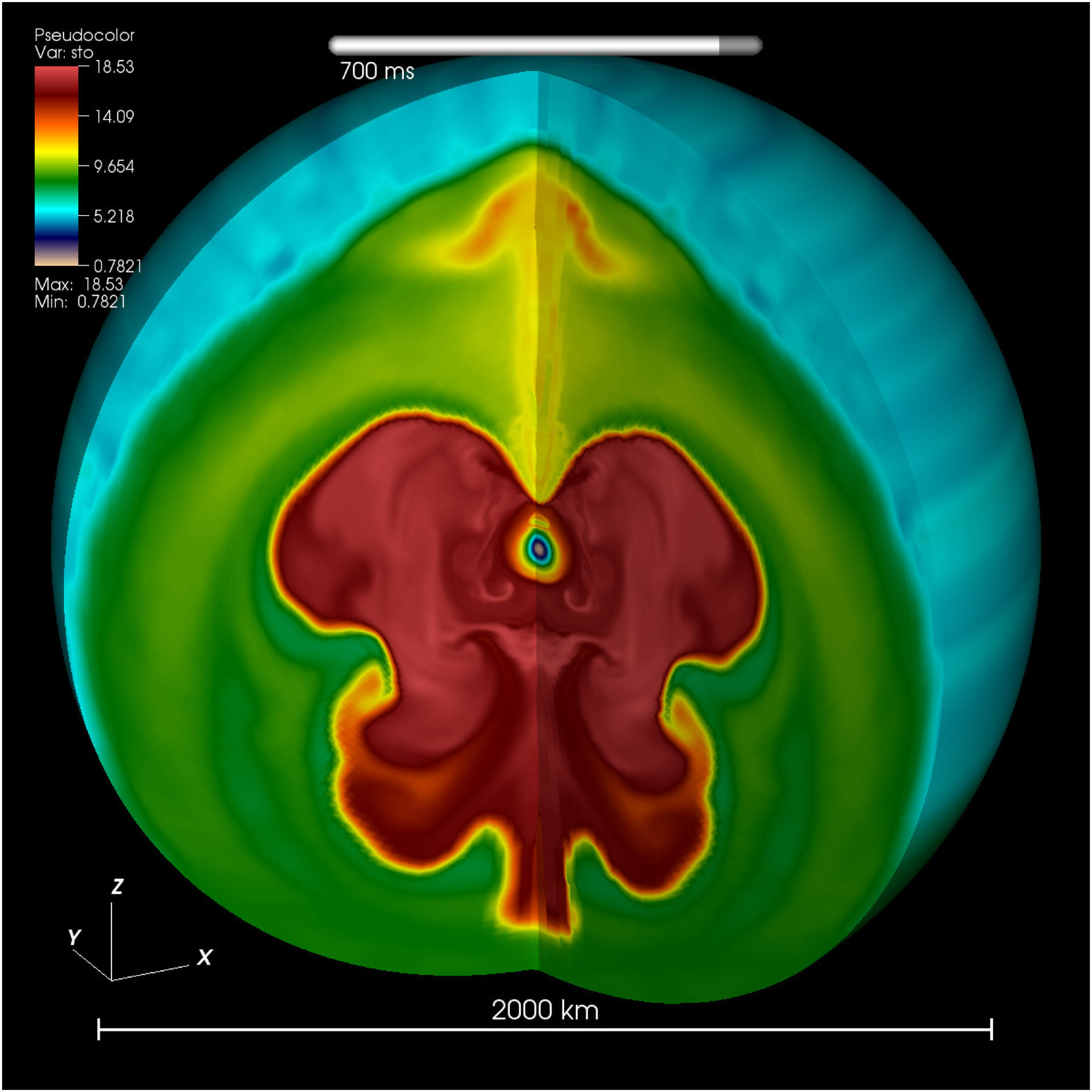}{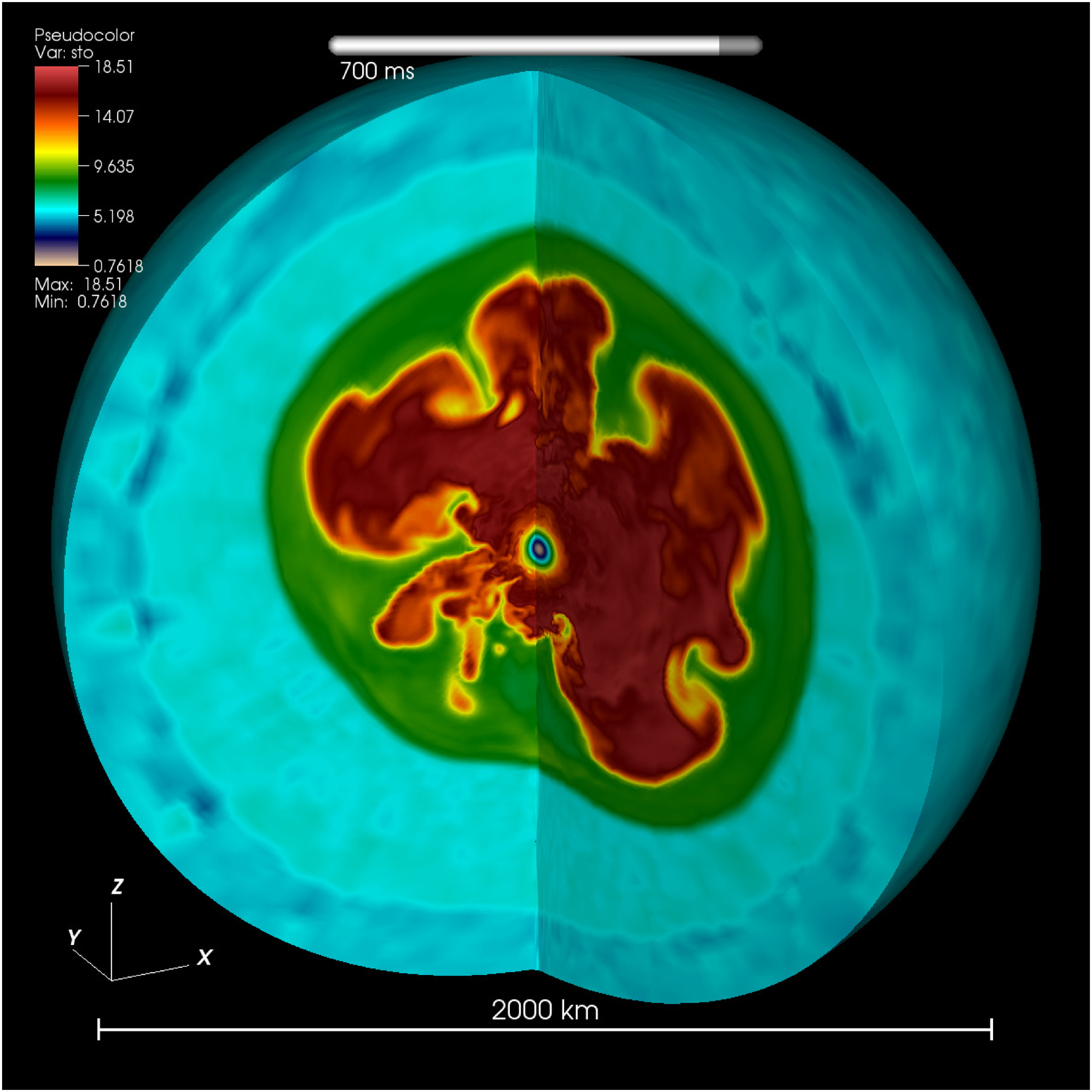}\\
 \plottwo{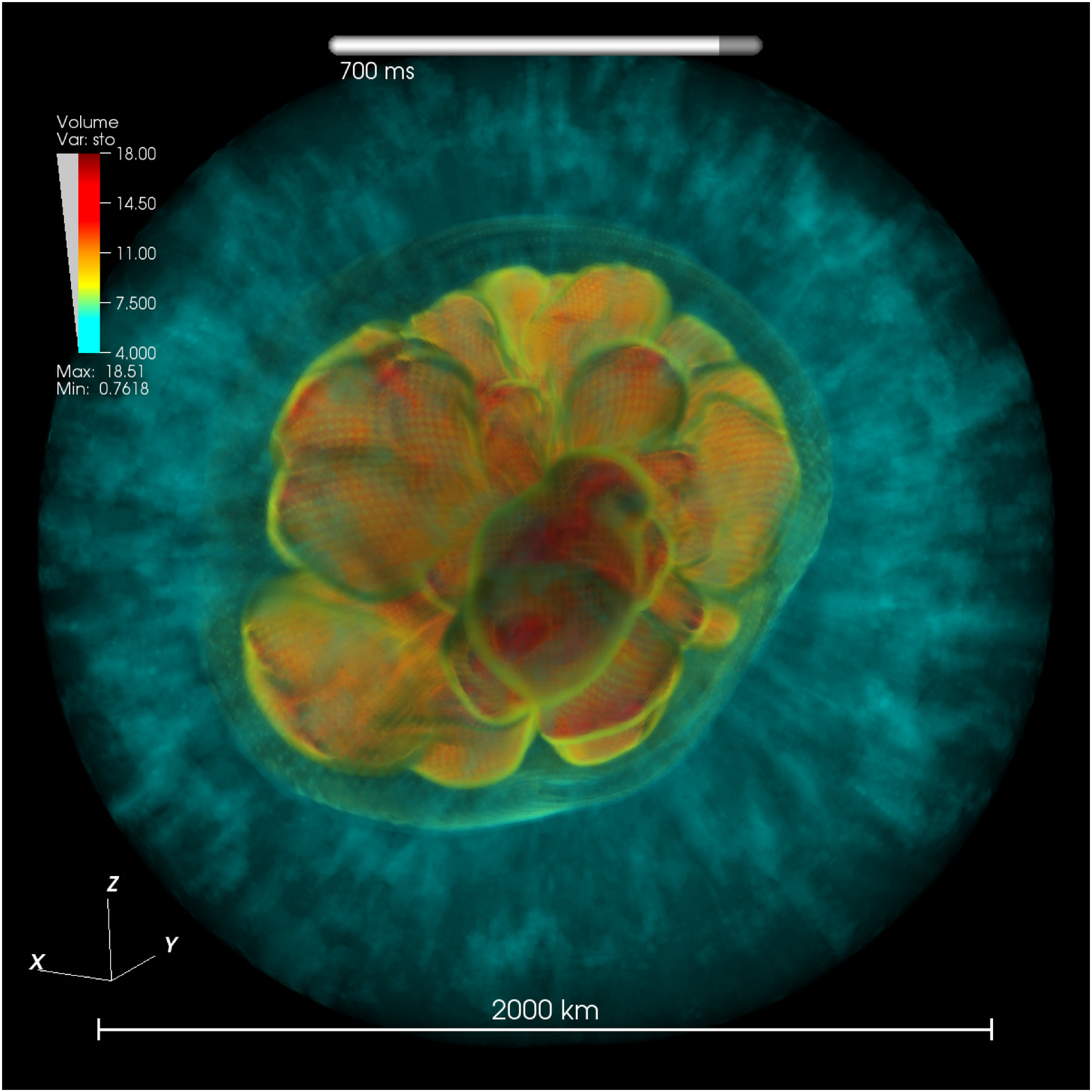}{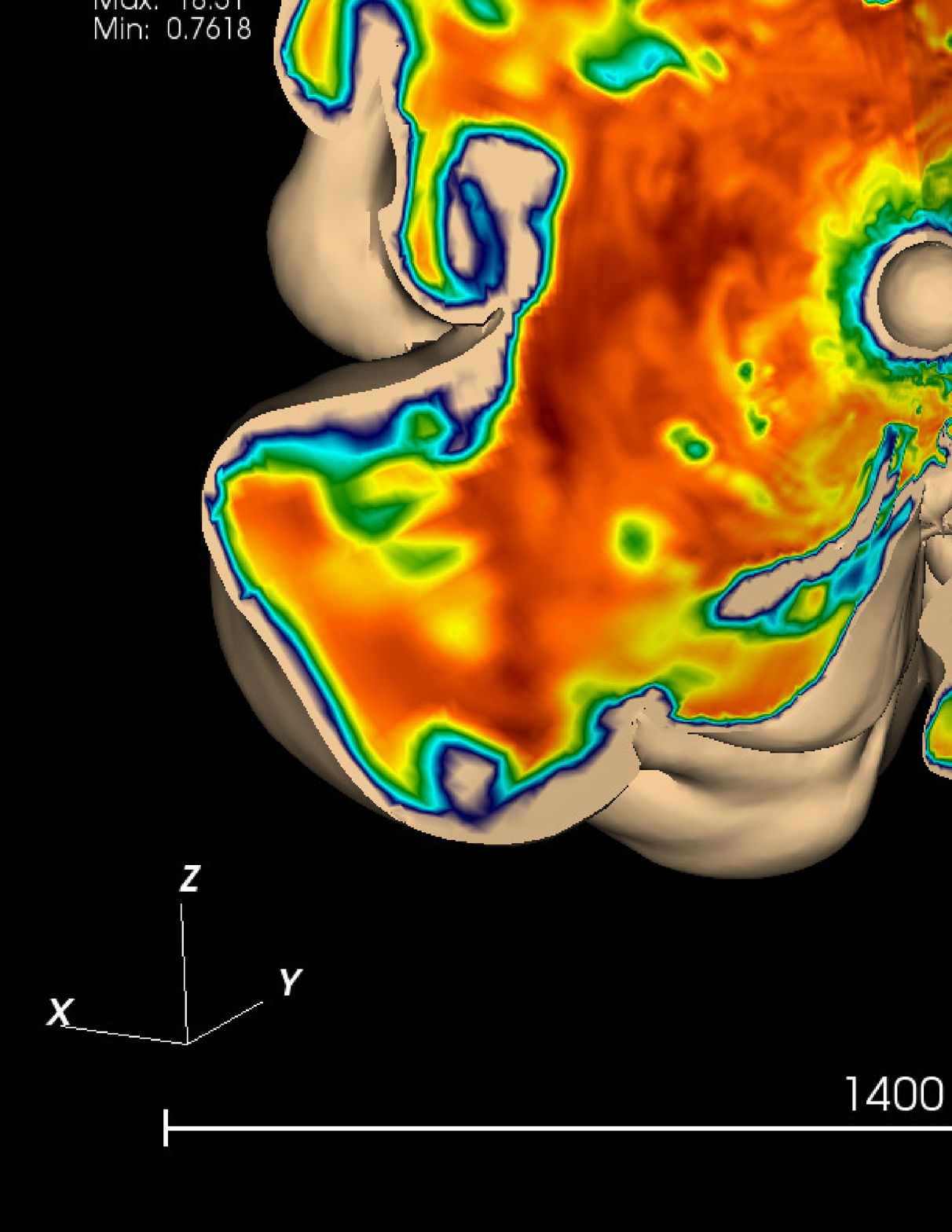}
 \caption{{\em Upper row:} 
Quasi-three-dimensional visualization of the 11.2\,$M_{\odot}$ simulations
in 2D (upper left panel) and 3D (upper right panel) with an
electron-neutrino luminosity of $L_{\nu_e} = 1.0\cdot10^{52}$\,erg\,s$^{-1}$
and an angular resolution of 2$^\circ$, comparing the structure at 
700\,ms p.b., roughly 150\,ms after the onset of the explosions. Since
the explosion started slightly earlier in the 2D model (see the upper panel
of Fig.~\ref{fig:spos_res} and Table~\ref{table:models_res})
the shock is more extended in the left image. While in this case 
the shock possesses a much stronger dipolar deformation component than in
3D (cf.\ Fig.~\ref{fig:sn_s11_hr}, lower right panel),
the distribution of accretion funnels and plumes of neutrino-heated matter
exhibits a hemispheric asymmetry in both cases. Because of the axisymmetry of
the 2D geometry this concerns the hemispheres above and below the $x$-$y$-plane 
in the upper left plot, whereas the virtual equator lies in the plane connecting
the upper left and lower right corners of the top right image and the lower
left and upper right corners of the bottom right picture.
Note that the jet-like axis feature in the upper left figure is a
consequence of the symmetry constraints of the 2D setup, which redirect
flows moving towards the polar grid axis. Such artifacts do not occur in
the 3D simulation despite the use of a polar coordinate grid there, too.
{\em Lower row:}
Ray-tracing and volume-rendering images of the three-dimensional 
explosion of the 11.2\,$M_{\odot}$ progenitor for the same simulation and
time displayed in the upper right image. 
The left lower panel visualizes the outer boundaries of the buoyant 
bubbles of neutrino-heated gas and the outward driven shock, which
can be recognized as a nearly transparent, enveloping surface. The 
visualization uses the fact that both are entropy discontinuities in the
flow. The infalling matter in the preshock region appears as diffuse,
nebular cloud. The right lower panel displays the interior structure by the
entropy per nucleon of the plasma (red, yellow, green, light blue, dark blue
correspond to decreasing values) within the volume formed by the high-entropy
bubbles, whose surface is cut open by removing a wide cone facing the 
observer. Note the clear dipolar anisotropy with stronger explosion towards
the north-west direction and more accretion at the south-east side of the
structure.}
\label{fig:viz}
\end{figure*}

\section{Models with higher resolution}
\label{sec:high_res}

In order to test whether our results for the multi-dimensional models
depend on the agreeably moderate 3$^\circ$ angular resolution used in
the standard runs, we performed a large
set of simulations with finer grid spacing especially in the
angular directions, but also in radial direction. For this purpose
we concentrated on cases around the minimum values of the driving
luminosity that triggered explosions of both progenitors in our 
standard runs. The results are listed in Table~\ref{table:models_res}.
They indicate a very interesting trend: 2D models with finer angular
zoning tend to explode more readily, whereas better angular resolution 
in 3D simulations turns out to have the opposite effect.

In the case of the 11.2\,$M_\odot$ progenitor, for example, the 3D 
explosion found
to set in at about 730\,ms p.b.\ with an angular zone size of 3$^\circ$ 
for $L_{\nu_e} = 0.9\cdot10^{52}$\,erg\,s$^{-1}$ cannot be reproduced
with an angle binning of 2$^\circ$. Moreover, a luminosity of
$L_{\nu_e} = 1.0\cdot10^{52}$\,erg\,s$^{-1}$ leads to an explosion of
the 3D model at $\sim$540\,ms after bounce with a 3$^\circ$-grid,
but $\sim$35\,ms later when 2$^\circ$ are used.
The corresponding 2D models show the inverse trend as visible in
the top panel of Fig.~\ref{fig:spos_res}. 

Note that the average
shock radii plotted in Fig.~\ref{fig:spos_res} as well as
Fig.~\ref{fig:spos} exhibit alternating periods of increase and
decrease in particular in 2D simulations. Such features are a 
consequence of the strong sloshing motions of the accretion shock and
of the associated time-dependent, large global shock deformations,
which are typical of violent activity by the SASI.
In 3D the corresponding wiggles and local
maxima of the average shock trajectory are much less pronounced.
A measure of the degree of shock asphericity, irrespective of the
relative weights of different spherical harmonics components, is
the standard deviation of the shock radius defined by
$\sigma_\mathrm{S} \equiv 
\sqrt{\frac{1}{4\pi}\oint\mathrm{d}\Omega\,
[R_\mathrm{S}(\vec{\Omega})-\langle R_\mathrm{S}\rangle]^2}$.
The standard deviations corresponding to the average shock
radii of Fig.~\ref{fig:spos_res} are plotted in 
Fig.~\ref{fig:sigma_res}, which confirms the mentioned 
difference between 2D and 3D runs.

In spite of this 2D-3D difference of the shock asphericity,
an inspection of cross-sectional snapshots of 
our best resolved simulations of the 11.2\,$M_\odot$ progenitor
with an electron-neutrino luminosity of
$L_{\nu_e} = 1.0\cdot10^{52}$\,erg\,s$^{-1}$ reveals that the
sizes of the convective plumes and the structure of the 
neutrino-heated postshock layer are fairly similar in the 2D and 
3D cases before explosion (which in both models develops
shortly after 500\,ms): 
In Fig.~\ref{fig:sn_s11_hr} it is difficult to judge
by eye inspection whether the displayed simulation was conducted 
in 2D (left half-panels) or 3D (right half-panels). Even after the
explosions have taken off the global deformation of the shock
in both cases is not fundamentally different in the sense
that low-order spherical harmonics modes (dipolar and quadrupolar
components) determine the global asymmetry of 
the shock surface and in particular of the distribution of 
downdrafts and expanding bubbles in the gain region (see
Fig.~\ref{fig:sn_s11_hr}, lower right panel, and Fig.~\ref{fig:viz}).
At a closer inspection one can notice some secondary differences
in the morphology of the convective and downflow features. 
Despite the same angular resolution the images of 
Figs.~\ref{fig:sn_s11_hr} and \ref{fig:viz} reveal more small
structures in the 3D case compared to the 2D data, which appear
more coherent, smoother, and less fragmented into finer 
substructures and filaments. We will
refer to this observation in Sect.~\ref{sec:cond}.

\begin{figure*}
 \plottwo{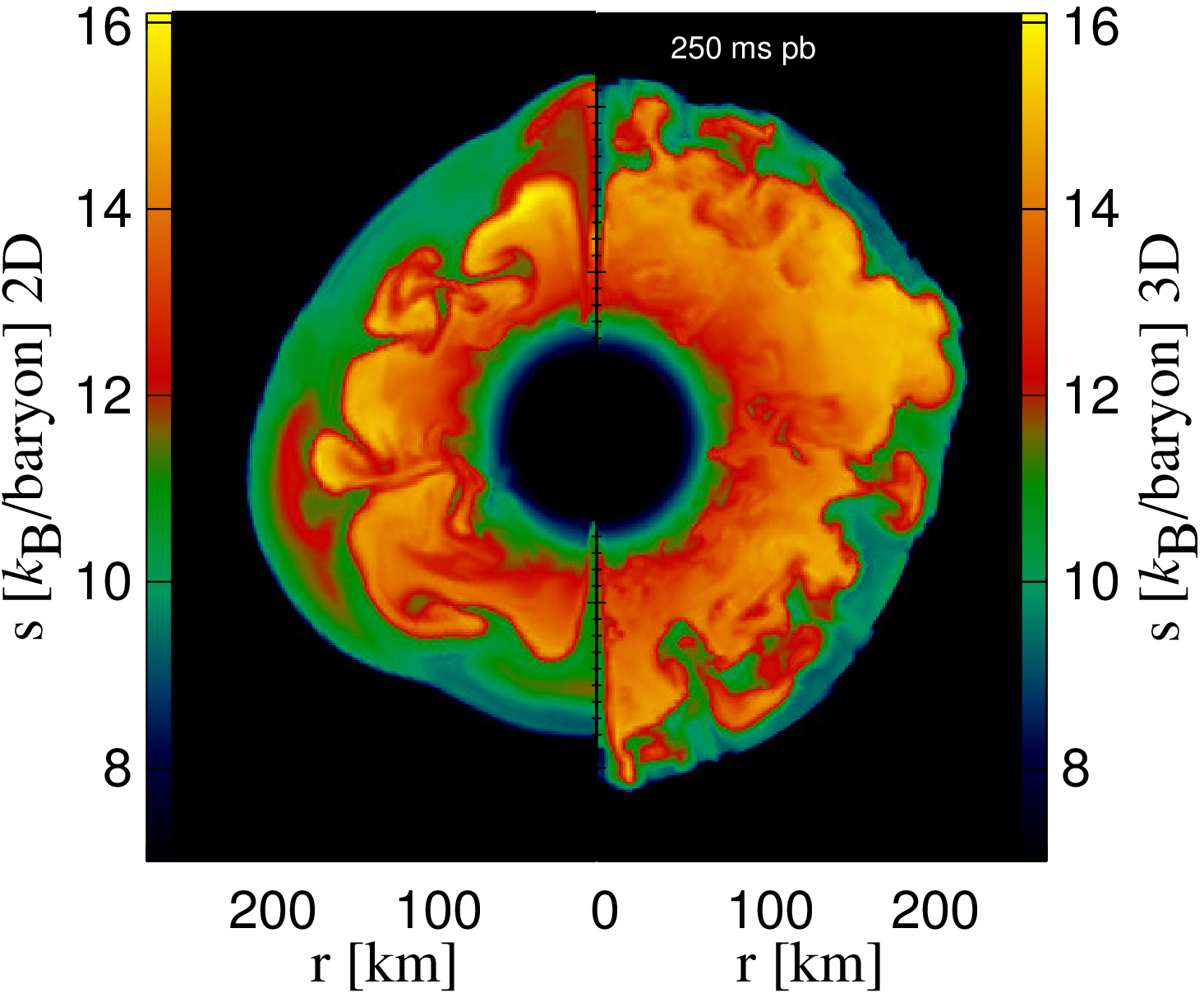}{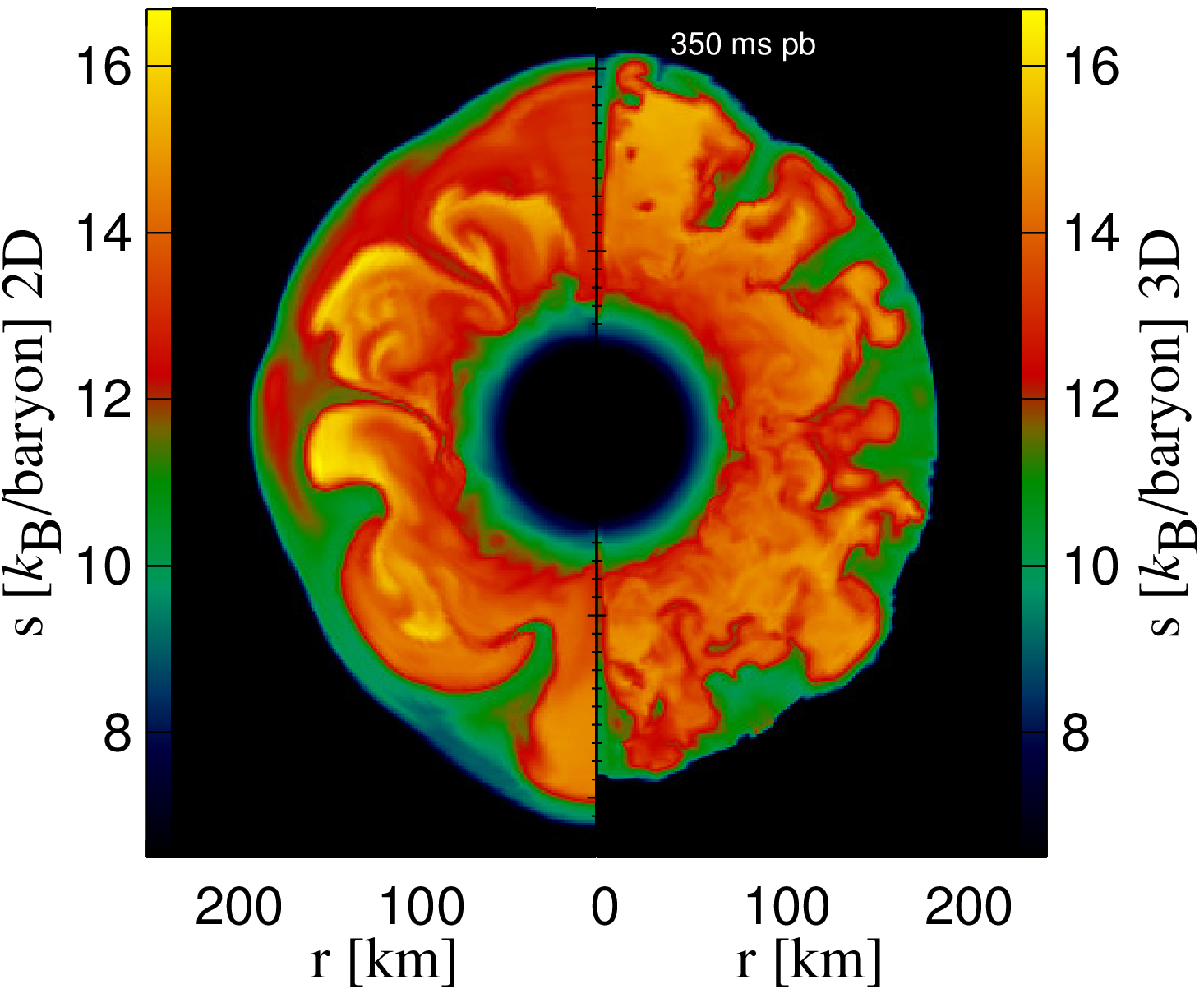}\\
 \plottwo{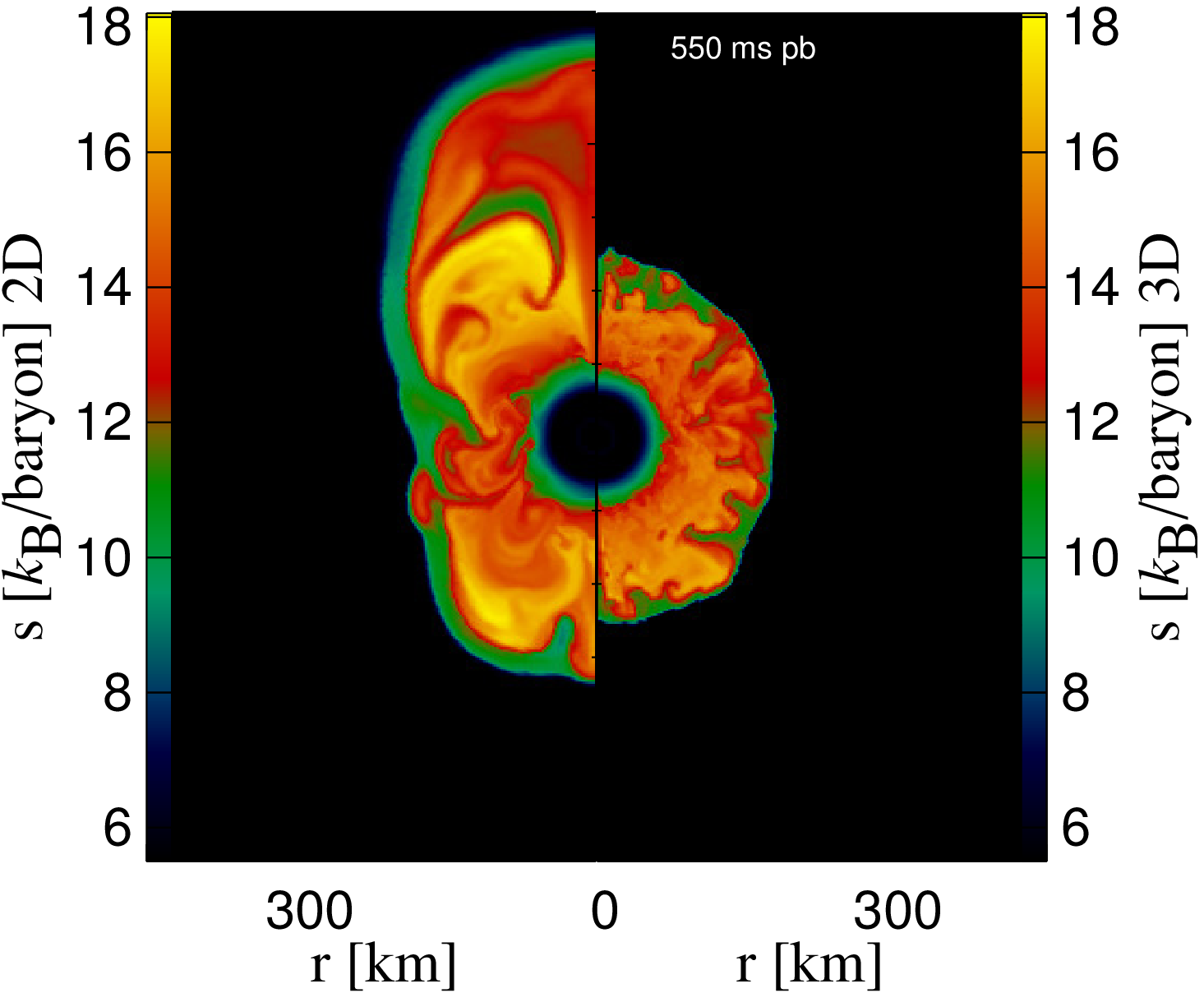}{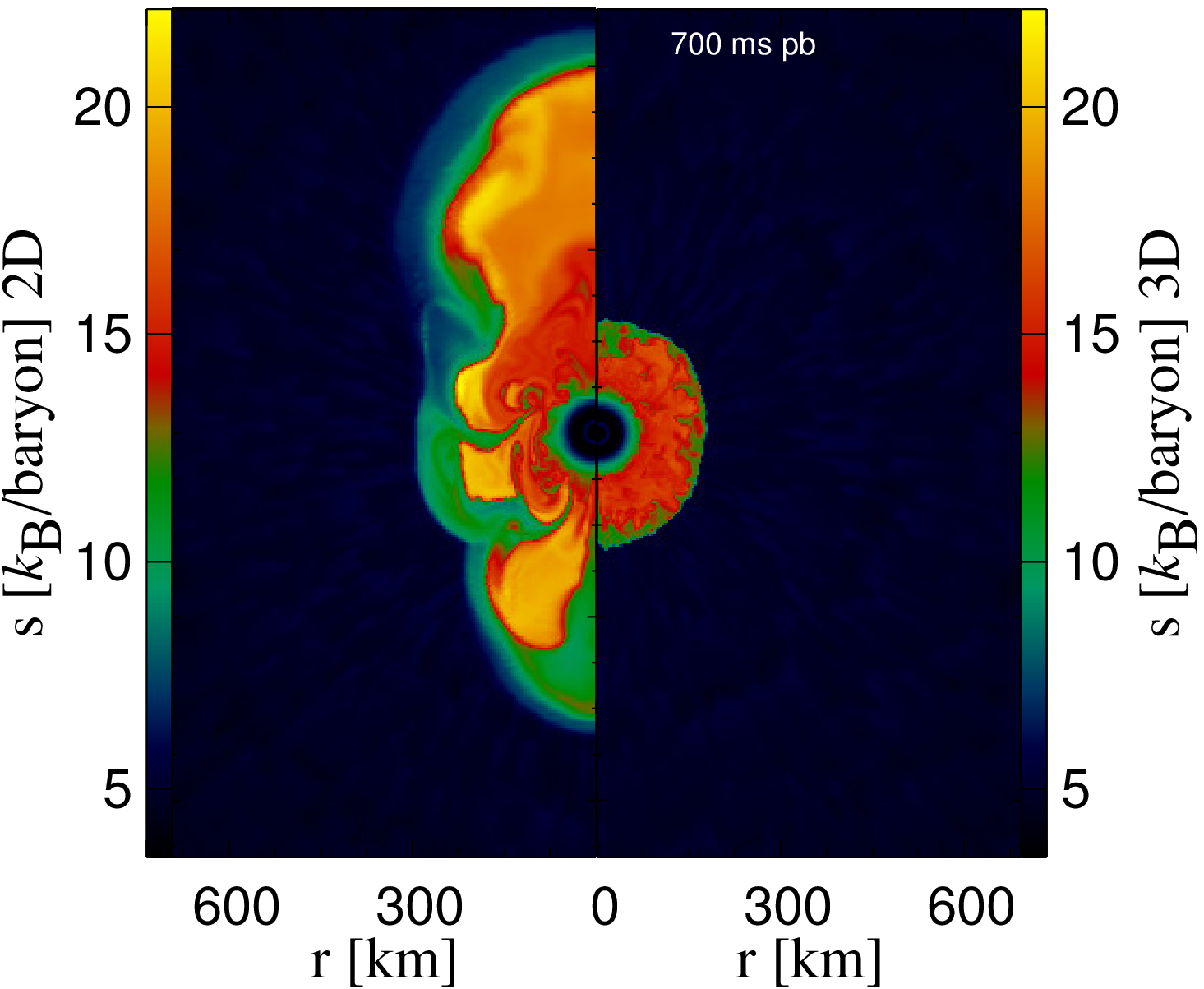}
 \caption{Snapshots of the post-bounce evolution of the
15\,$M_{\odot}$ model with an electron-neutrino luminosity of $L_{\nu_e} =
2.1\cdot10^{52}$\,erg\,s$^{-1}$ and angular resolution of 1.5$^\circ$
at $t_\mathrm{pb} =$\,250, 350, 550, and 700\,ms. The color coding represents
the entropy per nucleon of the stellar gas. The left half of each panel shows
the entropy distribution for a 2D simulation, and the right half displays a 
meridional cut from the corresponding 3D model. While the 2D run with
the given resolution leads to a highly prolate explosion, the 3D calculation
does not end in a successful blast (see Fig.~\ref{fig:spos_res}, middle panel,
and Table~\ref{table:models_res}). Note that the convective plumes are
considerably smaller and more fine-structured in the 3D simulation.}
\label{fig:sn_s15_hr}
\end{figure*}

Our 15\,$M_\odot$ runs with varied resolution confirm the trends
seen for the 11.2\,$M_\odot$ progenitor. For a neutrino luminosity of
$L_{\nu_e} = 2.1\cdot 10^{52}$\,erg\,s$^{-1}$, for which neither
2D nor 3D simulations with standard resolution produce an explosion,
we find that 2D models with angular binning of 1.5$^\circ$ or
better do explode, whereas explosions in 3D cannot be obtained with
angular zones in the range from
1.5$^\circ$ to 3$^\circ$ (Fig.~\ref{fig:spos_res}, middle panel;
Table~\ref{table:models_res}).
The 2D simulations exhibit violent SASI sloshing motions and the
quasi-periodic appearance of large
shock asymmetries (Fig.~\ref{fig:sigma_res}, middle panel),
and the 2D model with 1.5$^\circ$ angular zoning explodes with a 
huge prolate deformation (Fig.~\ref{fig:sn_s15_hr}).
A similar behavior is seen for the 15\,$M_\odot$ runs with
$L_{\nu_e} = 2.2\cdot 10^{52}$\,erg\,s$^{-1}$: While all 2D models
computed with angular zone sizes between 0.5$^\circ$ and 3$^\circ$
explode, we observe an explosion for the 3D calculation with 
3$^\circ$ but none for the case with 2$^\circ$ angular binning
(Table~\ref{table:models_res} and Fig.~\ref{fig:spos_res}, bottom 
panel). It is highly interesting that the 3$^\circ$ case, where the 
explosion occurs more readily in 3D than in 2D, is associated with
a large asphericity of the supernova shock at the time the 
3D run begins to develop the successful blast 
(Fig.~\ref{fig:sigma_res}, bottom panel).
Note again that the structures of the higher-resolved 3D model in 
Fig.~\ref{fig:sn_s15_hr} reveal finer details and fragmentation
into smaller filaments than the corresponding 2D simulation, 
despite both having the same zone sizes in the angular directions.

The data listed in Table~\ref{table:models_res} contain the
clear message that 2D models with better angular resolution 
usually develop explosions
earlier in contrast to 3D runs, which explode later or not at all
when the angular zoning is finer. There can be 2D exceptions 
to the general trend (e.g., the 15\,$M_\odot$ cases with 
$L_{\nu_e} = 2.1\cdot 10^{52}$\,erg\,s$^{-1}$ and 0.5$^\circ$ 
and 1$^\circ$ resolution for 400 radial zones; see also 
Fig.~\ref{fig:spos_res}, middle panel), which are either affected by
the difficulty to exactly determine the onset of the blast in cases
with a highly deformed shock, or which can be stochastic outliers 
associated with the chaotic processes leading to the explosion.
Note in this context that at late times $\dot M(t)$ is very flat
and therefore differences in $t_\mathrm{exp}$ correspond to only
small differences in $\dot M_\mathrm{exp}$. For this reason the
beginning of the explosion can be shifted by minor perturbations, 
e.g.\ connected to stochastic fluctuations.
It is also possible that for special circumstances 
the symmetry axis of the 2D geometry has 
an influence on such a non-standard behavior because of its effect 
to redirect converging flows outwards or inwards and thus to 
have a positive feedback on the violence of the SASI activity.

Improved radial resolution for fixed angular grid turns out 
to have a negative influence on the possibility of an explosion
also in multi-dimensional simulations.
2D runs with better radial zoning (600 or even 800 instead of 
400 radial zones) fail to develop explosions or explode
significantly later than their less well resolved counterparts (see 
the 15\,$M_\odot$ results for $L_{\nu_e} = 2.1\cdot 10^{52}$\,erg\,s$^{-1}$
and $2.2\cdot 10^{52}$\,erg\,s$^{-1}$ in Table~\ref{table:models_res}).
In general, in 1D, 2D, and 3D improved radial resolution
shifts the onset of the explosion to later times monotonically. 
The only successful 3D model in the set that
is useful for the present discussion, a 15\,$M_\odot$ run with 
$L_{\nu_e} = 2.2\cdot 10^{52}$\,erg\,s$^{-1}$ and 3$^\circ$ angular
resolution, supports our findings in 2D. It shows an increasingly
delayed explosion for better radial resolution, although the results
with 600 and 800 radial zones appear to be nearly converged. All
non-exploding 3D models do not yield successes also with higher 
radial resolution.

The conclusion that good radial resolution is very important for reliable 
results, in particular when the explosion is ``marginal'', would not
be a surprise, because \cite{Sato2009} have pointed
out the importance of the radial zoning close to the neutron star
and around the supernova shock in order to accurately capture
the entropy and vorticity production at the shock and to determine
growth times and oscillation frequencies of the SASI. The latter
is unquestionably an essential ingredient for the success of the 
neutrino-driven mechanism in our 2D runs and it may as well be 
a crucial component for the mechanism to work in 3D.
As mentioned in Sect.~\ref{sec:crit_lum}, however, the sensitive 
influence of the radial zoning in the discussed 
model set is mainly a consequence of an artificial
density peak developing in the neutrino-loss region because
of the use of the simplified neutrino-cooling treatment. The narrow
shape of this numerical artifact in the density structure, which
enhances the energy emission by neutrinos, can even cause a 
dependence of the results on the particular choice of the grid-cell
locations. Different from the 1D runs the results of multi-dimensional 
simulations with more than $\sim$600 radial zones do not seem to 
converge. Since the local density maximum lies between
$10^{12}$\,g\,cm$^{-3}$ and $10^{13}$\,g\,cm$^{-3}$ in the core that
is treated spherically symmetrically in our simulations, we interpret
this phenomenon as the consequence of a subtle feedback between 
higher zoning and cooling strength on the one hand and 
multi-dimensional processes in the accretion layer on the other hand.
Because of the artificial nature of the underlying density feature,
however, we did not further explore this finding.

Finally, we remark that prior to our present work \cite{Scheck2007} 
has already performed resolution studies with a large set of 
2D simulations, in which he varied the lateral zone width between
0.5$^\circ$ and 4$^\circ$. In addition, he conducted three 3D
simulations with angular bin sizes of 2--4$^\circ$. However, instead 
of the highly simplified heating and cooling description used by us
he employed the much more sophisticated approximation for grey
neutrino transport described in detail in \cite{Scheck2006}.
This approximation included, e.g., the feedback of accretion on the 
neutrino emission properties and on the corresponding energy and lepton
number transport by neutrinos and antineutrinos of all flavors, as 
well as a more elaborate description of neutrino-matter interactions
in detailed dependence on the thermodynamical state of the stellar 
plasma. \cite{Scheck2007} was not interested in a systematic 
exploration of the critical explosion condition, but his project was
focussed on investigating the possibility of hydrodynamic pulsar 
kicks by successful asymmetric explosions. 
Despite the grave differences of the neutrino treatments
and numerical setups, the results obtained by \cite{Scheck2007} 
are compatible with our present findings: 2D simulations with higher
resolution turned out to yield explosions significantly earlier 
and thus also more energetically than the low-resolution runs.
Within the tested range of angular resolutions \cite{Scheck2007}
did not observe any significant differences between his 3D models.
This, however, may just be a consequence of the fact that the
models were clearly above the threshold conditions for an explosion
and did not linger along the borderline between blast and failure.

\begin{figure}
 \plotone{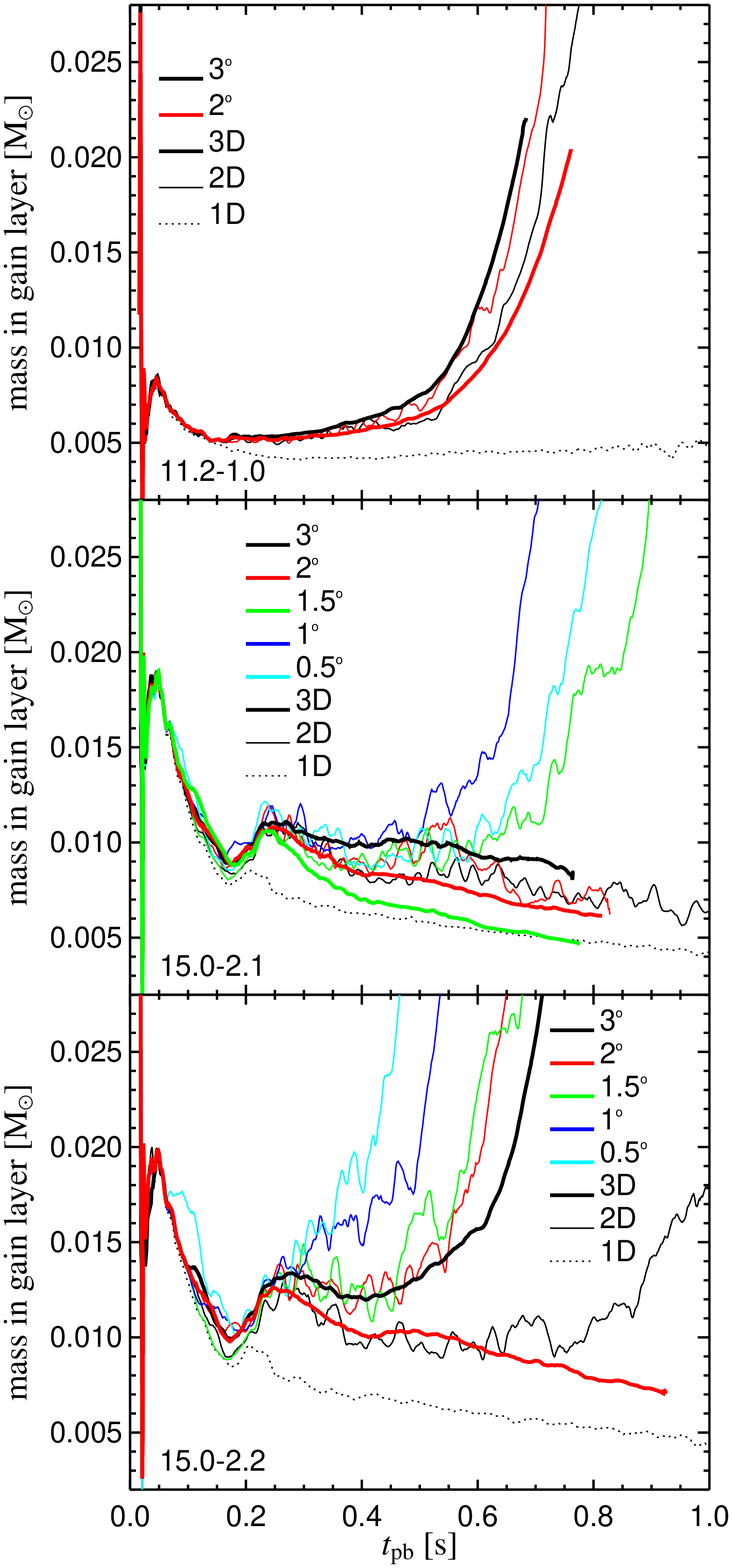}
 \caption{Time evolution of the mass in the gain region (in seconds after
bounce) for simulations in 1D (thin dotted line), 2D (thin solid lines), and
3D (thick lines). The multi-dimensional models are displayed for
all employed angular resolutions depicted by different colors. The top panel
shows the results for the 11.2\,$M_{\odot}$ star with an electron-neutrino
luminosity of $L_{\nu_e} = 1.0\cdot10^{52}$\,erg\,s$^{-1}$, the middle panel
the results for the 15\,$M_{\odot}$ runs with 
$L_{\nu_e} = 2.1\cdot10^{52}$\,erg\,s$^{-1}$, and the bottom panel the 
15\,$M_{\odot}$ models for $L_{\nu_e} = 2.2\cdot10^{52}$\,erg\,s$^{-1}$.
The different cases are the same as in
Figs.~\ref{fig:spos_res} and \ref{fig:sigma_res}.}
\label{fig:mass_in_gain}
\end{figure}

\begin{figure}
 \plotone{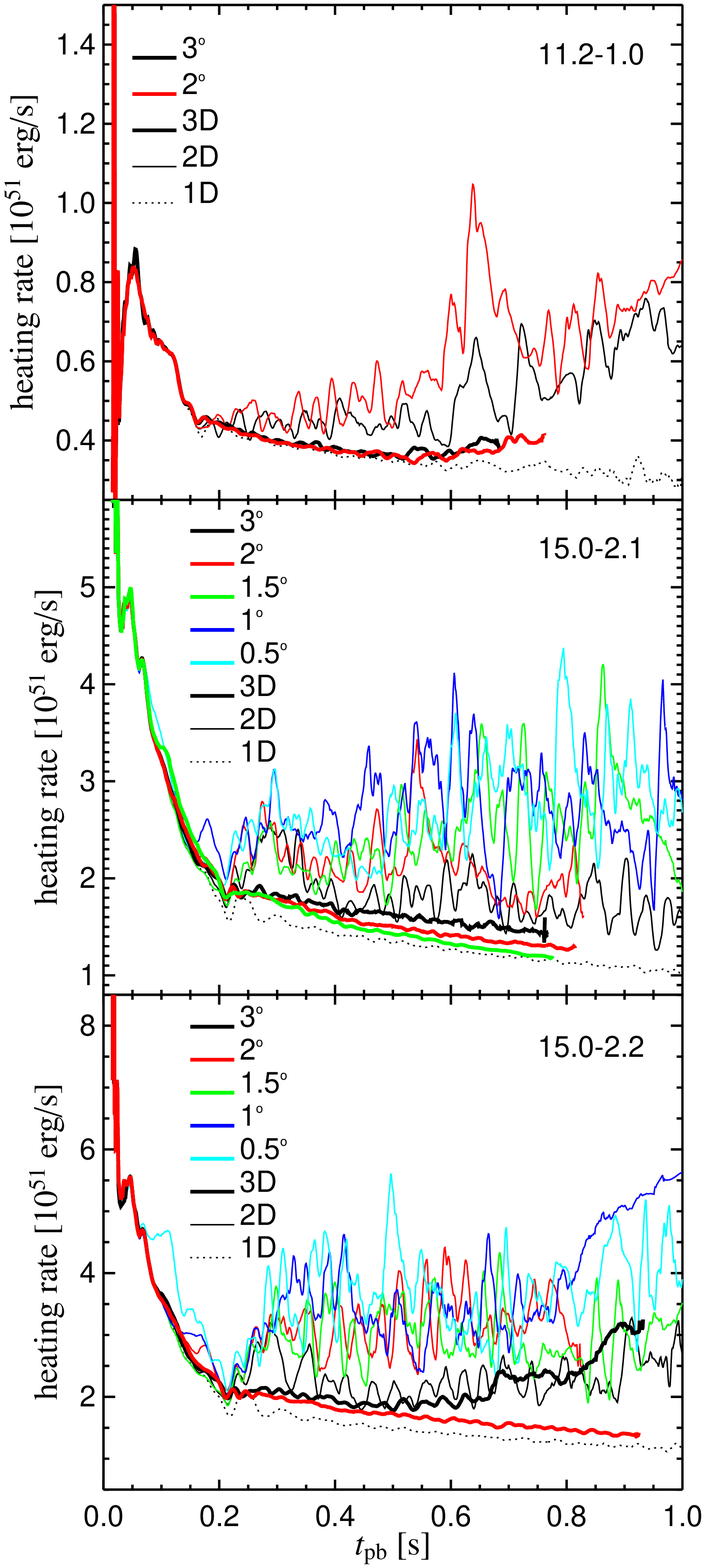}
 \caption{Analogous to Fig.~\ref{fig:mass_in_gain}, but for the time 
evolution of the total net rate of neutrino heating in the gain region.}
\label{fig:heating_rate}
\end{figure}

\begin{figure}
 \plotone{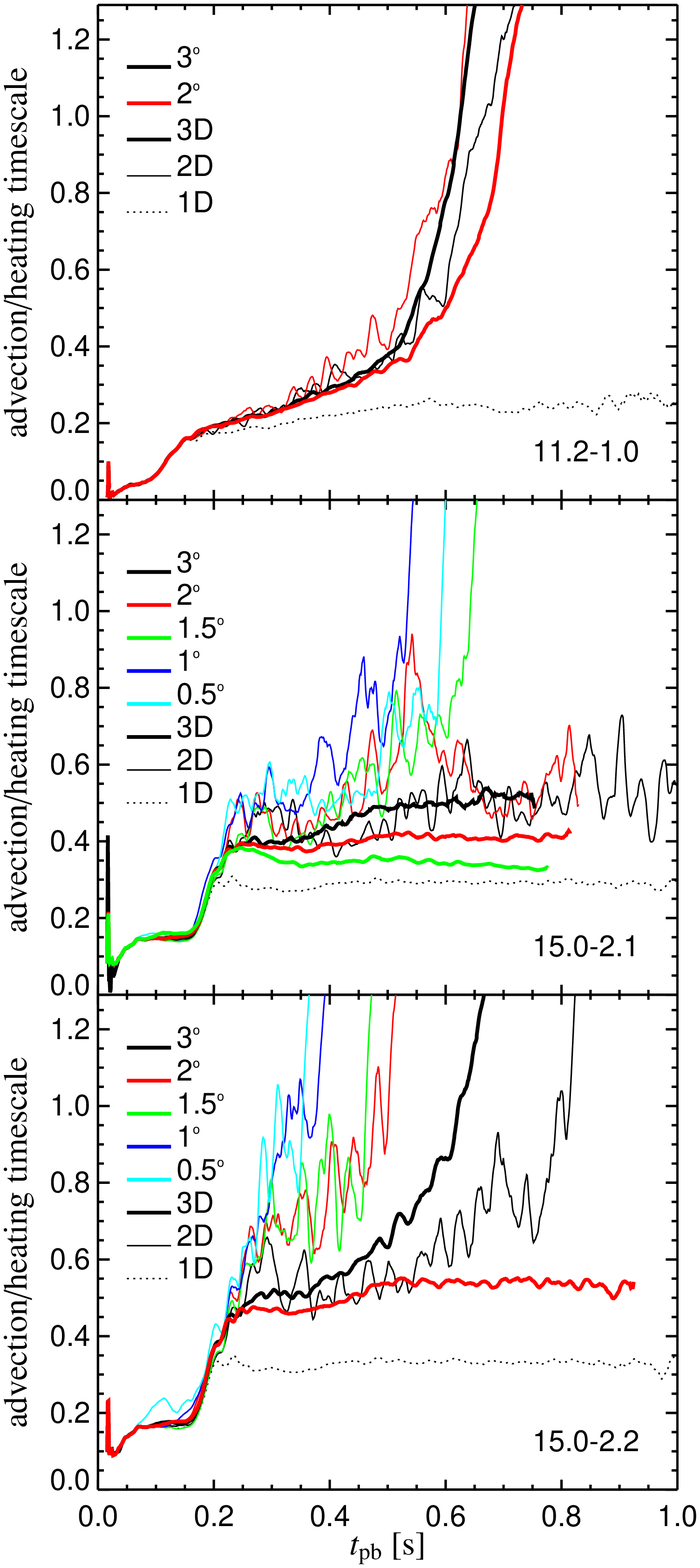}
 \caption{
Analogous to Fig.~\ref{fig:mass_in_gain}, but for the time 
evolution of the ratio of advection timescale to heating timescale in
the gain layer.}
\label{fig:timescales}
\end{figure}

\begin{figure}
 \plotone{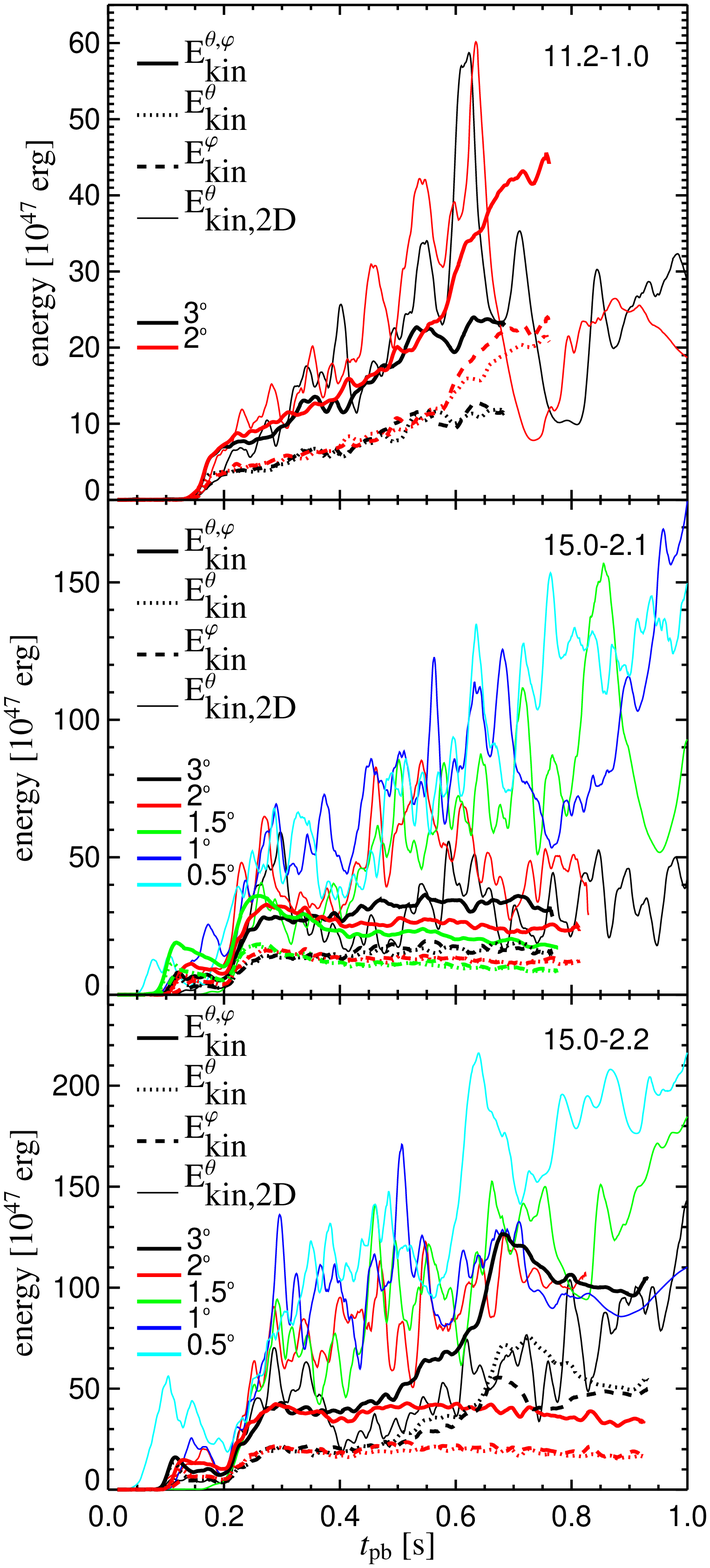}
 \caption{Kinetic energies of angular mass motions in the gain layer
as functions of time after bounce for the 11.2\,$M_{\odot}$ runs with
an electron-neutrino luminosity of 
$L_{\nu_e} = 1.0\cdot10^{52}$\,erg\,s$^{-1}$ (top panel) and the
15\,$M_{\odot}$ runs with $L_{\nu_e} = 2.1\cdot10^{52}$\,erg\,s$^{-1}$
(middle panel) and $L_{\nu_e} = 2.2\cdot10^{52}$\,erg\,s$^{-1}$
(bottom panel). Thin solid lines correspond to the lateral kinetic
energy of 2D models, while for 3D simulations (thick lines) the
lateral, azimuthal, and total kinetic energies are represented
by dotted, dashed,
and solid line styles, respectively. Both angular directions
contribute essentially equally to the total kinetic energy of
nonradial motions in the 3D case. As in Figs.~\ref{fig:spos_res},
\ref{fig:sigma_res}, \ref{fig:mass_in_gain}, and \ref{fig:heating_rate},
different colors depict different angular resolutions. It is
visible that for models closer to a success of the neutrino-driven 
mechanism the angular kinetic energy exhibits larger temporal
variations and an overall trend of increase as the onset of the
explosion is approached.}
\label{fig:ene_kin}
\end{figure}

\section{Interpretation and discussion} 
\label{sec:cond}

In this section we discuss the meaning of our results in
comparison to previous studies and present an interpretation
that could explain the main trends found in our multi-dimensional
simulations with varied resolution.

\subsection{Variation with dimension}
\label{sec:vardimen}

In Sects.~\ref{sec:crit_lum} and \ref{sec:high_res} we have reported
that our simulations do not support the central finding by
\cite{Nordhaus2010} that the tendency to explode is a monotonically
increasing function of dimension. While we confirm more
favorable explosion conditions in 2D than in 1D, we do not observe
that in 3D considerably lower driving luminosities are needed
for a success of the neutrino-driven mechanism than in 2D.
Moreover, we cannot confirm the finding by \cite{Nordhaus2010}
that the mass-weighted average of the entropy
per nucleon in the gain region, $\langle s(t)\rangle$, is
a quantity that is suitable as an indicator of the proximity of
models to an explosion and thus can serve as an explanation of
differences between 1D, 2D, and 3D simulations. In particular, our
3D models turned out to have slightly higher mean entropies than
corresponding 2D cases (Fig.~\ref{fig:ave_sto_res}) without 
developing better explosion conditions.
This raises the question why our models,
and multi-dimensional simulations in general,
have produced successful explosions by the
neutrino-heating mechanism when corresponding 1D models fail?

It is by no means obvious that 
$\langle s(t)\rangle$ should increase in the gain layer in the
multi-dimensional case.  While neutrino energy deposition naturally
leads to a rise of the entropy of the heated gas, the averaging
process over the volume of the gain layer also encompasses the
downdrafts carrying cool matter from the postshock region towards
the gain radius and the neutron star. These downdrafts are much 
denser, they are hardly heated by neutrinos because of their extremely
rapid infall, and they can contain more mass than the surrounding, 
dilute bubbles that are inflated by the expanding, neutrino-heated
plasma. It is therefore not clear that the
spatial (mass-weighted) average $\langle s(t)\rangle$ grows
in multi-dimensions compared to 1D runs.

Moreover, it is not even clear that convective overturn
in the gain layer must lead to an average entropy of the
neutrino-heated gas itself that is higher than in 1D simulations. 
Different from the 1D case high-entropy matter becomes buoyant
and begins to float in the multi-dimensional environment. 
Thus the heated gas is quickly carried away
from the vicinity of the gain radius, where neutrino-energy
deposition is maximal, to larger radii.
Such dynamics of the gas can well limit the
amount of energy and entropy that is stored in individual 
chunks of matter. Little, if any of the gas is subject to
multiple overturn cycles bringing the gas close to the gain 
radius more than once as suggested by the ``convective 
engine'' picture of \cite{Herant1994} but 
questioned by \cite{Burrows1995}. Instead, the majority of the
heated gas either expands to larger distances, pushing shock
expansion, or, in the disadvantageous case, is swept back
below the gain radius (e.g.\ by large-amplitude sloshing
motions of the shock), where it loses its energy again by
efficiently reradiating 
neutrinos\footnote{The real multi-dimensional
situation is even more complicated. The mentioned violent
sloshing motions of the shock can cause strong shock-heating
of the postshock matter as discussed in detail by 
\cite{Scheck2008}, \cite{Blondin2003}, and \cite{Blondin2006b}, 
thus not only massively affecting the postshock flow but also
providing an additional entropy source besides 
neutrino-energy deposition.}.

Our results imply that the dominant effect that makes the
multi-dimensional case more favorable for an explosion than
spherical symmetry is associated with an inflation of the shock
radius and postshock layer, driven by the buoyant rise and
expansion of the plumes of neutrino-heated plasma. In course 
of the postshock volume becoming more extended, the integrated
mass $M_\mathrm{gain}$ in the heating layer increases compared to
the 1D case. This can be seen in Fig.~\ref{fig:mass_in_gain}, 
which displays the mass in the gain layer as function of 
post-bounce time for the 11.2 and 15\,$M_\odot$ runs with 
the different neutrino luminosities and resolutions already
shown in Figs.~\ref{fig:spos_res} and \ref{fig:sigma_res}. 
While the mass-averaged entropy $\langle s\rangle$
in the gain region hardly changes, the integral value of 
the entropy, $M_\mathrm{gain} \langle s\rangle$, clearly
increases with models coming closer to explosion. This 
dependence is particularly well visible when 2D and 3D
models with different resolutions are compared with each
other. 

The longer dwell times of matter in the gain layer of 2D
simulations observed by \cite{Murphy2008}, which correspond
to the advection times $\tau_\mathrm{adv}$ evaluated by
\cite{Buras2006b} and \cite{Marek2009} (though different 
ways of calculation have been considered, in particular for
the multi-dimensional case), are a manifestation
that a growing mass accumulates in the postshock region 
to get energy-loaded by neutrino absorption and to
finally drive the successful supernova blast. In 
near-steady-state conditions the mass accretion rate through
the gain layer is equal to the mass infall rate $\dot M$
ahead of the shock, where it is determined by the core 
structure of the progenitor star.
Since $\tau_\mathrm{adv} \approx M_\mathrm{gain}/\dot M$
(cf.\ \citealp{Marek2009}) a larger value of $\tau_\mathrm{adv}$ 
correlates with a higher mass $M_\mathrm{gain}$.
Accordingly, the total net heating rate 
$Q_\mathrm{gain}$ and thus the heating efficiency
$\epsilon \equiv Q_\mathrm{gain}/(L_{\nu_e}+L_{\bar\nu_e})
= Q_\mathrm{gain}/(2L_\nu)$ of the gas residing
in the gain layer is also higher for models that develop an
explosion (see Fig.~\ref{fig:heating_rate} and \citealt{Murphy2008}).
Figure~\ref{fig:timescales} shows the ratio of 
the advection (dwell) timescale to the neutrino-heating timescale
in the gain layer (our evaluation employs the Newtonian 
analog of the formulas given in \citealp{Mueller2012}) for
the models also displayed in Figs.~\ref{fig:mass_in_gain} and
\ref{fig:heating_rate}. The same trends as in the previous
images can be seen. Models closer to an explosion exhibit 
higher values of the timescale ratio. The ratio approaches unity,
indicating the proximity to a runaway instability, roughly
around the time when we define the onset of an explosion 
(i.e., when the average shock radius of Fig.~\ref{fig:spos_res}
passes 400\,km). The quality of the coincidence of these moments,
however, differs from model to model and depends on the exact
definition used for the timescales and the accuracy at which
the relevant quantities can be evaluated in highly perturbed
flows. Nevertheless, the evolutionary trend of the timescale
ratio (increasing or decreasing) is suggestive for whether
a simulation run leads to a successful explosion or failure.

A larger mass in the gain layer and higher total net energy
deposition rate are therefore better indicators of the proximity
of our models to explosion than the mean entropy of the 
gas in this region, which does not exhibit the 1D-2D-3D 
hierarchy with dimension found previously by \cite{Nordhaus2010}. 
As discussed in Sect.~\ref{sec:crit_lum} the main reason for this
discrepancy are most probably the different treatments of neutrino
lepton number losses and our consequential recalibration of the 
energy source terms. This leads to significantly higher energy 
drain from the cooling layer in our simulations. While this
hypothesis is supported by tests that we conducted in 1D, we
cannot be absolutely certain that no other effects play a role
for the discrepancies between our results and those of 
\cite{Nordhaus2010}, because detailed cross-comparisons
are not available and our knowledge of the details of the 
implementation of neutrino effects by \cite{Nordhaus2010}
may be incomplete. Other potential reasons for differences
may be connected to the hydrodynamics scheme ({\sc Prometheus} 
with a higher-order Godunov solver and directional splitting vs.\
{\sc Castro} with unsplit methodology, \citealp{Almgren2010}), 
the employed grid (polar coordinates vs.\ structured grid with 
adaptive refinement by a nested hierarchy of rectangular grids),
potentially ---though not very likely--- the use of a 1D core
above $10^{12}$\,g\,cm$^{-3}$ in our simulations, or differences
in the exact structure and properties of the infall region upstream
of the stalled shock as a consequence of different treatments
of the collapse phase until 15\,ms after core bounce (due to full 
neutrino transport plus a nuclear-burning approximation in the
{\sc Prometheus-Vertex} code vs.\ the simple deleptonization 
scheme of \citealp{Liebendorfer2005}) or of different seeding of
nonradial hydrodynamic instabilities (in our case by imposed, small
random seed perturbations of the density), or linked to differences
of the low-density EoS outside of the application regime of the 
\cite{Shen1998} EoS.

Despite these uncertainties about the exact cause of the 
differences, whose ultimate elimination will require systematic
and time-consuming studies, our results, as they are, send a
clear message: The outcome of the 1D-2D-3D comparison and the
effects of the third dimension advertised by \cite{Nordhaus2010}
``as a key to the neutrino mechanism of core-collapse
supernova explosions'' are not at all robust results. Instead,
the exact slope of the critical explosion condition $L_\nu(\dot M)$, 
its location, and its shift with dimension, as well as the 
existence of a 1D-2D-3D hierarchy of the mass-averaged entropy
in the gain layer seem to depend sensitively on subtle details
of the neutrino treatment or other numerical aspects of the
simulations. 

\begin{figure*}
 \plottwo{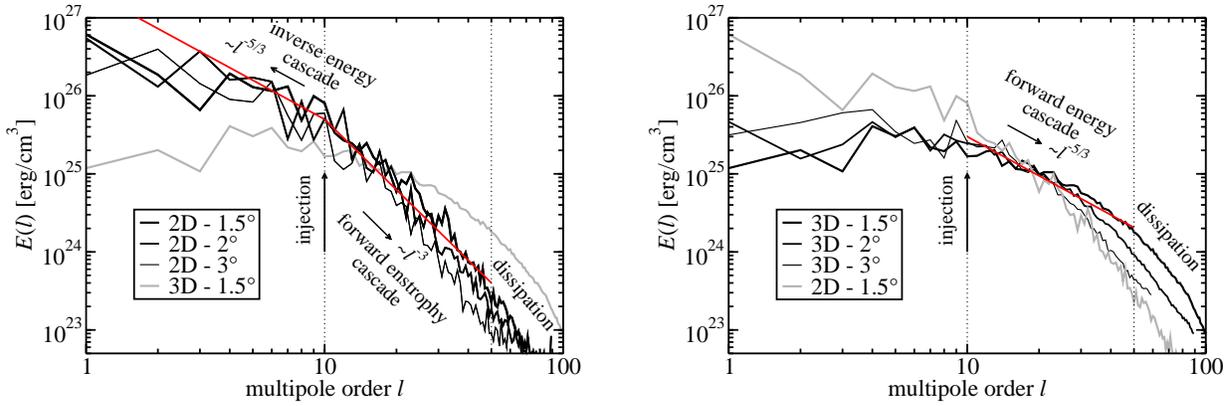}{f16b.eps}
 \caption{
 Turbulent energy spectra $E(l)$ as functions of the multipole order 
 $l$ for different angular resolution. The spectra are based on a decomposition
 of the azimuthal velocity $v_{\theta}$ into spherical harmonics at radius 
 $r=150$\,km and 400\,ms post-bounce time for 15\,$M_{\odot}$ runs with an 
 electron-neutrino luminosity of $L_{\nu_e} = 2.2\cdot10^{52}$\,erg\,s$^{-1}$. 
 {\em Left:} 
 2D models with different angular resolution (black, different thickness) 
 and, for comparison, the 3D model with the highest employed angular resolution 
 (grey). 
 {\em Right:} 3D models with different angular resolution and, for 
 comparison, the 2D model with the highest employed angular resolution (grey).
 The power-law dependence and direction of the energy and enstrophy cascades
 (see text) are indicated by red lines and labels for 2D in the left panel 
 and 3D in the right panel. The left vertical, dotted line roughly marks
 the energy-injection scale, and the right vertical, dotted line denotes 
 the onset of dissipation at high $l$ for the best displayed resolution.}
\label{fig:turbulence}
\end{figure*}

\subsection{Resolution dependence}
\label{sec:resoldep}
Let us now turn to the second, highly interesting question in
connection with our set of models, namely to the resolution 
dependence of our results. Our set of simulations performed
with different angular binnings reveals that quantities
that turned out to diagnose healthy conditions
for an explosion, i.e.\ the growth of the average shock radius,
the degree of shock deformation, or the mass and total heating
in the gain layer (but not the mass-averaged entropy of the matter
in the gain region), show a clear dependence on the angular
zoning (see Figs.~\ref{fig:spos_res}, \ref{fig:sigma_res},
\ref{fig:mass_in_gain}, and \ref{fig:heating_rate} in contrast
to Fig.~\ref{fig:ave_sto_res}).
In particular, 2D models with better angular resolution
exhibit more favorable conditions and explode more readily
(in agreement with results obtained by \citealp{Scheck2007}
with a more sophisticated treatment of neutrino transport 
than the simple heating and cooling source terms applied in 
our investigation),
whereas 3D models obey the opposite behavior. What does the 
resolution dependence of our simulations tell us about the 
mechanism leading to explosions in our models? And how can we 
understand the puzzling finding that 2D and 3D runs follow
opposite trends when the angular resolution is refined? 

We interpret this as a manifestation of two aspects or facts: 
\begin{itemize}
\item[(1)]
The success of our models, at least in the neighborhood of 
the explosion threshold, is fostered mainly by large-scale mass
flows as associated with strong SASI activity, but not
by enhanced fragmentation of structures and vortex motions on 
small spatial scales.
\item[(2)]
Our resolution study reflects the consequences of the turbulent
energy cascade, which redistributes energy fed into
the flow by external sources in opposite directions in 2D and
3D: While in 3D the turbulent energy flow goes from large
to small scales, it pumps energy from small to large spatial
scales in 2D.
\end{itemize}
Point~(1) is supported by the kinetic energies of nonradial 
mass motions in the gain layer of the 2D and 3D models plotted
in Fig.~\ref{fig:ene_kin}. From this picture it is obvious
that in the case of successful models the angular kinetic
energy is higher and shows an overall trend of growth in time
until the blast has taken off. Moreover, the spiky maxima and
minima of quasi-periodic variations, which are indicative of 
the presence of low-order 
SASI modes, are significantly larger for exploding models.
This does not only hold for 2D models, whose lateral 
kinetic energies exhibit variation amplitudes of several
10\% and partially up to even $\sim$50\% of the 
time-averaged value. It is also true for 3D models, although
in this case the amplitudes are generally smaller and the 
nonradial kinetic energy is split essentially equally into 
lateral and azimuthal contributions.
When comparing successful runs in 2D with those in 3D, 
our studies suggest that in both cases the shock exhibits
a growing degree of asphericity (expressed by the standard
deviation of the shock deformation plotted in
Fig.~\ref{fig:sigma_res}) when the explosion is approached,
and the kinetic energy of nonradial mass motions reaches roughly
the same magnitude (Fig.~\ref{fig:ene_kin}), at least for
models near the explosion threshold.

Actually a variety of observations can be interpreted
as support of the hypothesis that flows on the largest 
possible scales rather than on small scales play a crucial
role for the success of the neutrino-heating mechanism in
our simulations:
\begin{itemize}
\item
The strength of low-mode SASI activity in 2D models as 
indicated by growing fluctuations of the angular kinetic
energy and of the shock deformation (Figs.~\ref{fig:ene_kin},
\ref{fig:sigma_res}) increases with higher resolution in
clear correlation with an earlier
onset of the explosion (Fig.~\ref{fig:spos_res}). 
Stronger SASI activity obviously facilitates explosions, 
which is visible by a growing average shock 
radius as well as larger mass and higher total heating rate 
in the gain layer.
\item
Exploding models in 2D as well as in 3D exhibit large shock 
deformation at the time of explosion (although the relative
asphericity $\sigma_\mathrm{S}/R_\mathrm{S}$ of the shock 
surface is somewhat smaller in 3D than in 2D;
see Figs.~\ref{fig:spos_res} and \ref{fig:sigma_res}).
\item
More fine structure on small spatial scales, which can be
seen in 3D models computed with higher resolution 
in Figs.~\ref{fig:sn_s11_hr}--\ref{fig:sn_s15_hr}, 
does not imply improved conditions for an explosion.
\item
Exploding 2D models are {\em not} connected with the highest
mean entropies in the gain region (Fig.~\ref{fig:ave_sto_res}). 
This fuels doubts in a random-walk picture where turbulent 
vortex motions on small scales enlarge the residence time of 
matter in the gain layer \citep{Murphy2008} and thus could
allow for more energy absorption of such mass elements from
neutrinos. If this effect occurred, it does not concern a 
major fraction of the mass in the gain region.
\end{itemize}

Point~(2) is the only plausible argument we can give
for explaining the opposite response to higher angular 
resolution that we discovered in our 2D and 3D simulations.
The sequence of 2D runs with gradually reduced lateral 
zone sizes reflects the growing violence of large-scale
flows by higher fluctuation amplitudes of the kinetic
energy in the gain layer (Fig.~\ref{fig:ene_kin}) and 
larger temporal variations of the average shock radius
and shock deformation 
(Figs.~\ref{fig:spos_res}, \ref{fig:sigma_res}).  
In contrast, more energy on small spatial scales 
in the 3D case manifests itself
by a progressing fragmentation of the flow, leading to
a growing richness of vortex structures and finer
filaments in the case of 3D models with smaller angle
bins (Figs.~\ref{fig:sn_s11_hr}--\ref{fig:sn_s15_hr}). As
a consequence, 3D models with higher angular resolution 
become more similar to the 1D case in various quantities
that we considered as explosion indicators,
see, e.g., Figs.~\ref{fig:spos_res}, \ref{fig:sigma_res},
\ref{fig:mass_in_gain}, and \ref{fig:heating_rate}.

Both the powerful coherent mass motions of the 
SASI layer in 2D and the vivid activity in small
vortex structures in the 3D environment
are fed by two external sources which
supply the postshock layer with an inflow of fresh
energy: (i) gravitational
potential energy that is released by the continuous 
stream of matter falling through the accretion shock and
(ii) energy deposition by neutrinos. The energy stored in
the fluid is then redistributed towards small or large
scales according to the turbulent cascades characteristic
of two- and three-dimensional environments.

Direct evidence for the action of different
turbulent energy cascades in 2D and 3D can be obtained by
considering the energy spectrum $E(k)$ of turbulent motions
as a function of wavenumber $k$ in the gain region.
The spectral shape of $E(k)$ can already be adequately established
by considering only the azimuthal velocity $v_{\theta}$ at a 
given radius using a decomposition into spherical harmonics 
$Y_{lm}(\theta,\phi)$:
\begin{equation}
E(l) = \sum^{l}_{m=-l}\left|\int_{\Omega}Y_{lm}^{*}(\theta,\phi) \, 
\sqrt{\rho} \, v_{\theta}(r,\theta,\phi) \, d\Omega\right|^{2} \,.
\label{eq:turbpower}
\end{equation}
Here, the velocity fluctuations have been expressed in
  terms of the multipole order $l$ instead of the wave number $k$.  A
  summation over the energies of modes with the same $l$ has been
  carried out\footnote{Note that a factor $\sqrt{\rho}$ has been
    introduced to ensure that the integrated energy of all modes sums
    up to the total kinetic energy contained in azimuthal motions at
    radius $r=150$\,km (modulo a normalization factor)
    (cp.\ \citealp{Endeve2012}).}, and in order to obtain smoother spectra,
    we average $E(l)$ over 30~km in radius and over 10 timesteps.
One expects that the resulting spectrum $E(l)$
directly reflects the properties of $E(k)$ such as
the slopes in different regimes of the turbulent
cascade\footnote{For the precise relation between Fourier and spherical
harmonics power spectra, the reader may consult Chapter~21 of \citet{Peebles1993}.
For a power-law spectrum $E(k) \propto k^{\alpha}$, one obtains
$E(l) \propto (2l+1) \Gamma(l+\alpha/2+1/2) / \Gamma(l-\alpha/2+3/2)$,
or $E(l) \propto l ^{\alpha}$ in the limit of large $l$. In practice,
the power-law indices of $E(l)$ and $E(k)$ appear to
correspond well to each other already for $l \gtrsim 4$. Empirically,
broken power laws transform in a similar manner.}. The computed spectra $E(l)$ (Fig.~\ref{fig:turbulence}) 
indeed confirm the predictions from 3D and (planar) 2D turbulence theory,
at least for sufficiently high multipole order $l$.

In 3D (right panel), a power-law spectrum with $E(l)\propto l^{-5/3}$
\citep{Landau1959} develops at intermediate wavenumbers
as the resolution is increased, reflecting the transfer of energy
from large to small scales in a forward cascade until dissipation
takes over at large $l$. At high resolution, the energy contained 
in small-scale disturbances increases, as the dissipation range
moves to larger $l$. One observes that the $5/3$-power-law is 
broken at low $l$ ($l \lesssim 10$), suggesting that kinetic energy
is injected into the flow at wavenumbers $l\approx 10$, i.e.
at scales typical for growing convective plumes.

By contrast, the power-law dependence $E(l)\propto l^{-5/3}$
approximately holds for $l\lesssim10$ in 2D as a result of 
the reverse energy cascade \citep{Kraichnan1967}. The energy
injected at $l\approx 10$ is therefore not transferred to the
dissipative range; only enstrophy (the squared vorticity
of the velocity field) is transported in a forward
cascade with a different power-law index ($E(l)\propto l^{-3}$).

This appears to be a natural explanation for the predominance
of large-scale and small-scale structures in 2D and 3D, respectively.
Moreover, this picture suggests that as dissipation affects the
``injection scale'' at $l\approx 10$ less with increasing resolution,
the differences between 2D and 3D become more pronounced with finer
grid zoning.

Our analysis of the spectral properties of turbulence thus 
further strengthens our view that
the trends seen in our simulations strongly
suggest that nonradial kinetic energy available on large
scales, not on small
scales, assists the development of an explosion
by the neutrino-heating mechanism. This explains why 2D
models with higher angular resolution tend to explode 
earlier and thus at higher values of the mass-accretion
rate than less resolved models. On the other hand, the 
energy ``drain'' by vortex motions on ever smaller scales
---with the same reservoir of pumping energy per unit mass
being available from accretion and neutrino heating---
disfavors explosions in better resolved 3D models.

We therefore conclude that the key to the mechanism of
core-collapse supernova explosions seems intrinsically and
tightly linked to the question how much kinetic energy
of the matter in the gain region can be accumulated in
nonradial fluid motions on the largest possible scales, 
i.e., in the lowest-order spherical harmonics modes of
nonradial hydrodynamic instabilities.
The predominant growth of such
flows is typical of SASI activity, whose lowest-order
spherical harmonics modes possess the highest growth rates
\citep{Blondin2003,Blondin2006b,Foglizzo2006,Foglizzo2007,Ohnishi2006}.
Strong SASI motions drive shock
expansion, increase the gain layer and its mass content,
allow a larger fraction of the accreted matter to stay in
the gain layer and be exposed to efficient neutrino heating,
and thus aid the development of an explosion 
\citep{Scheck2008,Marek2009}.
However, our models do not show a systematic trend of higher
average entropies of the matter in the gain layer
for models closer to explosion. Instead, we find that such 
models have larger mass, larger nonradial kinetic energy,
larger total neutrino-heating rate, and larger total entropy 
in the gain layer.

\section{Summary and conclusions} 
\label{sec:con}

We have performed a systematic study of the post-bounce evolution
of supernova cores of 11.2 and 15\,$M_\odot$ and their explosion 
by the neutrino-heating mechanism in 1D, 2D, and 3D, employing
simple neutrino cooling and heating terms with varied values of 
the driving luminosity. We conceptually followed 
previous studies by \cite{Murphy2008} and \cite{Nordhaus2010},
but did not apply the deleptonization treatment that they adopted 
from \cite{Liebendorfer2005}, who introduced it for an approximative
description of neutrino losses during the infall phase until core 
bounce. We argued (Sect.~\ref{sec:num}) that this approximation
---with or without the source term proposed by 
\cite{Liebendorfer2005} to account for entropy generation in
neutrino-electron scatterings--- does not provide a suitable
treatment of the evolution of the electron abundance after core 
bounce. Therefore we did not consider changes of the net electron
fraction $Y_e$ of the stellar plasma at times later than 15\,ms
after bounce, up to which the collapse was followed with the 
{\sc Prometheus-Vertex} code including full neutrino transport.
While ignoring $Y_e$ changes subsequently is certainly not a 
good approximation, it is not necessarily more unrealistic than
describing the lepton-number evolution during the accretion phase
of the stalled shock by the scheme of \cite{Liebendorfer2005}.
As a consequence, we had to replace an exponential factor
$e^{-\tau_{\mathrm{eff}}}$, which was introduced in an ad hoc
way by \cite{Murphy2009} and \cite{Nordhaus2010}
to damp the neutrino source terms at high optical
depths $\tau_{\mathrm{eff}}$, by $e^{-\tau_{\mathrm{eff}}/2.7}$
in order to reproduce the minimum luminosity found to yield 
explosions in the 1D simulations by \cite{Murphy2008} and
\cite{Nordhaus2010}. This modification led to enhanced neutrino
losses in the cooling layer, which were better compatible with 
total energy loss rates found in simulations with detailed
neutrino transport, e.g., in \cite{Buras2006a},
and is responsible for some of the findings and differences
discussed in Sect.~\ref{sec:crit_lum}.

Our results and conclusions can be briefly summarized as follows:
\begin{itemize}
\item[1.]
We cannot reproduce the exact slopes and relative locations of the
critical curves $L_\nu(\dot M)$ of 1D, 2D, and 3D simulations
found by \cite{Nordhaus2010}. While our results confirm the
well-known fact that explosions in 2D occur for a lower 
driving luminosity $L_\nu$ than in 1D when the mass accretion
rate $\dot M$ is fixed, we cannot discover any significant
further reduction when we go from 2D to 3D.
\item[2.]
We cannot confirm that the mass-averaged entropy of the matter 
in the gain region, $\langle s \rangle$, 
is a good diagnostic quantity for the 
proximity to an explosion. As we argued in Sect.~\ref{sec:vardimen},
it is neither clear nor necessary that $\langle s \rangle$ is
higher for cases where explosions are obtained more readily. 
Our successful 2D models do not exhibit larger mean entropies
than the corresponding 1D cases, which fail to explode.
Instead, we observed that the
total mass, total entropy, total neutrino-heating rate, and
the nonradial kinetic energy in
the gain layer are higher in cases that develop an explosion.
\item[3.]
We conclude that the tendency for an explosion as a monotonically
increasing function of dimension as well as the 1D-2D-3D 
hierarchy of $\langle s(t) \rangle$ found by \cite{Nordhaus2010}
are not robust results. They seem to be sensitive to subtle 
differences of the approximations of neutrino effects (and/or
to other differences in the numerical treatments of the models).
It is therefore unclear how far studies with radical
simplifications of the neutrino physics (without detailed energy
and lepton-number source terms and transport; no feedback
between accretion and neutrino properties) can yield results
that are finally conclusive for the explosion-triggering 
processes in real supernova cores.
\item[4.]
Increasing the angular resolution we observed a clear tendency
of 2D models to explode earlier, in agreement with previous
results by \cite{Scheck2007}, who employed a more sophisticated
treatment of neutrino effects based on the transport 
approximation described in \cite{Scheck2006}. In contrast,
3D models show the opposite trend and in a variety of quantities
and aspects become more similar to their 1D counterparts. The
easier explosion of the 2D models is connected to an enhanced
violence of large-scale mass motions in the postshock region
due to SASI activity, whereas 3D models with better angular
resolution appear to develop less strength in low-order SASI 
modes.
\item[5.]
We interpret this finding as a consequence of the opposite
turbulent energy cascades in 2D and 3D. In 2D the energy
continuously pumped into the gain layer by
neutrino heating and the release of gravitational binding
energy flows from small to large scales and thus helps to
power coherent mass motions on the largest possible spatial
scales. In contrast, in 3D this energy seems to instigate 
flow vorticity and fragmentation of structures on
small scales. Evidence for this interpretation
is provided by Fig.~\ref{fig:turbulence}.
\item[6.]
We also conclude from our resolution studies that the 
presence of violent mass motions connected to low-order SASI
modes is favorable for an explosion (in agreement with 
arguments given by \citealp{Marek2009} and \citealp{Scheck2008}). 
This is supported by the fact that 2D and 3D models that are
closer to explosion show signs of growing power in large-scale
mass motions (signalled by growing fluctuations of the 
kinetic energy of nonradial velocity components)
and in particular develop significant shock 
deformation and global ejecta asymmetries when the explosion 
sets in.
\item[7.]
We found that higher radial resolution makes explosions more
difficult with the setup chosen for the investigated set of
models. Higher resolution turned out to prevent explosions or 
to let them occur later in simulations in 1D, 2D, and 3D.
This result could be diagnosed to be a consequence
of a local density maximum in the neutrino-cooling layer,
which grows with higher resolution and enhances the energy
loss by neutrino emission. This density peak, however, is 
a numerical artifact of the employed simple neutrino-cooling
treatment by an analytic source term which is exponentially
damped at high densities.
\end{itemize}

The lack of very precise information on the physics ingredients
and their exact implementation, e.g., details of 
the treatment of neutrino source terms, low-density EoS,
and progenitor data when mapped into the simulation and
seeded with small random perturbations, as well
as a variety of methodical differences like the hydrodynamics 
scheme, numerical grid, and the use of a 1D core at high
densities or not, prevent us from presenting a rigorous
proof that could causally link the discrepancies between
our results and those of \cite{Nordhaus2010} to one or more
well understood reasons. We think that the nagging uncertainties
in this context demand a future, involved, collaborative
code-comparison project. This will also require considerable
amounts of computer time for further 3D simulations, in 
particular with high resolution, thus needing more computer
resources than available to us for the described project.

Despite this deficiency, however, our results suggest that
the differences of 3D compared to 2D simulations observed by 
\cite{Nordhaus2010} are unlikely to be a robust outcome but 
seem to depend on relevant aspects of the modeling
(most probably the neutrino physics but potentially, and not
finally excluded, also technical aspects).

We therefore conclude that the influence of 3D effects on
the supernova mechanism is presently not clear. We strongly
emphasize, however, that the fact that our results do not 
corroborate improved explosion conditions in 3D compared to 2D
{\em cannot} be used as an argument that 3D effects do not
facilitate the supernova explosion mechanism or are of minor 
importance. We just think that in the context of the
neutrino-driven mechanism the relevance and exact role of 
3D fluid dynamics are not understood yet. We therefore have the
opinion that the results obtained by \cite{Nordhaus2010} do not
justify their claims that 3D hydrodynamics offers the key
to a fundamental understanding of the neutrino mechanism 
while other physics in the supernova core, like general 
relativity or the properties of the nuclear EOS, are only
of secondary importance. Though this may well be right, such 
statements at the present time are premature and not 
supported by solid facts and results.

Our study, however, raises further important questions. How
far can our understanding be developed on grounds of modeling
approaches that employ radical simplifications of the neutrino
physics? Which aspects of the complex interplay between
different components of the problem are linked to the essence 
of how the explosion is triggered by the combination of 
neutrino energy supply and nonradial hydrodynamic instabilities?
Examples for such mutually dependent components are the
neutrino transport and hydrodynamics, the neutron star
core evolution and fluid motions around the neutron star, or
the mass flux from the accretion shock to the deceleration
layer (both being the coupling regions for the advective-acoustic
cycle that is thought to be responsible for the SASI growth;
e.g., \citealp{Scheck2008}) 
and the conditions in the neutrino heating and cooling 
layers. Much more work needs
to be done to find the answers of these questions.

A major shortcoming of the setup
applied in previous works and adopted also in our investigation
is the neglect of neutrino cooling and deleptonization inside
the proto-neutron star. The employed simple neutrino source 
terms and their exponential suppression at high optical depths
do not allow the neutron star to evolve. Underestimating the 
neutron star contraction, however, slows down the infall velocities
in the postshock layer and thus has disfavorable consequences
for the growth of the SASI (\citealp{Scheck2008}), similar 
to the effects of reduced neutrino losses in the cooling layer 
(\citealp{Scheck2008}) or increased nuclear photodissociation
behind the stalled shock (\citealp{Fernandez2009}). On the contrary,
a slower postshock flow improves the conditions for the growth
of convective instability, whose development is supported by a
high ratio (larger than $\sim$3 signals a linearly unstable 
situation) of the advection timescale through the gain region
divided by the inverse of the Brunt-V\"ais\"al\"a frequency as
discussed by \cite{Foglizzo2006}, \cite{Buras2006b}, and
\cite{Scheck2008}. 
It is therefore well possible that the weak SASI activity 
diagnosed in recent 2D and 3D simulations by \cite{Burrows2012}
---leading them to the conclusion that dipolar asymmetries are
caused by convection rather than the SASI--- 
is an artifact of the approximative treatment of neutrino cooling
and the disregard of neutron star contraction in their models.
In full-scale supernova models with 
sophisticated neutrino transport, \cite{Buras2006b} observed
differences in the growth conditions for convection compared to
the SASI in collapsing stellar cores of a variety of progenitor
stars. Also the high-density equation of state and general 
relativity, influencing the contraction behavior of
the nascent neutron star, can make a difference
(\citealp{Marek2009,Mueller2012}). Claims, based on highly 
simplified models, 
that SASI is less important than convection and at most a minor 
feature of the supernova dynamics (\citealp{Burrows2012}) are
therefore certainly premature.

Finally, our resolution study suggests that the action
of the turbulent cascade in 3D extracts energy from
coherent large-scale modes of fluid motion and instead fuels
fragmentation and enhanced vortex flows on small spatial
scales. At least in our 3D models with better grid
zoning the appearance of finer structures in the postshock 
flow was connected with a tendency of damping
the development of explosions. While a finally convincing
proof of such a negative feedback may require much better
resolved simulations than we presently can afford to 
conduct (in order to minimize numerical dissipation on
small scales), this result implies that good resolution 
---considerably higher than recently used by \cite{Takiwaki2011}, 
whose 3D simulation
had only 32 azimuthal zones (corresponding to a cell size of
11.25$^\circ$)--- is indispensable to clarify the 3D effects on
the explosion mechanism. Moreover, our result 
points to an interesting direction. It is possible
that the success of the neutrino-driven mechanism in 3D
is tightly coupled to the presence of violent SASI activity,
a connection that was found before ---and is confirmed by our 
present study--- to foster explosions in 2D?  
If so, what is the key to instigate such violent SASI
motions of the supernova core in three dimensions? 
Will they occur with a better (more realistic) treatment
of the neutrino transport and correspondingly altered
conditions in the heating and cooling layers and in the
contracting core of the proto-neutron star? Or are they
associated with stellar rotation, which even with a slow 
rate can initiate the faster growth of spiral 
(non-axisymmetric) SASI modes
\citep{Blondin2006a,Yamasaki2008,Iwakami2009,Fernandez2010}?
Or is strong SASI activity in the supernova core triggered
by large-scale inhomogeneities in the three-dimensional
progenitor star \citep{Arnett2011}, which could provide a more
efficient seed for SASI growth than the random cell-to-cell
small-amplitude perturbations employed in our simulations?
Should the presence of large-amplitude SASI mass motions indeed 
turn out to be the key to the neutrino mechanism in 3D, it
would mean that neutrino-driven explosions are not only a
generically multi-dimensional phenomenon, but one that is
generically associated with dominant low-order modes of
asymmetry and deformation from the very beginning.

While this paper raises many more questions than it is
able to answer, it definitely makes clear that our understanding
of the supernova physics in the third dimension is still in its
very infancy. A virgin territory with distant horizons lies
ahead of us and awaits to be explored.

\acknowledgements
  We are grateful to Lorenz H\"udepohl for his valuable input to different 
  aspects of the reported project and thank Elena Erastova and Markus Rampp
  (Max-Planck-Rechenzentrum Garching) for their help in the visualization 
  of our 3D data. HTJ would like to thank Rodrigo Fern{\'a}ndez, Thierry
  Foglizzo, Jerome Guilet, and Christian Ott for stimulating and 
  informative discussions.
  This work was supported by the Deutsche Forschungsgemeinschaft through
  Sonderforschungsbereich/Transregio~27 ``Neutrinos and Beyond'',
  Sonderforschungsbereich/Transregio~7 ``Gravitational-Wave Astronomy'',
  and the Cluster of Excellence EXC~153 ``Origin and Structure of the
  Universe''. The computations were performed on the
  Juropa cluster at the John von Neumann Institute for Computing (NIC) in
  J\"ulich, partially through a DECI-6 grant of the DEISA initiative, on 
  the IBM p690 at Cineca in Italy  through a DECI-5 grant of the DEISA
  initiative, and on the IBM p690 at the Rechenzentrum  Garching.
%


\appendix

\begin{figure}
\plotone{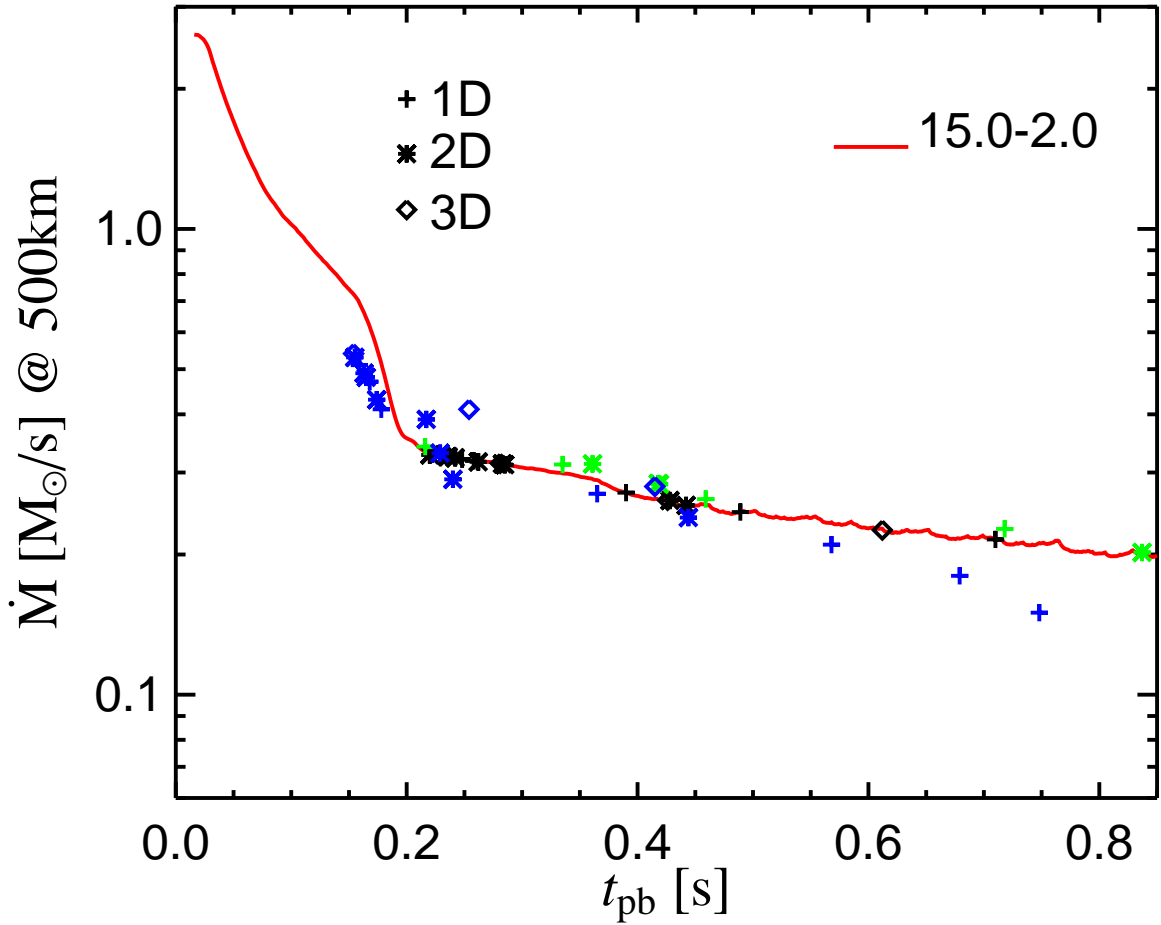}
\caption{Mass accretion rate for the 15\,$M_{\odot}$ progenitor. The red curve
shows the line of Fig.~\ref{fig:mdot_time}. The black symbols represent the
values extracted from our simulations at the time $t_\mathrm{exp}$
when the explosion sets in. Green symbols are data from \cite{Murphy2008}
and blue symbols those from \cite{Nordhaus2010}. Different symbols
are used for results of 1D, 2D, and 3D simulations.}
\label{fig:mdot_time_cmp}
\end{figure}

\begin{figure}
\plotone{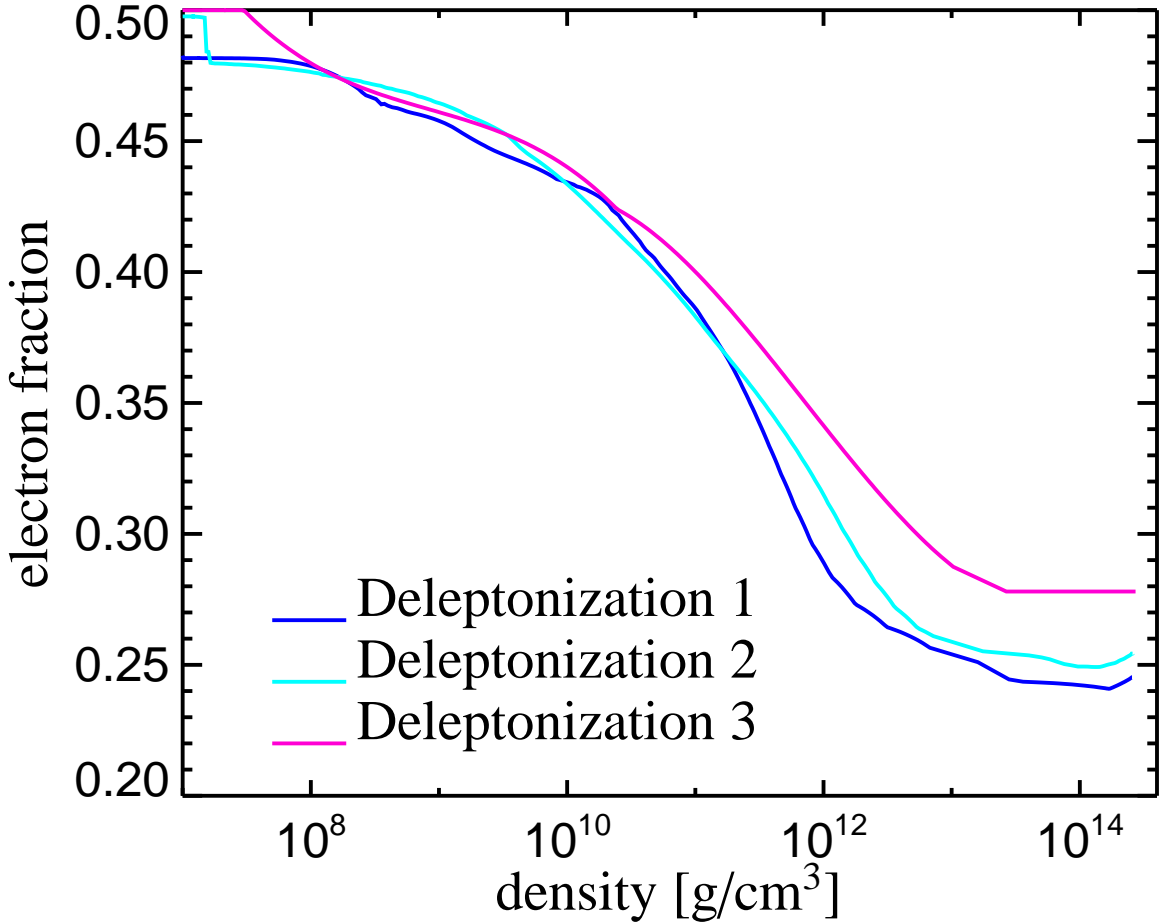}
\caption{Trajectories of the electron fraction with density deduced from
different core-collapse studies (see text for details) and employed in
our 1D simulations for parametrizing lepton losses by neutrino emission
in the stellar core according to \cite{Liebendorfer2005}.}
\label{fig:delep}
\end{figure}

\begin{figure}
\plotone{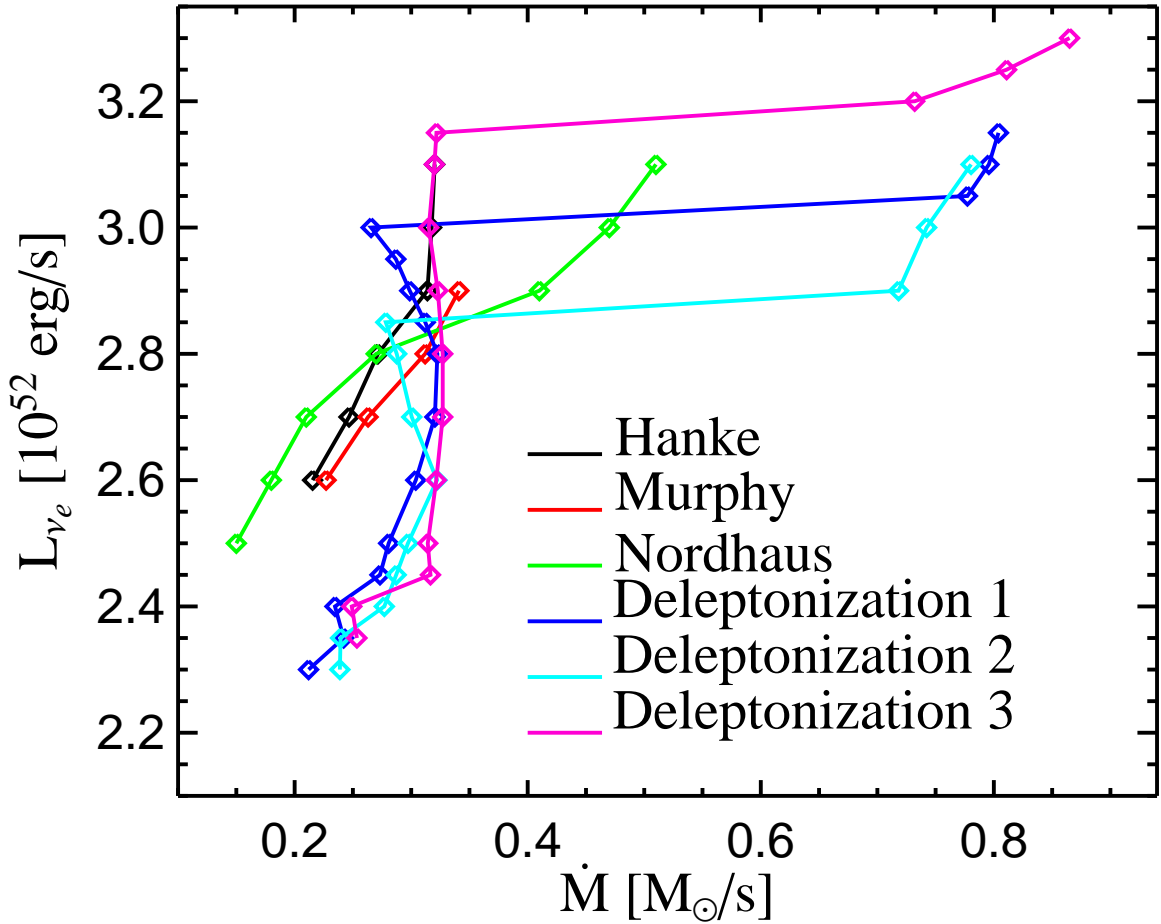}
\caption{Critical curves for the electron-neutrino luminosity ($L_{\nu_e}$)
versus mass accretion rate ($\dot{M}$), representing the explosion threshold
for different sets of 1D simulations of the 15\,$M_\odot$ progenitor.
The black line corresponds to our results shown as black curve in
Fig.~\ref{fig:lum_mdot} (see also Table~\ref{tbl:models_std}), red are
results of \cite{Murphy2008}, green of \cite{Nordhaus2010}, and the
three additional curves (dark blue, light blue, and pink) correspond to
different sets of simulations that we performed with the deleptonization
treatment of \cite{Liebendorfer2005} for the core-collapse phase and 
the different electron-fraction trajectories of Fig.~\ref{fig:delep} in our
effort to reproduce the 1D results of \cite{Murphy2008} and 
\cite{Nordhaus2010}.}
\label{fig:lum_mdot_cmp}
\end{figure}

\begin{figure}
\plotone{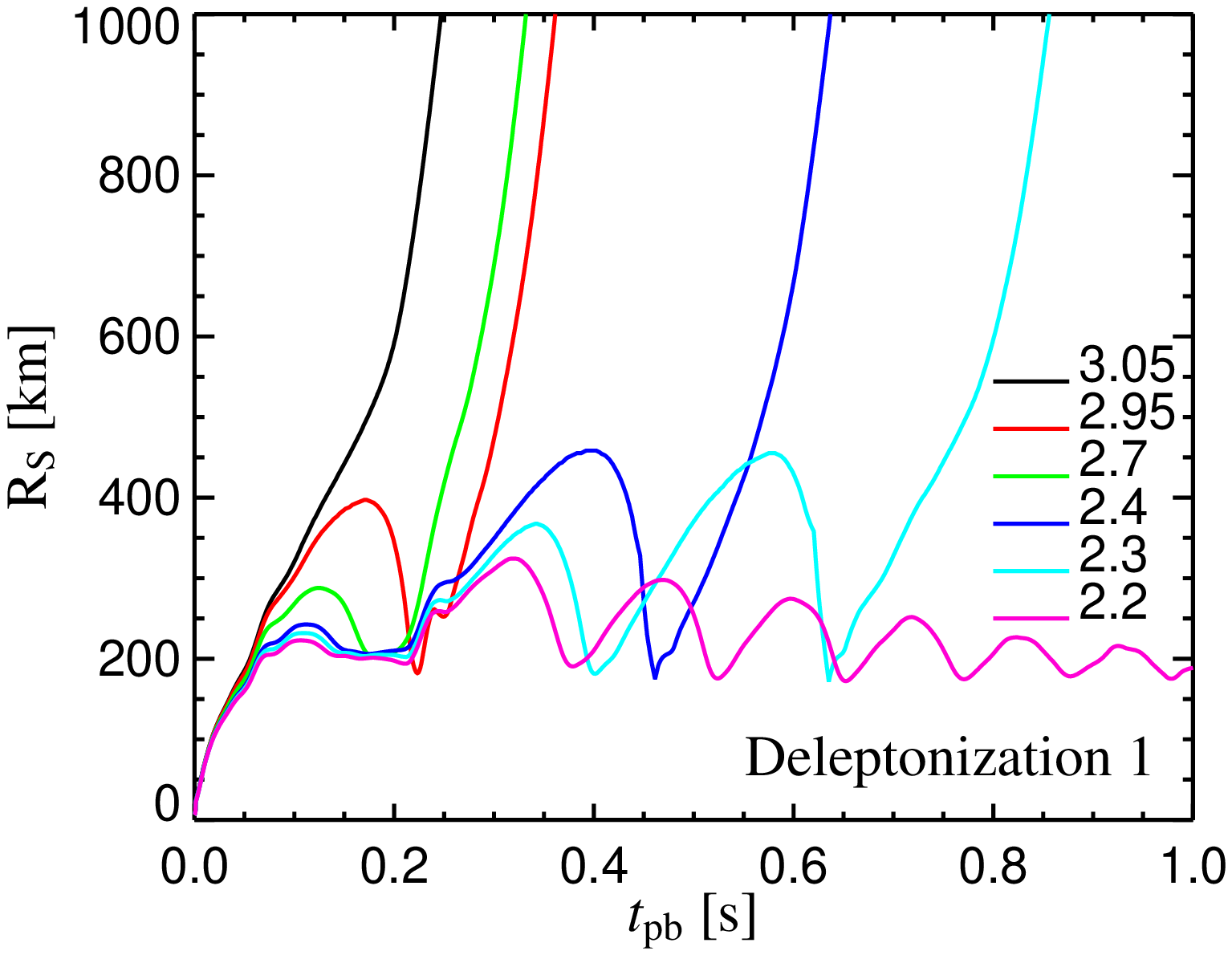}
\caption{Time evolution of the shock radius as a function of 
post-bounce time, $t_\mathrm{pb}$, for 1D simulations performed with 
the deleptonization scheme of \cite{Liebendorfer2005} for electron-fraction
trajectory ``Deleptonization~1'' of Fig.~\ref{fig:delep}. The colors 
correspond to different electron-neutrino luminosities, which are labeled 
in the plot in units of $10^{52}$\,erg\,s$^{-1}$.}
\label{fig:spos_delep}
\end{figure}

\section{Simulations with parametrized deleptonization treatment 
for the core-collapse phase}
\label{sec:appendix}

In this Appendix we briefly report on our efforts to reproduce the
1D results of \cite{Murphy2008} and \cite{Nordhaus2010} for
the critical explosion conditions of the 15\,$M_\odot$ progenitor,
applying a neutrino treatment that was intended to copy the
procedure outlined in these publications as closely as possible.

For this purpose we retained the exponential suppression factor
$e^{-\tau_\mathrm{eff}}$ of Eqs.~(\ref{eq:heat}) and 
(\ref{eq:cool})
without a reduction factor of 2.7 in the exponent, and the lepton 
evolution before and after core bounce was described by employing
a predefined $Y_e(\rho)$ relation. We also aimed at reproducing
the core infall of the previous works as closely as possible,
because the density structure of the infall region ahead of the
stalled shock determines the mass-infall rate $\dot M(t)$ at the
shock, and some differences became visible when we compared our
values with those given by \cite{Murphy2008} and \cite{Nordhaus2010}
(Fig.~\ref{fig:mdot_time_cmp}).
We therefore recomputed the collapse phase from the onset of
gravitational instability of the progenitor core through core 
bounce with the deleptonization scheme of \cite{Liebendorfer2005}.
Entropy changes were taken into account as suggested by 
\cite{Liebendorfer2005}, but were switched off after core bounce 
following \cite{Murphy2008} and \cite{Nordhaus2010}. 

We tested three different cases for the functional relation
$Y_e(\rho)$: First, we used a tabulated result for the $Y_e(\rho)$
evolution as obtained with the {\sc Prometheus-Vertex} code and
state-of-the-art electron-capture rates \citep{Langanke2003} 
(``Deleptonization~1''). Second, we applied a $Y_e(\rho)$-table
provided by Christian Ott as a co-developer of the {\sc CoCoNuT} code
(http://www.mpa-garching.mpg.de/hydro/COCONUT/). These data are based
on collapse simulations with the {\sc Vulcan/2D} code \citep{Livne2004}
(``Deleptonization~2''). Third, we employed a fitting formula given 
by \cite{Liebendorfer2005} for the parameters of model G15 
(``Deleptonization~3''). All three $Y_e(\rho)$ trajectories are 
depicted in Fig.~\ref{fig:delep}.

The three sets of 1D simulations conducted for this Appendix 
were performed with 800 radial zones.
The corresponding critical luminosity curves $L_{\nu_e}(\dot M)$
are displayed in Fig.~\ref{fig:lum_mdot_cmp} in comparison to those
of \cite{Murphy2008} and \cite{Nordhaus2010} and to our
results of Fig.~\ref{fig:lum_mdot} (for 400 zones, because for this
resolution the calibration of the exponential suppression factor for
best agreement with the critical curve of \cite{Murphy2008} was done).
The overall slopes of all three curves are similar but none of 
them is quantitatively or qualitatively in good agreement with
those of \cite{Murphy2008} and \cite{Nordhaus2010}. Explosions in 
our simulations occurred significantly more readily (i.e., for lower
$L_{\nu_e}$) than in the previous works. This suggests less cooling
in our runs, although we made all possible efforts to exactly follow
the description of the neutrino treatment in those papers.
The steep rise and partly backward bending of our curves for $\dot M$
values around 0.2--0.3\,$M_\odot$\,s$^{-1}$ can be understood by
an inspection of Fig.~\ref{fig:spos_delep}, which shows the 
time evolution of the shock radius for simulations with prescription
``Deleptonization~1'' for a selection of neutrino luminosity values.
One can see that in the case of
$L_{\nu_e} = 2.95\cdot 10^{52}$\,erg\,s$^{-1}$ the shock makes
a larger excursion before it returns again. Its reexpansion, leading
to an explosion, therefore happens later than in the model with 
$L_{\nu_e} = 2.7\cdot 10^{52}$\,erg\,s$^{-1}$, where the first
shock expansion is much less strong. Correspondingly, the explosion
in the former case sets in at a later time and lower mass accretion
rate than in the latter case, explaining the backward bending of
$L_{\nu_e}(\dot M)$ in this regime of luminosities and $\dot M$.
The nearly horizontal parts of the critical curves can be understood
by the fact that for such high values of the luminosities the 
neutrino cooling (with the unmodified $e^{-\tau_\mathrm{eff}}$
suppression factor) is so weak that the explosion sets in very
early (see the black line in Fig.~\ref{fig:spos_delep}) and
therefore for large values of the mass accretion rate.
The region between $\dot M \approx 0.3\,M_\odot$\,s$^{-3}$ and 
$\dot M \approx 0.8\,M_\odot$\,s$^{-3}$ is difficult to probe
with a stepwise increase of $L_{\nu_e}$, because the mass-accretion
rate there changes so rapidly that the shock shows time-dependent
dynamics instead of settling into a quasi-steady state.

None of the critical curves obtained with the direct
implementation of the neutrino treatment described in 
\cite{Murphy2008} and \cite{Nordhaus2010} can reproduce the
critical luminosity curves reported in these papers reasonably
well. We therefore decided to proceed with the modifications
described in Sect.~\ref{sec:num}.

\clearpage




\end{document}